\documentclass[longauth]{aa}
\usepackage{threeparttable}
\usepackage{graphicx}
\usepackage{amsmath}
\usepackage{amssymb}
\usepackage{multicol}
\usepackage{booktabs}
\usepackage{txfonts}
\usepackage{url}
\usepackage{xcolor}
\usepackage{longtable} 
\usepackage{pdflscape} 
\usepackage[breaklinks=true]{hyperref} 
\bibpunct{(}{)}{;}{a}{}{,}             

\newcommand{\gaia}{{Gaia}\xspace}

\newcommand{\tess}{{TESS}\xspace}
\newcommand{\cheops}{{CHEOPS}\xspace}

\newcommand{\kepler}{{Kepler}\xspace}

\newcommand{\wasp}{{WASP}\xspace}
\newcommand{\harps}{{HARPS}\xspace}
\newcommand{\soar}{{SOAR}\xspace}

\newcommand{\terra}{\texttt{TERRA}\xspace}

\newcommand{\sname}{{TOI-1203}\xspace}

\newcommand{\cms}{$\mathrm{cm\,s^{-1}}$}
\newcommand{\ms}{$\mathrm{m\,s^{-1}}$}
\newcommand{\kms}{$\mathrm{km\,s^{-1}}$}
\newcommand{\gccm}{$\mathrm{g\,cm^{-3}}$}

\newcommand\vsini{$v$\,sin\,$i_\star$}

\newcommand\vmic{$v_{\rm mic}$}
\newcommand\vmac{$v_{\rm mac}$}

\newcommand\teff{$T_{\rm eff}$}

\newcommand\logg{log\,{\it g$_\star$}}

\newcommand{\logrhk}{$\log\,\mathrm{R^\prime_{HK}}$}

\newcommand{\Tzerob}[1][days] {$58545.8916_{-0.0012}^{+0.0013}$~#1}
\newcommand{\Pb}[1][days] {$4.1572888_{-0.0000049}^{+0.0000050}$~#1}
\newcommand{\esinb}[1][ ] {$0.16_{-0.22}^{+0.16}$~#1}
\newcommand{\ecosb}[1][ ] {$-0.10_{-0.12}^{+0.15}$~#1}
\newcommand{\bb}[1][ ] {$0.847_{-0.029}^{+0.017}$~#1}

\newcommand{\rrb}[1][ ] {$0.01182_{-0.00034}^{+0.00033}$~#1}
\newcommand{\kb}[1][${\rm m\,s^{-1}}$] {$1.51\pm0.13$~#1}
\newcommand{\mpb}[1][$M_{\oplus}$] {$3.51_{-0.32}^{+0.33}$~#1}
\newcommand{\rpb}[1][$R_{\oplus}$] {$1.520_{-0.046}^{+0.045}$~#1}
\newcommand{\Tperib}[1][days] {$58546.16_{-0.42}^{+0.76}$~#1}
\newcommand{\eb}[1][ ] {$0.067_{-0.045}^{+0.063}$~#1}
\newcommand{\wb}[1][deg] {$125.4_{-39.0}^{+84.7}$~#1}
\newcommand{\ib}[1][deg] {$84.29_{-0.32}^{+0.24}$~#1}
\newcommand{\arb}[1][ ] {$8.88\pm0.15$~#1}
\newcommand{\ab}[1][AU] {$0.04869_{-0.00095}^{+0.00094}$~#1}

\newcommand{\insolationb}[1][${\rm F_{\oplus}}$] {$572.2_{-30.3}^{+32.3}$~#1}
\newcommand{\denstrb}[1][${\rm g\,cm^{-3}}$] {$0.767\pm0.039$~#1}

\newcommand{\Teqb}[1][K] {$1361.2_{-18.4}^{+18.8}$~#1}
\newcommand{\ttotb}[1][hours] {$1.911\pm0.027$~#1}
\newcommand{\tfulb}[1][hours] {$1.758\pm0.031$~#1}

\newcommand{\denpb}[1][${\rm g\,cm^{-3}}$] {$5.48_{-0.67}^{+0.74}$~#1}

\newcommand{\Tzeroc}[1][days] {$59198.81_{-0.32}^{+0.36}$~#1}
\newcommand{\Pc}[1][days] {$13.0766_{-0.0072}^{+0.0079}$~#1}
\newcommand{\esinc}[1][ ] {$0.03_{-0.22}^{+0.21}$~#1}
\newcommand{\ecosc}[1][ ] {$-0.20_{-0.16}^{+0.22}$~#1}
\newcommand{\kc}[1][${\rm m\,s^{-1}}$] {$1.62\pm0.14$~#1}
\newcommand{\mpc}[1][$M_{\oplus}$] {$5.46_{-0.50}^{+0.51}$~#1}
\newcommand{\Tperic}[1][days] {$59201.10_{-2.92}^{+1.98}$~#1}
\newcommand{\ec}[1][ ] {$0.092_{-0.063}^{+0.081}$~#1}
\newcommand{\wc}[1][deg] {$172.8_{-59.6}^{+68.7}$~#1}

\newcommand{\Tzerod}[1][days] {$58553.0711\pm0.0012$~#1}
\newcommand{\Pd}[1][days] {$25.502672_{-0.000028}^{+0.000031}$~#1}
\newcommand{\esind}[1][ ] {$-0.01_{-0.17}^{+0.16}$~#1}
\newcommand{\ecosd}[1][ ] {$-0.03_{-0.12}^{+0.13}$~#1}
\newcommand{\bd}[1][ ] {$0.407_{-0.088}^{+0.077}$~#1}
\newcommand{\rrd}[1][ ] {$0.02269\pm0.00028$~#1}
\newcommand{\kd}[1][${\rm m\,s^{-1}}$] {$1.74\pm0.014$~#1}
\newcommand{\mpd}[1][$M_{\oplus}$] {$7.39\pm0.62$~#1}
\newcommand{\rpd}[1][$R_{\oplus}$] {$2.918_{-0.045}^{+0.046}$~#1}
\newcommand{\Tperid}[1][days] {$58554.96_{-9.84}^{+7.90}$~#1}
\newcommand{\ed}[1][ ] {$0.032_{-0.023}^{+0.034}$~#1}
\newcommand{\wdd}[1][deg] {$186.4_{-113.7}^{+94.3}$~#1}
\newcommand{\id}[1][deg] {$89.22_{-0.13}^{+0.16}$~#1}
\newcommand{\ard}[1][ ] {$29.77_{-0.51}^{+0.50}$~#1}
\newcommand{\ad}[1][AU] {$0.1632_{-0.0032}^{+0.0031}$~#1}

\newcommand{\insolationd}[1][${\rm F_{\oplus}}$] {$51.0_{-2.7}^{+2.9}$~#1}

\newcommand{\Teqd}[1][K] {$743.6_{-10.1}^{+10.3}$~#1}
\newcommand{\ttotd}[1][hours] {$6.153_{-0.048}^{+0.051}$~#1}
\newcommand{\tfuld}[1][hours] {$5.824_{-0.054}^{+0.056}$~#1}

\newcommand{\denpd}[1][${\rm g\,cm^{-3}}$] {$1.63_{-0.15}^{+0.16}$~#1}

\newcommand{\Tzeroe}[1][days] {$58915.45_{-2.07}^{+2.00}$~#1}
\newcommand{\Pe}[1][days] {$204.59_{-0.63}^{+0.67}$~#1}
\newcommand{\esine}[1][ ] {$-0.066_{-0.072}^{+0.071}$~#1}
\newcommand{\ecose}[1][ ] {$-0.379_{-0.038}^{+0.044}$~#1}
\newcommand{\ke}[1][${\rm m\,s^{-1}}$] {$5.02_{-0.16}^{+0.17}$~#1}
\newcommand{\mpe}[1][$M_{\oplus}$] {$42.10_{-1.78}^{+1.83}$~#1}
\newcommand{\Tperie}[1][days] {$58962.29_{-5.89}^{+6.38}$~#1}
\newcommand{\ee}[1][ ] {$0.152\pm0.029$}
\newcommand{\we}[1][deg] {$189.7_{-10.8}^{+11.4}$~#1}

\newcommand{\qoneTESS}[1][] {$0.27\pm0.08$~#1}
\newcommand{\qtwoTESS}[1][] {$0.22_{-0.10}^{+0.09}$~#1}

\newcommand{\qoneCHEOPS}[1][] {$0.45_{-0.07}^{+0.08}$~#1}
\newcommand{\qtwoCHEOPS}[1][] {$0.32\pm0.09$~#1}

\newcommand{\gammaHARPS}[1][\ms] {$0.11\pm0.11$~#1}
\newcommand{\jHARPS}[1][${\rm m\,s^{-1}}$] {$0.83\pm0.10$~#1}

\definecolor{BlueGreen}{rgb}{0.0, 0.87, 0.87}

\begin{document}

   \title{A four-planet system orbiting the old thick disk star TOI-1203\thanks{Based on observations performed with the 3.6\,m telescope at the European Southern Observatory (La Silla, Chile) under programs 1102.C-0923, 106.21TJ.001, and 60.A-9709. This study uses \cheops\ data observed as part of the Guaranteed Time Observervation (GTO) programs CH\_PR00024, CH\_PR00031, and CH\_PR00045.}}



\author{
D.~Gandolfi\inst{\ref{inst:1},\thanks{Corresponding author: \href{mailto:davide.gandolfi@unito.it}{davide.gandolfi@unito.it}.}}\,$^{\href{https://orcid.org/0000-0001-8627-9628}{\protect\includegraphics[height=0.22cm]{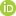}}}$\and 
A.~Alnajjarine\inst{\ref{inst:2}}\,$^{\href{https://orcid.org/0009-0005-5366-3371}{\protect\includegraphics[height=0.22cm]{figures/orcid.jpg}}}$\and 
L.~M.~Serrano\inst{\ref{inst:1}}\,$^{\href{https://orcid.org/0000-0001-9211-3691}{\protect\includegraphics[height=0.22cm]{figures/orcid.jpg}}}$\and 
J.~A.~Egger\inst{\ref{inst:3}}\,$^{\href{https://orcid.org/0000-0003-1628-4231}{\protect\includegraphics[height=0.22cm]{figures/orcid.jpg}}}$\and 
K.~W.~F.~Lam\inst{\ref{inst:4}}\,$^{\href{https://orcid.org/0000-0002-9910-6088}{\protect\includegraphics[height=0.22cm]{figures/orcid.jpg}}}$\and 
J.~Cabrera\inst{\ref{inst:4}}\,$^{\href{https://orcid.org/0000-0001-6653-5487}{\protect\includegraphics[height=0.22cm]{figures/orcid.jpg}}}$\and 
A.~P.~Hatzes\inst{\ref{inst:5}}\,$^{\href{https://orcid.org/0000-0002-3404-8358}{\protect\includegraphics[height=0.22cm]{figures/orcid.jpg}}}$\and 
M.~Fridlund\inst{\ref{inst:6},\ref{inst:7}}\,$^{\href{https://orcid.org/0000-0002-0855-8426}{\protect\includegraphics[height=0.22cm]{figures/orcid.jpg}}}$\and 
M.~Garbaccio~Gili\inst{\ref{inst:1}}\,$^{\href{https://orcid.org/0009-0006-8094-1362}{\protect\includegraphics[height=0.22cm]{figures/orcid.jpg}}}$\and 
T.~G.~Wilson\inst{\ref{inst:8}}\,$^{\href{https://orcid.org/0000-0001-8749-1962}{\protect\includegraphics[height=0.22cm]{figures/orcid.jpg}}}$\and 
W.~D.~Cochran\inst{\ref{inst:9},\ref{inst:10}}\,$^{\href{https://orcid.org/0000-0001-9662-3496}{\protect\includegraphics[height=0.22cm]{figures/orcid.jpg}}}$\and 
A.~Brandeker\inst{\ref{inst:11}}\,$^{\href{https://orcid.org/0000-0002-7201-7536}{\protect\includegraphics[height=0.22cm]{figures/orcid.jpg}}}$\and 
E.~Goffo\inst{\ref{inst:1}}\,$^{\href{https://orcid.org/0000-0001-9670-961X}{\protect\includegraphics[height=0.22cm]{figures/orcid.jpg}}}$\and 
S.~G.~Sousa\inst{\ref{inst:12}}\,$^{\href{https://orcid.org/0000-0001-9047-2965}{\protect\includegraphics[height=0.22cm]{figures/orcid.jpg}}}$\and 
G.~Nowak\inst{\ref{inst:13}}\,$^{\href{https://orcid.org/0000-0002-7031-7754}{\protect\includegraphics[height=0.22cm]{figures/orcid.jpg}}}$\and 
A.~Heitzmann\inst{\ref{inst:14}}\,$^{\href{https://orcid.org/0000-0002-8091-7526}{\protect\includegraphics[height=0.22cm]{figures/orcid.jpg}}}$\and 
C.~Hellier\inst{\ref{inst:15}}\,$^{\href{https://orcid.org/0000-0002-3439-1439 }{\protect\includegraphics[height=0.22cm]{figures/orcid.jpg}}}$\and 
J.~Venturini\inst{\ref{inst:14}}\,$^{\href{https://orcid.org/0000-0001-9527-2903}{\protect\includegraphics[height=0.22cm]{figures/orcid.jpg}}}$\and 
J.~Livingston\inst{\ref{inst:16},\ref{inst:17},\ref{inst:18}}\,$^{\href{https://orcid.org/0000-0002-4881-3620}{\protect\includegraphics[height=0.22cm]{figures/orcid.jpg}}}$\and 
A.~Bonfanti\inst{\ref{inst:19}}\,$^{\href{https://orcid.org/0000-0002-1916-5935}{\protect\includegraphics[height=0.22cm]{figures/orcid.jpg}}}$\and 
O.~Barragán\inst{\ref{inst:20}}\,$^{\href{https://orcid.org/0000-0003-0563-0493}{\protect\includegraphics[height=0.22cm]{figures/orcid.jpg}}}$\and 
V.~Adibekyan\inst{\ref{inst:12}}\,$^{\href{https://orcid.org/0000-0002-0601-6199}{\protect\includegraphics[height=0.22cm]{figures/orcid.jpg}}}$\and 
E.~Knudstrup\inst{\ref{inst:21},\ref{inst:22}}\,$^{\href{https://orcid.org/0000-0001-7880-594X}{\protect\includegraphics[height=0.22cm]{figures/orcid.jpg}}}$\and 
Y.~Alibert\inst{\ref{inst:23},\ref{inst:3}}\,$^{\href{https://orcid.org/0000-0002-4644-8818}{\protect\includegraphics[height=0.22cm]{figures/orcid.jpg}}}$\and 
S.~Grziwa\inst{\ref{inst:24}}\,$^{\href{https://orcid.org/0000-0003-3370-4058}{\protect\includegraphics[height=0.22cm]{figures/orcid.jpg}}}$\and 
C.~Ziegler\inst{\ref{inst:25}}\,$^{\href{https://orcid.org/0000-0002-0619-7639}{\protect\includegraphics[height=0.22cm]{figures/orcid.jpg}}}$\and 
L.~Fossati\inst{\ref{inst:19}}\,$^{\href{https://orcid.org/0000-0003-4426-9530}{\protect\includegraphics[height=0.22cm]{figures/orcid.jpg}}}$\and 
F.~Murgas\inst{\ref{inst:26},\ref{inst:42}}\,$^{\href{https://orcid.org/0000-0001-9087-1245}{\protect\includegraphics[height=0.22cm]{figures/orcid.jpg}}}$\and 
A.~C.~M.~Correia\inst{\ref{inst:27}}\,$^{\href{https://orcid.org/0000-0002-8946-8579}{\protect\includegraphics[height=0.22cm]{figures/orcid.jpg}}}$\and 
S.~H.~Albrecht\inst{\ref{inst:22}}\,$^{\href{https://orcid.org/0000-0003-1762-8235}{\protect\includegraphics[height=0.22cm]{figures/orcid.jpg}}}$\and 
J.~Laskar\inst{\ref{inst:2}}\,$^{\href{https://orcid.org/0000-0003-2634-789X}{\protect\includegraphics[height=0.22cm]{figures/orcid.jpg}}}$\and 
E.~W.~Guenther\inst{\ref{inst:5}}\,$^{\href{https://orcid.org/0000-0002-9130-6747}{\protect\includegraphics[height=0.22cm]{figures/orcid.jpg}}}$\and 
S.~J.A.J.~Salmon\inst{\ref{inst:28}}\and 
S.~Redfield\inst{\ref{inst:29}}\,$^{\href{https://orcid.org/0000-0003-3786-3486}{\protect\includegraphics[height=0.22cm]{figures/orcid.jpg}}}$\and 
N.~Billot\inst{\ref{inst:14}}\,$^{\href{https://orcid.org/0000-0003-3429-3836}{\protect\includegraphics[height=0.22cm]{figures/orcid.jpg}}}$\and 
H.~J.~Deeg\inst{\ref{inst:26},\ref{inst:42}}\,$^{\href{https://orcid.org/0000-0003-0047-4241}{\protect\includegraphics[height=0.22cm]{figures/orcid.jpg}}}$\and 
L.~Delrez\inst{\ref{inst:32},\ref{inst:33},\ref{inst:34}}\,$^{\href{https://orcid.org/0000-0001-6108-4808}{\protect\includegraphics[height=0.22cm]{figures/orcid.jpg}}}$\and 
L.~Palethorpe\inst{\ref{inst:35},\ref{inst:36},\ref{inst:37}}\,$^{\href{https://orcid.org/0000-0002-1664-4105}{\protect\includegraphics[height=0.22cm]{figures/orcid.jpg}}}$\and 
V.~Van~Eylen\inst{\ref{inst:37}}\,$^{\href{https://orcid.org/0000-0001-5542-8870}{\protect\includegraphics[height=0.22cm]{figures/orcid.jpg}}}$\and 
F.~Rodler\inst{\ref{inst:38}}\,$^{\href{https://orcid.org/0000-0003-0650-5723}{\protect\includegraphics[height=0.22cm]{figures/orcid.jpg}}}$\and 
J.~Alarcon\inst{\ref{inst:38}}\and 
J.~M.~Jenkins\inst{\ref{inst:39}}\,$^{\href{https://orcid.org/0000-0002-4715-9460}{\protect\includegraphics[height=0.22cm]{figures/orcid.jpg}}}$\and 
J.~D.~Twicken\inst{\ref{inst:40},\ref{inst:39}}\,$^{\href{https://orcid.org/0000-0002-6778-7552}{\protect\includegraphics[height=0.22cm]{figures/orcid.jpg}}}$\and 
A.~W.~Mann\inst{\ref{inst:41}}\,$^{\href{https://orcid.org/0000-0003-3654-1602}{\protect\includegraphics[height=0.22cm]{figures/orcid.jpg}}}$\and 
R.~Alonso\inst{\ref{inst:26},\ref{inst:42}}\,$^{\href{https://orcid.org/0000-0001-8462-8126}{\protect\includegraphics[height=0.22cm]{figures/orcid.jpg}}}$\and 
J.~Asquier\inst{\ref{inst:43}}\and 
T.~Bárczy\inst{\ref{inst:44}}\,$^{\href{https://orcid.org/0000-0002-7822-4413}{\protect\includegraphics[height=0.22cm]{figures/orcid.jpg}}}$\and 
D.~Barrado\inst{\ref{inst:45}}\,$^{\href{https://orcid.org/0000-0002-5971-9242}{\protect\includegraphics[height=0.22cm]{figures/orcid.jpg}}}$\and 
S.~C.~C.~Barros\inst{\ref{inst:12},\ref{inst:46}}\,$^{\href{https://orcid.org/0000-0003-2434-3625}{\protect\includegraphics[height=0.22cm]{figures/orcid.jpg}}}$\and 
W.~Baumjohann\inst{\ref{inst:19}}\,$^{\href{https://orcid.org/0000-0001-6271-0110}{\protect\includegraphics[height=0.22cm]{figures/orcid.jpg}}}$\and 
W.~Benz\inst{\ref{inst:3},\ref{inst:23}}\,$^{\href{https://orcid.org/0000-0001-7896-6479}{\protect\includegraphics[height=0.22cm]{figures/orcid.jpg}}}$\and 
L.~Borsato\inst{\ref{inst:47}}\,$^{\href{https://orcid.org/0000-0003-0066-9268}{\protect\includegraphics[height=0.22cm]{figures/orcid.jpg}}}$\and 
C.~Broeg\inst{\ref{inst:3},\ref{inst:23}}\,$^{\href{https://orcid.org/0000-0001-5132-2614}{\protect\includegraphics[height=0.22cm]{figures/orcid.jpg}}}$\and 
M.~Buder\inst{\ref{inst:48}}\,$^{\href{https://orcid.org/0009-0006-9440-236X}{\protect\includegraphics[height=0.22cm]{figures/orcid.jpg}}}$\and 
P.~Chaturvedi\inst{\ref{inst:49}}\,$^{\href{https://orcid.org/0000-0002-1887-1192}{\protect\includegraphics[height=0.22cm]{figures/orcid.jpg}}}$\and 
A.~Collier~Cameron\inst{\ref{inst:50}}\,$^{\href{https://orcid.org/0000-0002-8863-7828}{\protect\includegraphics[height=0.22cm]{figures/orcid.jpg}}}$\and 
Sz.~Csizmadia\inst{\ref{inst:4}}\,$^{\href{https://orcid.org/0000-0001-6803-9698}{\protect\includegraphics[height=0.22cm]{figures/orcid.jpg}}}$\and 
P.~E.~Cubillos\inst{\ref{inst:19},\ref{inst:51}}\and 
M.~B.~Davies\inst{\ref{inst:52}}\,$^{\href{https://orcid.org/0000-0001-6080-1190}{\protect\includegraphics[height=0.22cm]{figures/orcid.jpg}}}$\and 
M.~Deleuil\inst{\ref{inst:53}}\,$^{\href{https://orcid.org/0000-0001-6036-0225}{\protect\includegraphics[height=0.22cm]{figures/orcid.jpg}}}$\and 
A.~Deline\inst{\ref{inst:14}}\and 
O.~D.~S.~Demangeon\inst{\ref{inst:12},\ref{inst:46}}\,$^{\href{https://orcid.org/0000-0001-7918-0355}{\protect\includegraphics[height=0.22cm]{figures/orcid.jpg}}}$\and 
B.-O.~Demory\inst{\ref{inst:23},\ref{inst:3}}\,$^{\href{https://orcid.org/0000-0002-9355-5165}{\protect\includegraphics[height=0.22cm]{figures/orcid.jpg}}}$\and 
A.~Derekas\inst{\ref{inst:54}}\and 
B.~Edwards\inst{\ref{inst:55}}\and 
D.~Ehrenreich\inst{\ref{inst:14},\ref{inst:56}}\,$^{\href{https://orcid.org/0000-0001-9704-5405}{\protect\includegraphics[height=0.22cm]{figures/orcid.jpg}}}$\and 
A.~Erikson\inst{\ref{inst:4}}\and 
A.~Fortier\inst{\ref{inst:3},\ref{inst:23}}\,$^{\href{https://orcid.org/0000-0001-8450-3374}{\protect\includegraphics[height=0.22cm]{figures/orcid.jpg}}}$\and 
K.~Gazeas\inst{\ref{inst:57}}\,$^{\href{https://orcid.org/0000-0002-8855-3923}{\protect\includegraphics[height=0.22cm]{figures/orcid.jpg}}}$\and 
M.~Gillon\inst{\ref{inst:32}}\,$^{\href{https://orcid.org/0000-0003-1462-7739}{\protect\includegraphics[height=0.22cm]{figures/orcid.jpg}}}$\and 
M.~Güdel\inst{\ref{inst:58}}\,$^{\href{https://orcid.org/0000-0001-9818-0588}{\protect\includegraphics[height=0.22cm]{figures/orcid.jpg}}}$\and 
M.~N.~Günther\inst{\ref{inst:43}}\,$^{\href{https://orcid.org/0000-0002-3164-9086}{\protect\includegraphics[height=0.22cm]{figures/orcid.jpg}}}$\and 
Ch.~Helling\inst{\ref{inst:19},\ref{inst:59}}\and 
K.~G.~Isaak\inst{\ref{inst:43}}\,$^{\href{https://orcid.org/0000-0001-8585-1717}{\protect\includegraphics[height=0.22cm]{figures/orcid.jpg}}}$\and 
L.~L.~Kiss\inst{\ref{inst:60},\ref{inst:61}}\and 
J.~Korth\inst{\ref{inst:14},\ref{inst:62}}\,$^{\href{https://orcid.org/0000-0002-0076-6239}{\protect\includegraphics[height=0.22cm]{figures/orcid.jpg}}}$\and 
N.~Law\inst{\ref{inst:41}}\,$^{\href{https://orcid.org/0000-0001-9380-6457}{\protect\includegraphics[height=0.22cm]{figures/orcid.jpg}}}$\and 
A.~Lecavelier~des~Etangs\inst{\ref{inst:63}}\,$^{\href{https://orcid.org/0000-0002-5637-5253}{\protect\includegraphics[height=0.22cm]{figures/orcid.jpg}}}$\and 
A.~Leleu\inst{\ref{inst:14},\ref{inst:3}}\,$^{\href{https://orcid.org/0000-0003-2051-7974}{\protect\includegraphics[height=0.22cm]{figures/orcid.jpg}}}$\and 
M.~Lendl\inst{\ref{inst:14}}\,$^{\href{https://orcid.org/0000-0001-9699-1459}{\protect\includegraphics[height=0.22cm]{figures/orcid.jpg}}}$\and 
P.~Leonardi\inst{\ref{inst:64},\ref{inst:71}}\,$^{\href{https://orcid.org/0000-0001-6026-9202}
{\protect\includegraphics[height=0.22cm]{figures/orcid.jpg}}}$
\and 
D.~Magrin\inst{\ref{inst:47}}\,$^{\href{https://orcid.org/0000-0003-0312-313X}{\protect\includegraphics[height=0.22cm]{figures/orcid.jpg}}}$\and 
G.~Mantovan\inst{\ref{inst:84},\ref{inst:47}}\,$^{\href{https://orcid.org/0000-0002-6871-6131}{\protect\includegraphics[height=0.22cm]{figures/orcid.jpg}}}$\and  
L.~Marafatto\inst{\ref{inst:47}}\,$^{\href{https://orcid.org/0000-0002-8822-6834}{\protect\includegraphics[height=0.22cm]{figures/orcid.jpg}}}$\and 
P.~F.~L.~Maxted\inst{\ref{inst:66}}\,$^{\href{https://orcid.org/0000-0003-3794-1317}{\protect\includegraphics[height=0.22cm]{figures/orcid.jpg}}}$\and 
M.~Mecina\inst{\ref{inst:58}}\,$^{\href{https://orcid.org/0000-0002-3258-7526}{\protect\includegraphics[height=0.22cm]{figures/orcid.jpg}}}$\and 
B.~Merín\inst{\ref{inst:67}}\,$^{\href{https://orcid.org/0000-0002-8555-3012}{\protect\includegraphics[height=0.22cm]{figures/orcid.jpg}}}$\and 
C.~Mordasini\inst{\ref{inst:3},\ref{inst:23}}\,$^{\href{https://orcid.org/0000-0002-1013-2811}{\protect\includegraphics[height=0.22cm]{figures/orcid.jpg}}}$\and 
V.~Nascimbeni\inst{\ref{inst:47}}\,$^{\href{https://orcid.org/0000-0001-9770-1214}{\protect\includegraphics[height=0.22cm]{figures/orcid.jpg}}}$\and 
A.~Nigioni\inst{\ref{inst:14}}\and 
G.~Olofsson\inst{\ref{inst:11}}\,$^{\href{https://orcid.org/0000-0003-3747-7120}{\protect\includegraphics[height=0.22cm]{figures/orcid.jpg}}}$\and 
H.~P.~Osborn\inst{\ref{inst:23},\ref{inst:68}}\,$^{\href{https://orcid.org/0000-0002-4047-4724}{\protect\includegraphics[height=0.22cm]{figures/orcid.jpg}}}$\and 
R.~Ottensamer\inst{\ref{inst:58}}\,$^{\href{https://orcid.org/0000-0001-5684-5836}{\protect\includegraphics[height=0.22cm]{figures/orcid.jpg}}}$\and 
I.~Pagano\inst{\ref{inst:69}}\,$^{\href{https://orcid.org/0000-0001-9573-4928}{\protect\includegraphics[height=0.22cm]{figures/orcid.jpg}}}$\and 
E.~Pallé\inst{\ref{inst:26},\ref{inst:42}}\,$^{\href{https://orcid.org/0000-0003-0987-1593}{\protect\includegraphics[height=0.22cm]{figures/orcid.jpg}}}$\and 
C.~M.~Persson\inst{\ref{inst:7}}\,$^{\href{https://orcid.org/0000-0003-1257-5146}{\protect\includegraphics[height=0.22cm]{figures/orcid.jpg}}}$\and 
G.~Peter\inst{\ref{inst:48}}\,$^{\href{https://orcid.org/0000-0001-6101-2513}{\protect\includegraphics[height=0.22cm]{figures/orcid.jpg}}}$\and 
D.~Piazza\inst{\ref{inst:4}}\and 
G.~Piotto\inst{\ref{inst:47},\ref{inst:71}}\,$^{\href{https://orcid.org/0000-0002-9937-6387}{\protect\includegraphics[height=0.22cm]{figures/orcid.jpg}}}$\and 
D.~Pollacco\inst{\ref{inst:8}}\,$^{\href{https://orcid.org/0000-0001-9850-9697}{\protect\includegraphics[height=0.22cm]{figures/orcid.jpg}}}$\and 
D.~Queloz\inst{\ref{inst:68},\ref{inst:72}}\,$^{\href{https://orcid.org/0000-0002-3012-0316}{\protect\includegraphics[height=0.22cm]{figures/orcid.jpg}}}$\and 
R.~Ragazzoni\inst{\ref{inst:47},\ref{inst:71}}\,$^{\href{https://orcid.org/0000-0002-7697-5555}{\protect\includegraphics[height=0.22cm]{figures/orcid.jpg}}}$\and 
N.~Rando\inst{\ref{inst:43}}\and 
H.~Rauer\inst{\ref{inst:73},\ref{inst:74}}\,$^{\href{https://orcid.org/0000-0002-6510-1828}{\protect\includegraphics[height=0.22cm]{figures/orcid.jpg}}}$\and 
I.~Ribas\inst{\ref{inst:75},\ref{inst:76}}\,$^{\href{https://orcid.org/0000-0002-6689-0312}{\protect\includegraphics[height=0.22cm]{figures/orcid.jpg}}}$\and 
G.~R.~Ricker\inst{\ref{inst:77}}\,$^{\href{https://orcid.org/0000-0003-2058-6662}{\protect\includegraphics[height=0.22cm]{figures/orcid.jpg}}}$\and 
N.~C.~Santos\inst{\ref{inst:12},\ref{inst:46}}\,$^{\href{https://orcid.org/0000-0003-4422-2919}{\protect\includegraphics[height=0.22cm]{figures/orcid.jpg}}}$\and 
G.~Scandariato\inst{\ref{inst:69}}\,$^{\href{https://orcid.org/0000-0003-2029-0626}{\protect\includegraphics[height=0.22cm]{figures/orcid.jpg}}}$\and 
S.~Seager\inst{\ref{inst:77},\ref{inst:78},\ref{inst:79}}\,$^{\href{https://orcid.org/0000-0002-6892-6948}{\protect\includegraphics[height=0.22cm]{figures/orcid.jpg}}}$\and 
D.~Ségransan\inst{\ref{inst:14}}\,$^{\href{https://orcid.org/0000-0003-2355-8034}{\protect\includegraphics[height=0.22cm]{figures/orcid.jpg}}}$\and 
A.~E.~Simon\inst{\ref{inst:3},\ref{inst:23}}\,$^{\href{https://orcid.org/0000-0001-9773-2600}{\protect\includegraphics[height=0.22cm]{figures/orcid.jpg}}}$\and 
A.~M.~S.~Smith\inst{\ref{inst:4}}\,$^{\href{https://orcid.org/0000-0002-2386-4341}{\protect\includegraphics[height=0.22cm]{figures/orcid.jpg}}}$\and 
M.~Stalport\inst{\ref{inst:33},\ref{inst:32}}\and 
S.~Sulis\inst{\ref{inst:53}}\,$^{\href{https://orcid.org/0000-0001-8783-526X}{\protect\includegraphics[height=0.22cm]{figures/orcid.jpg}}}$\and 
Gy.~M.~Szabó\inst{\ref{inst:54},\ref{inst:80}}\,$^{\href{https://orcid.org/0000-0002-0606-7930}{\protect\includegraphics[height=0.22cm]{figures/orcid.jpg}}}$\and 
S.~Udry\inst{\ref{inst:14}}\,$^{\href{https://orcid.org/0000-0001-7576-6236}{\protect\includegraphics[height=0.22cm]{figures/orcid.jpg}}}$\and 
S.~Ulmer-Moll\inst{\ref{inst:81},\ref{inst:33}}\,$^{\href{https://orcid.org/0000-0003-2417-7006}{\protect\includegraphics[height=0.22cm]{figures/orcid.jpg}}}$\and 
V.~Van~Grootel\inst{\ref{inst:33}}\,$^{\href{https://orcid.org/0000-0003-2144-4316}{\protect\includegraphics[height=0.22cm]{figures/orcid.jpg}}}$\and 
E.~Villaver\inst{\ref{inst:26},\ref{inst:42}}\,$^{\href{https://orcid.org/0000-0003-4936-9418}{\protect\includegraphics[height=0.22cm]{figures/orcid.jpg}}}$\and 
N.~A.~Walton\inst{\ref{inst:82}}\,$^{\href{https://orcid.org/0000-0003-3983-8778}{\protect\includegraphics[height=0.22cm]{figures/orcid.jpg}}}$\and 
J.~N.~Winn\inst{\ref{inst:83}}\,$^{\href{https://orcid.org/0000-0002-4265-047X}{\protect\includegraphics[height=0.22cm]{figures/orcid.jpg}}}$\and 
S.~Wolf\inst{\ref{inst:4}}\and 
T.~Zingales\inst{\ref{inst:71},\ref{inst:47}}\,$^{\href{https://orcid.org/0000-0001-6880-5356}{\protect\includegraphics[height=0.22cm]{figures/orcid.jpg}}}$
}

\institute{
\label{inst:1} Dipartimento di Fisica, Università degli Studi di Torino, via Pietro Giuria 1, I-10125, Torino, Italy \and
\label{inst:2} LTE, UMR8255 CNRS, Observatoire de Paris, PSL Univ., Sorbonne Univ., 77 av. Denfert-Rochereau, 75014 Paris, France \and
\label{inst:3} Space Research and Planetary Sciences, Physics Institute, University of Bern, Gesellschaftsstra{\ss}e 6, 3012 Bern, Switzerland \and
\label{inst:4} Institute of Space Research, German Aerospace Center (DLR), Rutherfordstra{\ss}e 2, 12489 Berlin, Germany \and
\label{inst:5} Th\"uringer Landessternwarte Tautenburg, 07778 Tautenburg, Germany \and
\label{inst:6} Leiden Observatory, University of Leiden, PO Box 9513, 2300 RA Leiden, The Netherlands \and
\label{inst:7} Department of Space, Earth and Environment, Chalmers University of Technology, Onsala Space Observatory, 439 92 Onsala, Sweden \and
\label{inst:8} Department of Physics, University of Warwick, Gibbet Hill Road, Coventry CV4 7AL, United Kingdom \and
\label{inst:9} McDonald Observatory and Department of Astronomy, The University of Texas at Austin, USA \and
\label{inst:10} Center for Planetary Systems Habitability, The University of Texas at Austin, USA \and
\label{inst:11} Department of Astronomy, Stockholm University, AlbaNova University Center, 10691 Stockholm, Sweden \and
\label{inst:12} Instituto de Astrofisica e Ciencias do Espaco, Universidade do Porto, CAUP, Rua das Estrelas, 4150-762 Porto, Portugal \and
\label{inst:13} Institute of Astronomy, Faculty of Physics, Astronomy and Informatics, Nicolaus Copernicus University, Grudzi\c{a}dzka 5, 87-100 Toru\'n, Poland \and
\label{inst:14} Observatoire astronomique de l'Université de Genève, Chemin Pegasi 51, 1290 Versoix, Switzerland \and
\label{inst:15} Astrophysics Group, Keele University, Staffordshire, ST5 5BG, UK \and
\label{inst:16} Astrobiology Center, NINS, 2-21-1 Osawa, Mitaka, Tokyo 181-8588, Japan \and
\label{inst:17} National Astronomical Observatory of Japan, NINS, 2-21-1 Osawa, Mitaka, Tokyo 181-8588, Japan \and
\label{inst:18} Astronomical Science Program, Graduate University for Advanced Studies, SOKENDAI, 2-21-1, Osawa, Mitaka, Tokyo, 181-8588, Japan \and
\label{inst:19} Space Research Institute, Austrian Academy of Sciences, Schmiedlstra{\ss}e 6, A-8042 Graz, Austria \and
\label{inst:20} Department of Physics, Sub-department of Astrophysics, University of Oxford, Oxford, OX1 3RH, UK \and
\label{inst:21} Department of Space, Earth and Environment, Chalmers University of Technology, 412 93 Gothenburg, Sweden \and
\label{inst:22} Stellar Astrophysics Centre, Department of Physics and Astronomy, Aarhus University, Ny Munkegade 120, 8000 Aarhus C, Denmark \and
\label{inst:23} Center for Space and Habitability, University of Bern, Gesellschaftsstra{\ss}e 6, 3012 Bern, Switzerland \and
\label{inst:24} Rheinisches Institut f\"ur Umweltforschung an der Universit\"at zu K\"oln, Aachener Stra{\ss}e 209, 50931 K\"oln, Germany \and
\label{inst:25} Department of Physics, Engineering and Astronomy, Stephen F. Austin State University, 1936 North St, Nacogdoches, TX 75962, USA \and
\label{inst:26} Instituto de Astrofísica de Canarias, Vía Láctea s/n, 38200 La Laguna, Tenerife, Spain \and
\label{inst:42} Departamento de Astrofísica, Universidad de La Laguna, Astrofísico Francisco Sanchez s/n, 38206 La Laguna, Tenerife, Spain \and
\label{inst:27} CFisUC, Departamento de Física, Universidade de Coimbra, 3004-516 Coimbra, Portugal \and
\label{inst:28} Observatoire Astronomique de l’Université de Genève, Chemin Pegasi 51, 1290 Versoix, Switzerland \and
\label{inst:29} Astronomy Department and Van Vleck Observatory, Wesleyan University, Middletown, CT 06459, USA \and
\label{inst:32} Astrobiology Research Unit, Université de Liège, Allée du 6 Août 19C, B-4000 Liège, Belgium \and
\label{inst:33} Space sciences, Technologies and Astrophysics Research (STAR) Institute, Université de Liège, Allée du 6 Août 19C, 4000 Liège, Belgium \and
\label{inst:34} Institute of Astronomy, KU Leuven, Celestijnenlaan 200D, 3001 Leuven, Belgium \and
\label{inst:35} Institute for Astronomy, University of Edinburgh, Royal Observatory, Blackford Hill, Edinburgh EH9 3HJ, UK \and
\label{inst:36} Centre for Exoplanet Science, University of Edinburgh, Edinburgh EH9 3HJ, UK \and
\label{inst:37} Mullard Space Science Laboratory, University College London, Holmbury St Mary, Dorking, Surrey RH5 6NT, UK \and
\label{inst:38} European Southern Observatory, Alonso de Cordova 3107, Vitacura, Santiago de Chile, Chile \and
\label{inst:39} NASA Ames Research Center, Moffett Field, CA 94035, USA \and
\label{inst:40} SETI Institute, Mountain View, CA  94043, USA \and
\label{inst:41} Department of Physics and Astronomy, The University of North Carolina at Chapel Hill, Chapel Hill, NC 27599-3255, USA \and
\label{inst:43} European Space Agency (ESA), European Space Research and Technology Centre (ESTEC), Keplerlaan 1, 2201 AZ Noordwijk, The Netherlands \and
\label{inst:44} Admatis, 5. Kandó Kálmán Street, 3534 Miskolc, Hungary \and
\label{inst:45} Departamento de Astrofísica, Centro de Astrobiología (CSIC-INTA), ESAC campus, 28692 Villanueva de la Cañada (Madrid), Spain \and
\label{inst:46} Departamento de Fisica e Astronomia, Faculdade de Ciencias, Universidade do Porto, Rua do Campo Alegre, 4169-007 Porto, Portugal \and
\label{inst:47} INAF, Osservatorio Astronomico di Padova, Vicolo dell'Osservatorio 5, 35122 Padova, Italy \and
\label{inst:48} Institute of Optical Sensor Systems, German Aerospace Center (DLR), Rutherfordstra{\ss}e 2, 12489 Berlin, Germany \and
\label{inst:49} Department of Astronomy and Astrophysics, Tata Institute of Fundamental Research, Mumbai, India, 400005 \and
\label{inst:50} Centre for Exoplanet Science, SUPA School of Physics and Astronomy, University of St Andrews, North Haugh, St Andrews KY16 9SS, UK \and
\label{inst:51} INAF, Osservatorio Astrofisico di Torino, Via Osservatorio, 20, I-10025 Pino Torinese To, Italy \and
\label{inst:52} Centre for Mathematical Sciences, Lund University, Box 118, 221 00 Lund, Sweden \and
\label{inst:53} Aix Marseille Univ, CNRS, CNES, LAM, 38 rue Frédéric Joliot-Curie, 13388 Marseille, France \and
\label{inst:54} ELTE Gothard Astrophysical Observatory, 9700 Szombathely, Szent Imre h. u. 112, Hungary \and
\label{inst:55} SRON Netherlands Institute for Space Research, Niels Bohrweg 4, 2333 CA Leiden, Netherlands \and
\label{inst:56} Centre Vie dans l’Univers, Faculté des sciences, Université de Genève, Quai Ernest-Ansermet 30, 1211 Genève 4, Switzerland \and
\label{inst:57} National and Kapodistrian University of Athens, Department of Physics, University Campus, Zografos GR-157 84, Athens, Greece \and
\label{inst:58} Department of Astrophysics, University of Vienna, T\"urkenschanzstra{\ss}e 17, 1180 Vienna, Austria \and
\label{inst:59} Institute for Theoretical Physics and Computational Physics, Graz University of Technology, Petersgasse 16, 8010 Graz, Austria \and
\label{inst:60} Konkoly Observatory, Research Centre for Astronomy and Earth Sciences, 1121 Budapest, Konkoly Thege Miklós út 15-17, Hungary \and
\label{inst:61} ELTE E\"otv\"os Lor\'and University, Institute of Physics, P\'azm\'any P\'eter s\'et\'any 1/A, 1117 Budapest, Hungary \and
\label{inst:62} Lund Observatory, Division of Astrophysics, Department of Physics, Lund University, Box 118, 22100 Lund, Sweden \and
\label{inst:63} Institut d'astrophysique de Paris, UMR7095 CNRS, Université Pierre \& Marie Curie, 98bis blvd. Arago, 75014 Paris, France \and
\label{inst:64} Dipartimento di Fisica, Università di Trento, Via Sommarive 14, 38123 Povo \and
\label{inst:71} Dipartimento di Fisica e Astronomia ``Galileo Galilei'', Università degli Studi di Padova, Vicolo dell'Osservatorio 3, 35122 Padova, Italy \and
\label{inst:84} Centro di Ateneo di Studi e Attività Spaziali ``G. Colombo'', Università degli Studi di Padova, Via Venezia 15, 35131 Padova, Italy \and
\label{inst:66} Astrophysics Group, Lennard Jones Building, Keele University, Staffordshire, ST5 5BG, United Kingdom \and
\label{inst:67} European Space Agency, ESA - European Space Astronomy Centre, Camino Bajo del Castillo s/n, 28692 Villanueva de la Cañada, Madrid, Spain \and
\label{inst:68} ETH Zurich, Department of Physics, Wolfgang-Pauli-Stra{\ss}e 2, CH-8093 Zurich, Switzerland \and
\label{inst:69} INAF, Osservatorio Astrofisico di Catania, Via S. Sofia 78, 95123 Catania, Italy \and
\label{inst:72} Cavendish Laboratory, JJ Thomson Avenue, Cambridge CB3 0HE, UK \and
\label{inst:73} Institut f\"ur Geologische Wissenschaften, Freie Universit\"at Berlin, Maltheserstra{\ss}e 74-100, 12249 Berlin, Germany \and
\label{inst:74} German Aerospace Center (DLR), Markgrafenstra{\ss}e 37, 10117 Berlin, Germany \and
\label{inst:75} Institut de Ciencies de l'Espai (ICE, CSIC), Campus UAB, Can Magrans s/n, 08193 Bellaterra, Spain \and
\label{inst:76} Institut d'Estudis Espacials de Catalunya (IEEC), 08860 Castelldefels (Barcelona), Spain \and
\label{inst:77} Department of Physics and Kavli Institute for Astrophysics and Space Research, Massachusetts Institute of Technology, 77 Massachusetts Avenu, Cambridge, MA 02139, USA \and
\label{inst:78} Department of Earth, Atmospheric and Planetary Sciences, Massachusetts Institute of Technology,  77 Massachusetts Avenu, Cambridge, MA 02139, USA \and
\label{inst:79} Department of Aeronautics and Astronautics, Massachusetts Institute of Technology, 77 Massachusetts Avenue, Cambridge, MA 02139, USA \and
\label{inst:80} HUN-REN-ELTE Exoplanet Research Group, Szent Imre h. u. 112., Szombathely, H-9700, Hungary \and
\label{inst:81} Leiden Observatory, University of Leiden, Einsteinweg 55, 2333 CA Leiden, The Netherlands \and
\label{inst:82} Institute of Astronomy, University of Cambridge, Madingley Road, Cambridge, CB3 0HA, United Kingdom \and
\label{inst:83} Department of Astrophysical Sciences, Princeton University, Princeton, NJ 08544, USA
}

\date{Received March 19, 2025; accepted July 29, 2025}

\abstract{TOI-1203 is a bright (V\,=\,8.6) G3\,V star known to host a transiting warm sub-Neptune on a 25.5-day orbit. Here we report on an intensive high-precision radial velocity and photometric follow-up campaign carried out with the \harps spectrograph and the \cheops space telescope. We found that TOI-1203 has an effective temperature of \teff\,=\,5737\,$\pm$\,62~K, a mass of $M_\star$\,=\,0.886\,$\pm$\,0.036~$M_\odot$, a radius of $R_\star$\,=\,1.179\,$\pm$\,0.011~$R_\odot$, and an enhancement of $\alpha$ elements relative to iron of [$\alpha$/Fe]\,=\,0.21\,$\pm$\,0.04. With an age of $\sim$12.5~Gyr, TOI-1203 belongs to the old, $\alpha$-element enhanced stellar population of the galactic thick disk. We spectroscopically confirmed the planetary nature of the 25.5-day sub-Neptune TOI-1203~d, measured its mass ($M_\mathrm{d}$\,=\,\mpd) and refined its radius ($R_\mathrm{d}$\,=\,\rpd). We discovered the presence of an additional transiting super-Earth on a 4.2-day orbit (TOI-1203~b) with a mass of $M_\mathrm{b}$\,=\,\mpb\ and a radius of $R_\mathrm{b}$\,=\,\rpb. We also revealed the presence of two additional low-mass planets at 13.1~d and 204.6~d (TOI-1203~c and e), with minimum masses of $M_\mathrm{c}\,\mathrm{sin\,i}_\mathrm{c}$\,=\,\mpc\ and $M_\mathrm{e}\,\mathrm{sin\,i}_\mathrm{e}$\,=\,\mpe. We found that the outer planet TOI-1203~e lies on an eccentric orbit with $e_\mathrm{e}$\,=\,\ee. We performed a stability analysis of the system confirming that there are configurations consistent with the observed parameters that are dynamically stable over billion-year timescales. While analyzing the \harps time series, we discovered that the full width at half maximum of the \harps cross-correlation function shows a significant long-period signal ($\sim$615~d) that has no counterpart in the radial velocity data or in the remaining \harps ancillary time series. We significantly detected the same signal in the full width at half maximum of the Th-Ar calibration lines used to compute the nightly wavelength solution, and attributed this systematic effect to a long-term variation of the \harps instrumental profile.}
   

\maketitle

\nolinenumbers
\section{Introduction}

\begin{table}[!ht]
\centering
\begin{threeparttable}
\caption{Main identifiers, equatorial coordinates, distance, systemic radial velocity, optical and near-infrared magnitudes, and fundamental parameters of the star \sname.}
\label{tab:stellar_parameters}
\begin{tabular}{lrr}
\hline
\hline
\noalign{\smallskip}
Parameter & Value & Source \\
\noalign{\smallskip}
\hline
\noalign{\smallskip}
\multicolumn{3}{l}{\it Main identifiers}  \\
\noalign{\smallskip}
\multicolumn{2}{l}{HD}{97507}                        & ExoFOP\tnote{a} \\
\multicolumn{2}{l}{TOI}{1203}                        & ExoFOP \\
\multicolumn{2}{l}{TIC}{23434737}                    & TIC v8\tnote{b} \\
\multicolumn{2}{l}{TYC}{7209-1497-1}                 & ExoFOP \\
\multicolumn{2}{l}{2MASS}{J11125403-3424242}         & ExoFOP \\
\multicolumn{2}{l}{\gaia\ DR3}{5402390257832925952}  & \gaia\ DR3\tnote{c} \\
\noalign{\smallskip}
\hline
\noalign{\smallskip}
\multicolumn{3}{l}{\it Equatorial coordinates, distance, and radial velocity}  \\
\noalign{\smallskip}
R.A. (J2000.0)	&    11$^\mathrm{h}$\,12$^\mathrm{m}$\,54.05$^\mathrm{s}$	& \gaia\ DR3 \\
Dec. (J2000.0)	& $-$34$\degr$\,24$\arcmin$\,24.28$\arcsec$	                & \gaia\ DR3 \\
Distance (pc)   & $64.96 \pm 0.07$                                          & \gaia\ DR3 \\
Radial Velocity (\kms) & $72.657 \pm 0.003$ & This work \\ 
\noalign{\smallskip}
\hline
\noalign{\smallskip}
\multicolumn{3}{l}{\it Optical and near-infrared photometry} \\
\noalign{\smallskip}
\tess\                  & $8.000\pm0.006$     & TIC v8 \\
\noalign{\smallskip}
$B$                     & $9.210\pm0.030$      & UBV Phot. Cat.\tnote{d} \\
$V$                     & $8.590\pm0.030$      & TIC v8 \\
\noalign{\smallskip}
$J$ 			&  $7.411\pm0.021$     & TIC v8 \\
$H$			    &  $7.102\pm0.038$     & TIC v8 \\
$Ks$			    &  $6.997\pm0.021$     & TIC v8 \\
\noalign{\smallskip}
$W1$			& $6.905\pm0.057$      & All{\it WISE}\tnote{e} \\
$W2$			& $7.025\pm0.020$      & All{\it WISE} \\
$W3$            & $7.036\pm0.017$      & All{\it WISE} \\
$W4$            & $7.006\pm0.094$      & All{\it WISE} \\
\noalign{\smallskip}
\hline
\noalign{\smallskip}
\multicolumn{3}{l}{\it Fundamental parameters} \\
\noalign{\smallskip}
\teff\ (K)           &  $5737 \pm 62$                & This work \\
\logg\ (cgs; spec)          &  $4.34 \pm 0.10$              & This work \\
\logg\ (cgs; \gaia)        &  $4.26 \pm 0.03$              & This work \\
$[\mathrm{Fe/H}]$              & $-0.39 \pm 0.04$              & This work \\
$[\mathrm{Mg/H}]$              & $-0.15 \pm 0.05$              & This work \\
$[\mathrm{Si/H}]$              & $-0.22 \pm 0.03$              & This work \\
$[\mathrm{Ti/H}]$              & $-0.16 \pm 0.03$              & This work \\
$[\mathrm{\alpha/Fe}]$         &  $0.21 \pm 0.04$              & This work \\
$[\mathrm{M/H}]$               & $-0.23 \pm 0.05$              & This work \\
\vmic\ (\kms)                   & $1.02 \pm 0.03$               & This work \\ 
\vsini\ (\kms)                  & $1.4 \pm 0.5$                 & This work \\
$M_\star$ ($M_\odot$)          & $0.886 \pm 0.036$             & This work \\
$R_\star$ ($R_\odot$)          &  $1.179 \pm 0.011$            & This work \\
\logrhk                        & $-4.998 \pm 0.001$            & This work\tnote{f} \\
\noalign{\smallskip}
$t_\star$ (Gyr)                & $12.5_{-2.4}^{+1.3}$          & This work \\
\noalign{\smallskip}
Spectral Type                  & G3\,V                         & This work\tnote{g} \\
\noalign{\smallskip}
\hline
\end{tabular}
\begin{tablenotes}
\item[a] \href{https://exofop.ipac.caltech.edu/tess}{https://exofop.ipac.caltech.edu/tess.}
\item[b] \citet{Stassun2018}.
\item[c] \citet{GaiaCollaboration_DR3_2023}.
\item[d] Derived using the color index $B-V$\,=\,0.620, as reported in \citet{Mermilliod1987}. 
\item[e] \citet{Wright2010}.
\item[f] Median value extracted from Table~\ref{tab:RV-Table}.
\item[g] Based on the spectral type vs. effective temperature calibration table available at \url{https://www.pas.rochester.edu/~emamajek/EEM_dwarf_UBVIJHK_colors_Teff.txt}. \citep{Pecaut2012,Pecaut2013}.
\end{tablenotes}
\end{threeparttable}
\end{table}

The Transiting Exoplanet Survey Satellite \citep[\tess,][]{Ricker2015} has ushered in a new era of investigations of the composition and internal structure of super-Earths ($1\,\lesssim\,R_\mathrm{p}\,\lesssim\,2~R_\oplus$) and sub-Neptunes ($2\,\lesssim\,R_\mathrm{p}\,\lesssim\,4~R_\oplus$) in an unprecedented manner. While searching the whole sky for planets orbiting bright stars (V\,$\lesssim$\,11), \tess\ is discovering thousands of transiting candidates amenable to mass determination via high-precision Doppler observations \citep{Guerrero2021}. Intensive radial velocity (RV) follow-up observations of \tess\ candidates have confirmed the planetary nature of transit signals detected by \tess\ and provided masses for hundreds of small planets \citep[$R_\mathrm{p}$\,$\lesssim$\,4\,$R_\oplus$, e.g.,][]{Gandolfi2018,Nielsen2020,VanEylen2021,Barragan2022,Georgieva2023, Goffo2024, Mantovan2024}. The Doppler follow-up carried out by the community to characterize \tess\ targets has often led to the serendipitous discovery of additional planets \citep{Demangeon2021,Hatzes2022,Serrano2022,Lillo-Box2023,Cabrera2023,Goffo2023,Knudstrup2023,Osborne2024}, with a handful of RV planets later found to transit their host star by searching the \tess\ light curves for signals whose significance is below the multiple-event statistic \citep[MES;][]{Jenkins2002} detection threshold of 7.1 adopted by the \tess\ team \citep[see, e.g.,][]{Carleo2020}.

\sname\ is a bright (V\,=\,8.6) solar-type star (Table~\ref{tab:stellar_parameters}) found to host a 25.5-day sub-Neptune transiting candidate by the \tess team \citep[\sname.01;][]{Guerrero2021}. The candidate was later validated by \citet{Giacalone2021}. As part of our follow-up effort of \tess transiting planets, we intensively monitored the system with the \harps spectrograph to spectroscopically confirm the 25.5-day planet (\sname~d), determine its mass, and reveal the possible presence of additional orbiting companions. Thanks to the $\sim$3.4-year baseline of our \harps follow-up, we discovered three Doppler signals associated with the presence of three additional low-mass planets with periods of 4.2, 13.1, and 204.6~d (\sname~b, c, and e, respectively). Prompted by the \harps results, we searched the \tess light curve for additional transit signals and identified a possible signal at 4.2~d. To confirm the photometric detection of the 4.2-day planet and precisely measure the radius of the two transiting companions (\sname~b~and~d), we performed high-precision photometric follow-up with the \cheops space telescope \citep{Benz2021,Fortier2024}, as part of our guaranteed time observing (GTO) program.

The present paper is organized as follows. We report on the \tess and \harps observations in Sects.~\ref{Sec:TESS_Observations} and \ref{Sec:HARPS_Observations}. We present our search for the transit signal of \sname~b in the \tess light curve, its photometric confirmation with \cheops, and the high-resolution \soar imaging in Sects.~\ref{Sec:TransitSearch}, \ref{Sec:Cheops_Observations}, and \ref{Sec:SoarObservations}, respectively. The determination of the stellar fundamental parameters is described in Sect.~\ref{Stellar_Parameters}. The joint analysis of the \tess, \cheops, and \harps data is presented in Sect.~\ref{Sec:Joint_Analysis}. We describe our dynamical study of the planetary system in Sect.~\ref{sec:Dynamical_Analysis}. Our internal structure analysis of the two transiting planets \sname~b and d is given in Sect.~\ref{sec:InternalStructureAnalysis}. We summarize the properties of the four-planet system and conclude in Sect.~\ref{Sec:Conclusions}.

\section{\tess\ observations}
\label{Sec:TESS_Observations}

\begin{table*}[!h]
\centering
\caption{Log of the \tess observations of \sname.}
\label{tess_obs_log}
\begin{tabular}{rccccr}
\hline
\hline
\noalign{\smallskip}
Sector & Start date (UTC) & End date (UTC) & Camera & CCD & T$_\mathrm{exp}$ (s) \\
\noalign{\smallskip}
\hline
\noalign{\smallskip}
 9 & 2019-02-28T17:09:35.06 & 2019-03-25T23:23:34.17 & 2 & 3 & 120 \\  
10 & 2019-03-26T22:19:34.14 & 2019-04-22T04:19:33.17 & 2 & 4 & 120 \\
36 & 2021-03-07T09:45:30.68 & 2021-04-01T11:49:50.31 & 2 & 3 & 20$^\mathrm{a}$ \\
63 & 2023-03-10T20:55:09.25 & 2023-04-06T09:41:17.95 & 2 & 3 & 20$^\mathrm{a}$ \\
90 & 2025-03-12T15:54:03.49 & 2025-04-09T13:56:02.51 & 2 & 4 & 20$^\mathrm{a}$ \\
\noalign{\smallskip}
\hline
\end{tabular}
\tablefoot{$^\mathrm{a}$To increase the signal-to-noise (S/N) ratio of Sector 36, 63, and 90 photometric data points, we used the 120~s cadence light curves available at the MAST archive (\url{https://mast.stsci.edu}).}
\end{table*}

\begin{figure}[!t]
    \centering
    \includegraphics[trim= 0cm 0cm 0cm 0cm,clip, width=1.00\linewidth]{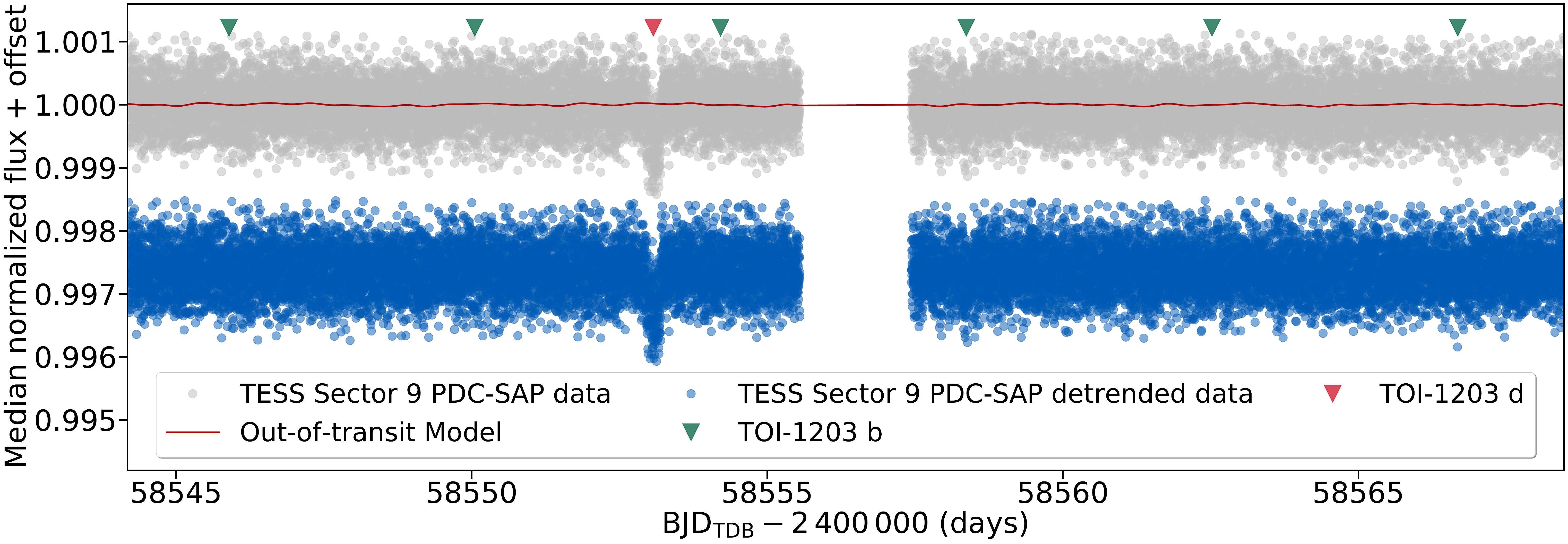}
    \includegraphics[trim= 0cm 0cm 0cm 0cm,clip, width=1.00\linewidth]{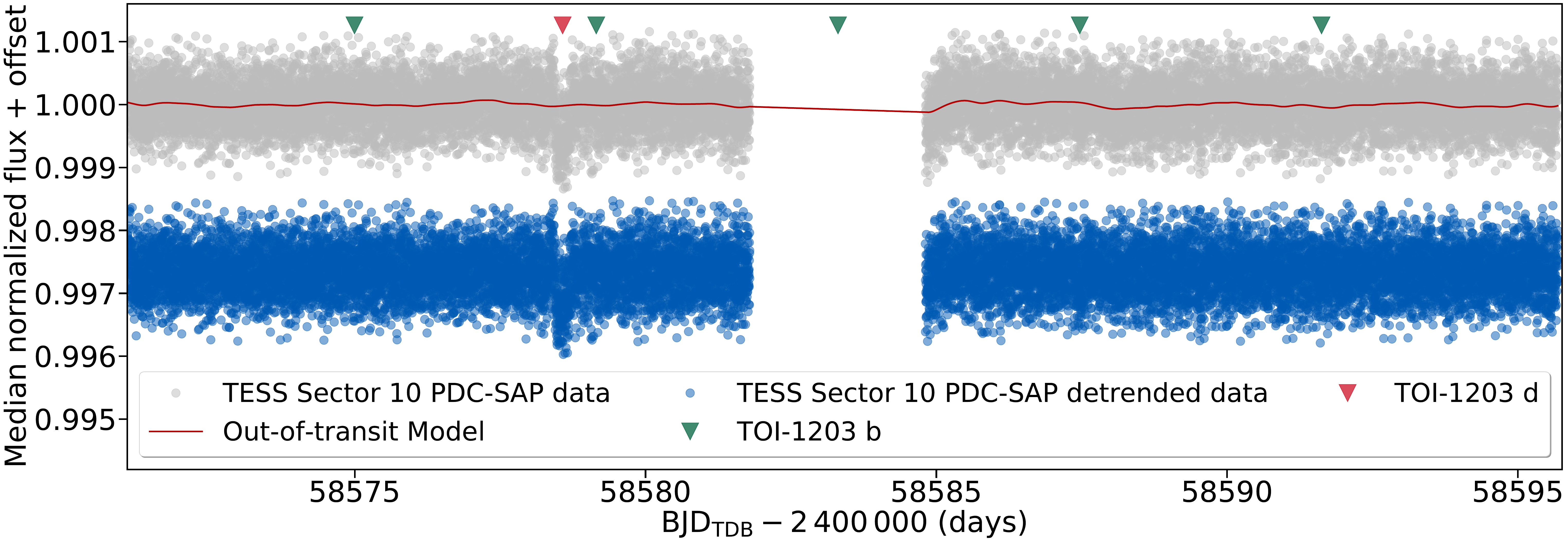}
    \includegraphics[trim= 0cm 0cm 0cm 0cm,clip, width=1.00\linewidth]{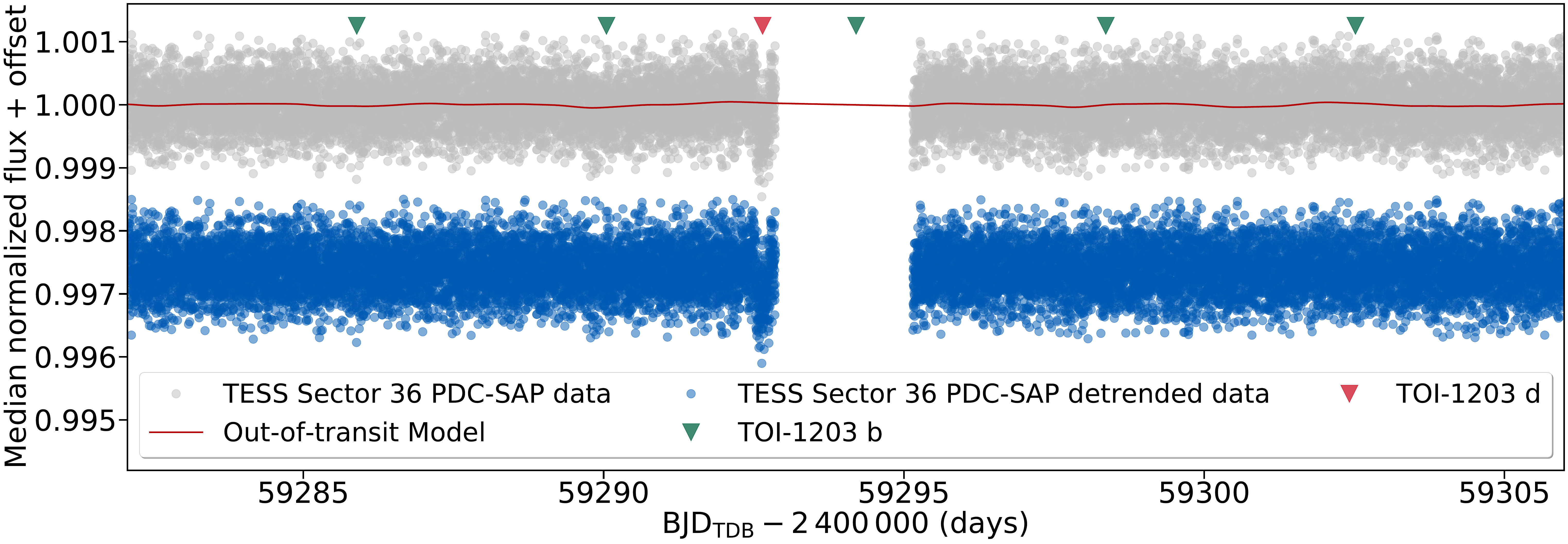}
    \includegraphics[trim= 0cm 0cm 0cm 0cm,clip, width=1.00\linewidth]{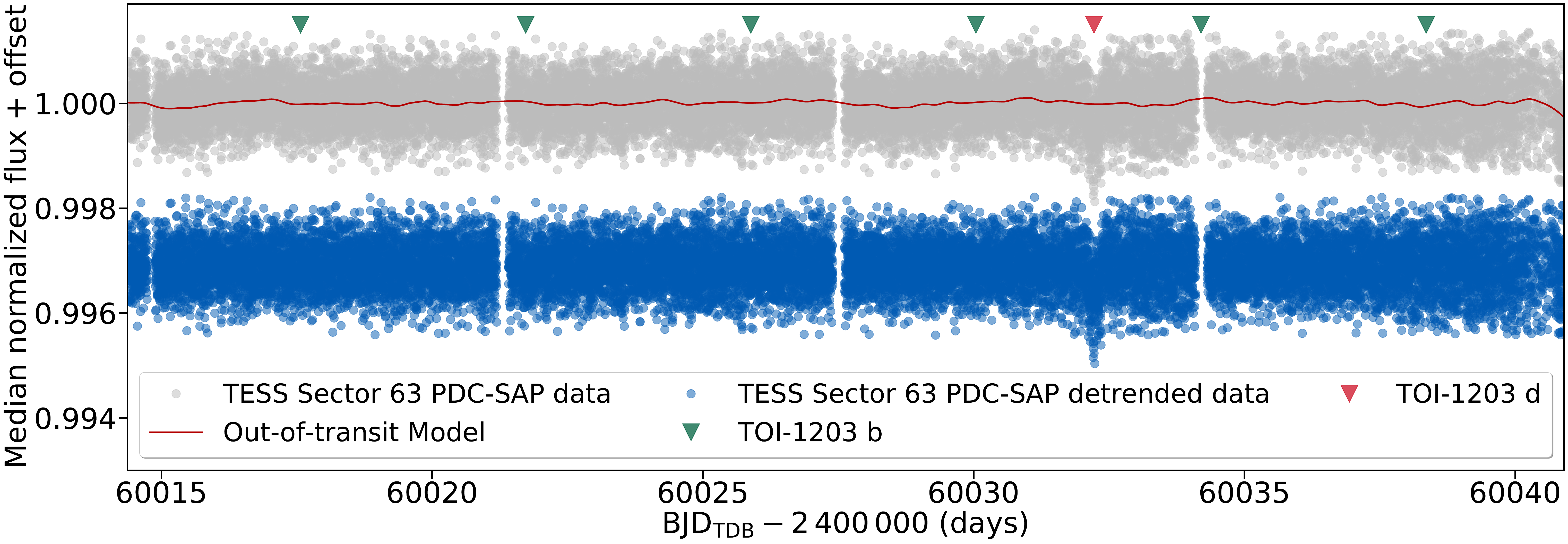}
    \includegraphics[trim= 0cm 0cm 0cm 0cm,clip, width=1.00\linewidth]{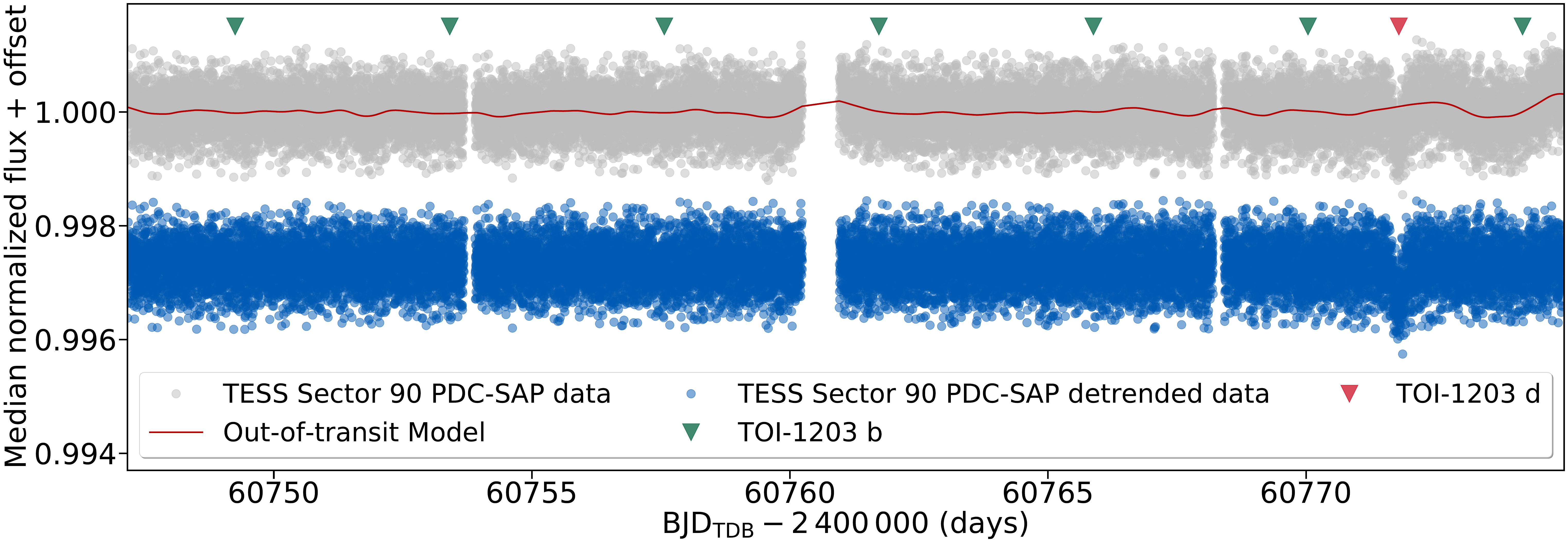}
    \caption{\tess Sector 9 (upper panel), Sector 10 (second panel), Sector 36 (third panel), Sector 63 (fourth panel), and Sector 90 (lower panel) PDC-SAP light curves of \sname. The 120-second data points are displayed as light gray circles, along with the out-of-transit Gaussian process model overplotted in red. The resulting detrended light curves are depicted as blue circles. The transit mid-times of \sname~b and d are marked with green and red triangles, respectively.} 
    \label{fig:TESS_LCs}
\end{figure}

NASA's \tess space telescope observed \sname (also known as TIC\,23434737 and HD\,97507; Table~\ref{tab:stellar_parameters}) in Sectors 9 and 10, during the first year of its nominal mission, from 28 February to 25 March 2019 (UTC; Sector~9), and from 26 March to 22 April 2019 (UTC; Sector 10). The star was reobserved by \tess in Sector 36, during the first year of its extended mission, from 7 March to 1 April 2021 (UTC), and two years later in Sector 63, during the first year of its second extended mission, from 10 March through 6 April 2023 (UTC). More recently,  while this paper was in the review stage, \tess reobserved \sname in Sector 90 during the third year of its second extended mission, from 12 March to 9 April 2025 (UTC). A log of the observations is reported in Table~\ref{tess_obs_log}. The photometric data have a 120~s cadence for Sectors 9 and 10, and a 20~s cadence for Sectors 36, 63, and 90. We note that \sname will be reobserved by \tess in Sectors 99 and 100 from 5 January to 2 March 2026, 2025.\footnote{See \url{https://heasarc.gsfc.nasa.gov/wsgi-scripts/TESS/TESS-point_Web_Tool/TESS-point_Web_Tool/wtv_v2.0.py/} and \url{https://tess.mit.edu/observations/sector-99/} and \url{https://tess.mit.edu/observations/sector-100/}.}

The \tess target pixel files were processed and calibrated by the Science Processing Operations Center \citep[SPOC;][]{Jenkins2016} at the NASA Ames Research Center. For each sector, the light curve was extracted using simple aperture photometry \citep[SAP;][]{Twicken2010, Morris2020} and processed using the Presearch Data Conditioning (PDC) algorithm, which uses a Bayesian maximum a posteriori approach to remove the majority of instrumental artifacts and systematic trends \citep{Smith2012, Stumpe2012, Stumpe2014}. We retrieved the PDC-SAP \tess light curves of \sname from the Mikulski Archive for Space Telescopes\footnote{\url{https://mast.stsci.edu}.} (MAST) and used them for the analyses presented in this paper. For the Sector 36, 63, and 90 \tess time series, we downloaded the 120~s cadence data to increase the signal-to-noise ratio (S/N) of the photometric data points.

In order to mitigate any systematic effects that the PDC algorithm may not have fully captured and remove possible low-frequency photometric variability induced by stellar activity, we flattened the \tess PDC-SAP light curves using the package\footnote{Available at \url{https://github.com/oscaribv/citlalicue}.} \texttt{citlalicue}. The code applies Gaussian processes (GP) as implemented in \texttt{george} \citep{Foreman-Mackey2014,Ambikasaran2015} to model out-of-transit photometric variability. We masked out the transits from the \tess PDC-SAP time series, binned the out-of-transit light curves using 3~h intervals, and modeled the binned photometry using a Mat\'ern 3/2 covariance function. We employed an iterative maximum likelihood optimization along with a 5$\sigma$ clipping algorithm to find the optimal model that describes the variability of the out-of-transit light curves and remove possible outliers. To obtain flattened light curves that contain only the transit signals, we divided the PDC-SAP light curves by the inferred GP model interpolated to the time stamps of the original data. Figure~\ref{fig:TESS_LCs} displays the PDC-SAP \tess light curves of \sname from Sectors 9, 10, 36, and 63, along with the GP models and the flattened time series.

The SPOC team searched the \tess light curves for transit-like signals using a pipeline that iteratively performs multiple transiting planet searches and stops when it fails to find subsequent transit-like signatures above the detection threshold of MES\,=\,7.1 \citep{Jenkins2002,Jenkins2010,Jenkins2020} On 5 August 2019, SPOC performed a transit search of the light curve from Sectors 9-10 and generated a Data Validation Report \citep[DVR;][]{Twicken2018, Li2019} with the search results. The candidate passed all the automatic validation tests from the Threshold Crossing Events (TCEs), such as odd-even transit depth variation and ghost diagnostic tests. The difference image centroiding test located the host star to within 1.0\,$\pm$\,3.0\arcsec\ of the transit source, significantly decreasing the likelihood that the signal was due to contamination by a nearby or background eclipsing binary. The \tess vetting team at the Massachusetts Institute of Technology (MIT) inspected the DVR to review the Threshold Crossing Events (TCEs), and announced the detection of a transiting planetary candidate \citep[\sname.01;][]{Guerrero2021} on 26 August 2019 with a period of $\sim$25.5~d, a depth of $\sim$500~ppm, and a duration of $\sim$6~h.

\section{\harps high-precision Doppler observations}
\label{Sec:HARPS_Observations}

With the aim of spectroscopically confirming the 25.5-day transit signal detected by the \tess team \citep{Guerrero2021} and later validated by \citet{Giacalone2021}, and deriving the mass of the planet, we carried out an intensive RV follow-up of~\sname. This was accomplished using the High Accuracy Radial velocity Planet Searcher \citep[\harps;][]{Mayor2003} spectrograph mounted at the ESO 3.6\,m telescope of La Silla observatory (Chile). We collected 190 high-resolution ($R$\,$\approx$\,115\,000) \harps\ spectra between 5 February 2020 and 24 May 2023 (UTC), covering a baseline of about 3.4 years, as part of our large observing programs 1102.C-0923 and 106.21TJ.001 (PI: D.\,Gandolfi), which are devoted to the Doppler follow-up of \tess small planets \citep[see, e.g.,][]{Esposito2019,Fridlund2020,Hoyer2021,Persson2022,Bonfanti2023,Alqasim2024,Subjak2025}. Six additional spectra were acquired in March 2020, December 2020, and January 2021, during technical and calibration nights (program 60.A-9709).

The \harps follow-up was interrupted shortly after its start on 23 March 2020 (UTC) due to the outbreak of the COVID-19 pandemic and resumed nearly eight months later on 15 November 2020 (UTC). A second interruption occurred between 24 March and 23 May 2021 (UTC) due to a second wave of COVID-19 in Chile. The exposure time was set to 900-1800~s, depending on sky conditions and scheduling constraints, leading to a median S/N of $\sim$124 per pixel at 550~nm. We reduced the \harps data using the dedicated Data Reduction Software (\texttt{DRS}) available at the telescope \citep{Pepe2002,Lovis2007} and extracted absolute RV measurements by performing a multi-order cross-correlation of the extracted Echelle spectra with a G2 numerical mask \citep{Baranne1996}. We also used the \texttt{DRS} to extract the Ca\,{\sc ii} H\,\&\,K lines activity indicator (\logrhk), and two profile diagnostics of the cross-correlation function (CCF), namely the full width at half maximum (FWHM) and the bisector inverse slope (BIS). 

Relative RV measurements were extracted using the Template-Enhanced Radial velocity Re-analysis Application \citep[\terra;][]{Anglada2012}, which is based on a template matching algorithm. We conducted a preliminary fit of both the \texttt{DRS} and the \terra Doppler time series utilizing the RV model described in Sect.~\ref{Sec:Joint_Analysis}. Our analysis revealed that the \terra measurements yields a lower root mean square residual, which led us to select them for the subsequent analyses presented in this paper. The \texttt{DRS} and \terra Doppler measurements, along with the activity indicator and line profile diagnostics, are listed in Table~\ref{tab:RV-Table}. Time stamps are barycentric Julian dates in barycentric dynamical time \citep[BJD$_\mathrm{TDB}$;][]{Eastman2010}.

\section{Frequency analysis of the \harps time series}
\label{Sec:HARPS_FrequencyAnalysis}

\begin{figure*}[!t]
    \centering
    \includegraphics[trim= 0cm 0cm 0cm 0cm,clip, width=\linewidth]{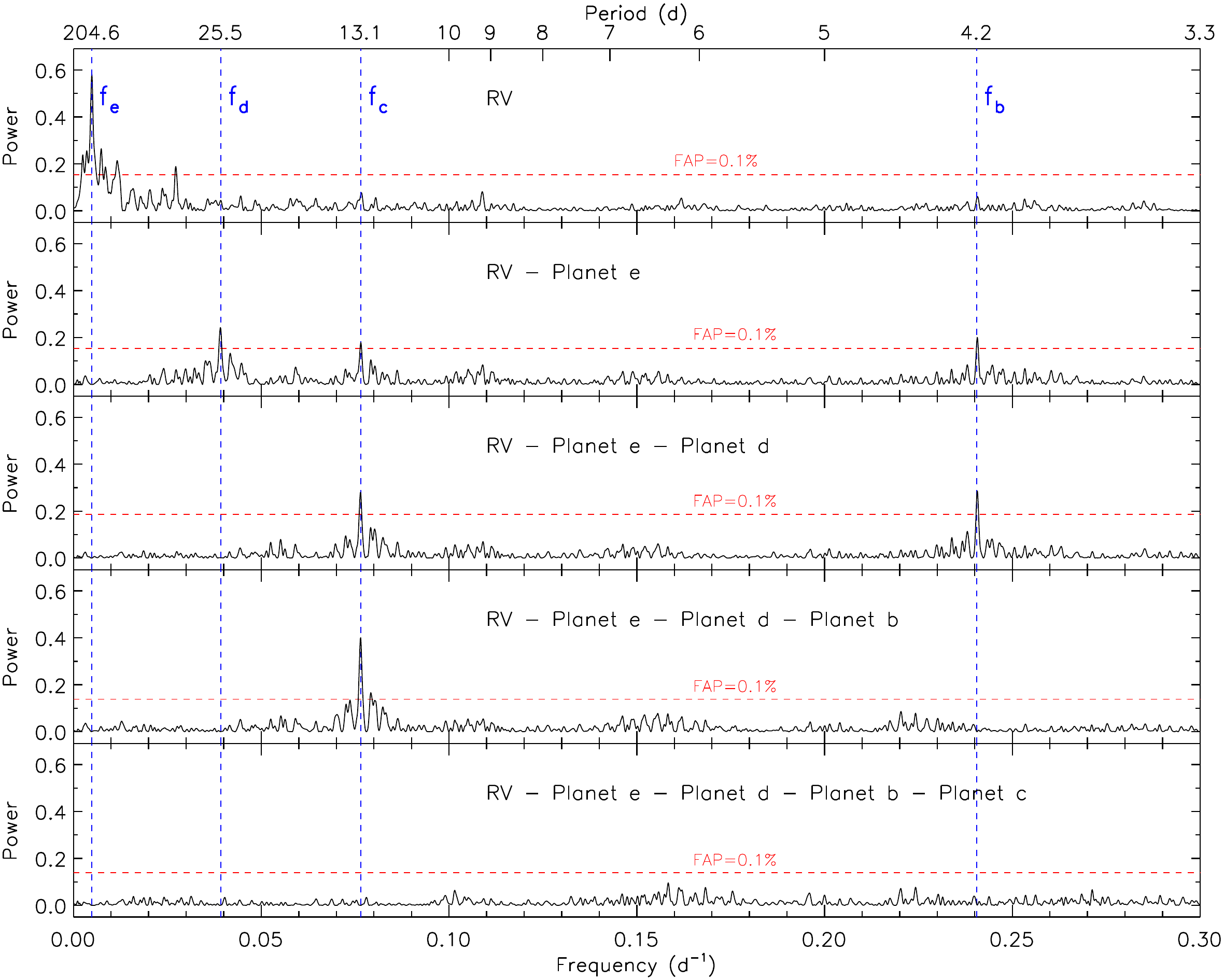}
    \caption{Generalized Lomb-Scargle periodograms of the \sname's \harps\ \terra RVs (upper panel) and of the RV residuals after subtracting the Doppler reflex motion(s) induced by \sname~e (second panel), \sname~e and d (third panel), \sname~e, d, and b (fourth panel), \sname~e, d, b, and c (bottom panel). The dashed horizontal red lines mark the 0.1\,\% false alarm probability (FAP) as computed using the bootstrap method. The orbital frequencies of \sname~b ($f_\mathrm{b}$\,$\approx$\,0.24054~d$^{-1}$; $P_\mathrm{b}$\,$\approx$\,4.2~d), \sname~c ($f_\mathrm{c}$\,$\approx$\,0.07647~d$^{-1}$; $P_\mathrm{c}$\,$\approx$\,13.1~d), \sname~d ($f_\mathrm{d}$\,=\,0.03921~d$^{-1}$; $P_\mathrm{d}$\,$\approx$\,25.5~d), and \sname~e ($f_\mathrm{e}$\,$\approx$\,0.00489~d$^{-1}$; $P_\mathrm{e}$\,$\approx$\,204.6~d) are marked with dashed vertical blue lines.} 
    \label{fig:GLS_DopplerData}
\end{figure*}

\begin{figure*}[!t]
    \centering
    \includegraphics[trim= 0cm 0cm 0cm 0cm,clip, width=\linewidth]{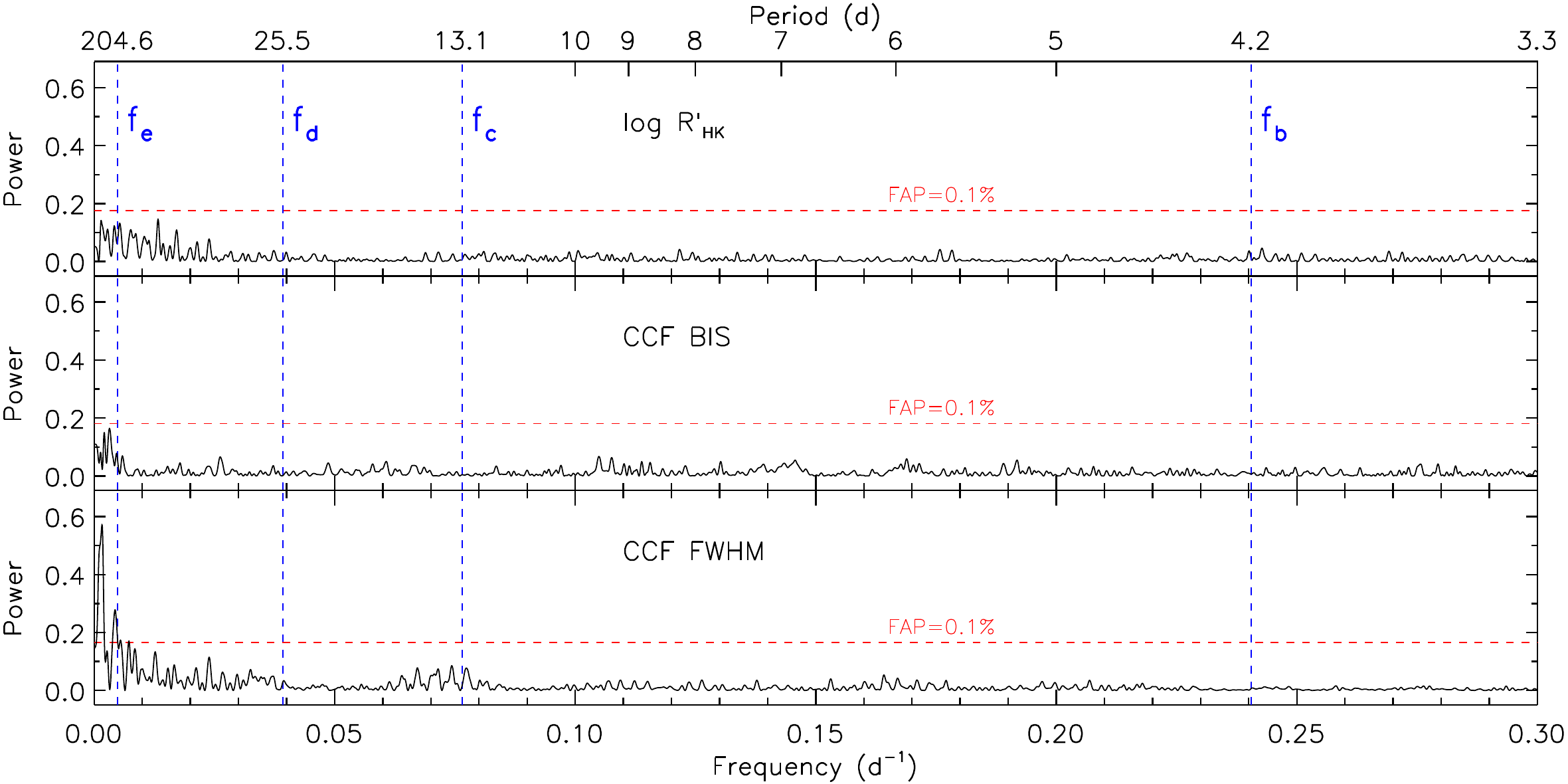}
    \includegraphics[trim= 0cm 0cm 0cm 0cm,clip, width=0.34\linewidth]{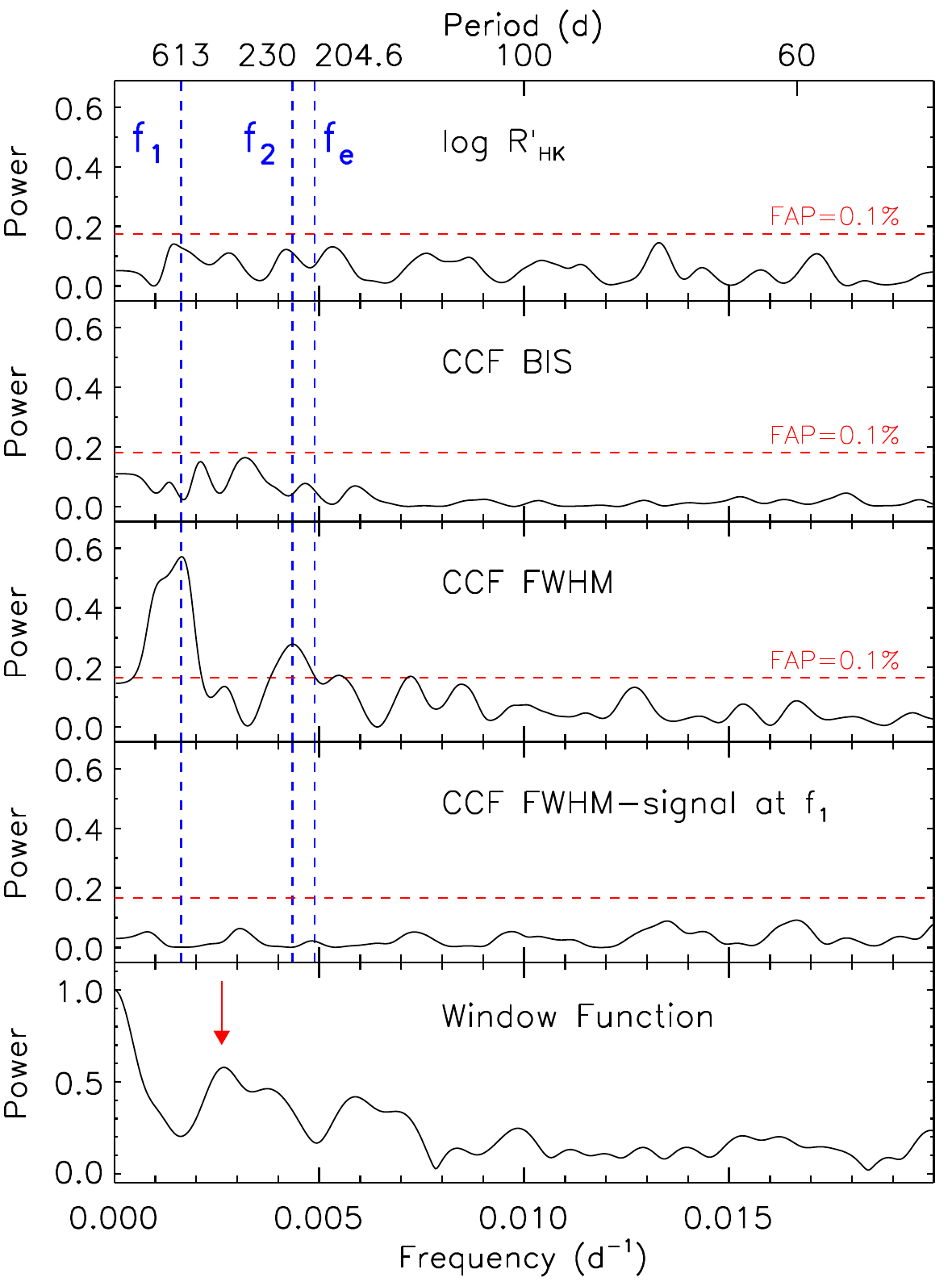}
    \includegraphics[trim= 0cm -2.1cm 0cm 0cm,clip, width=0.59\linewidth]{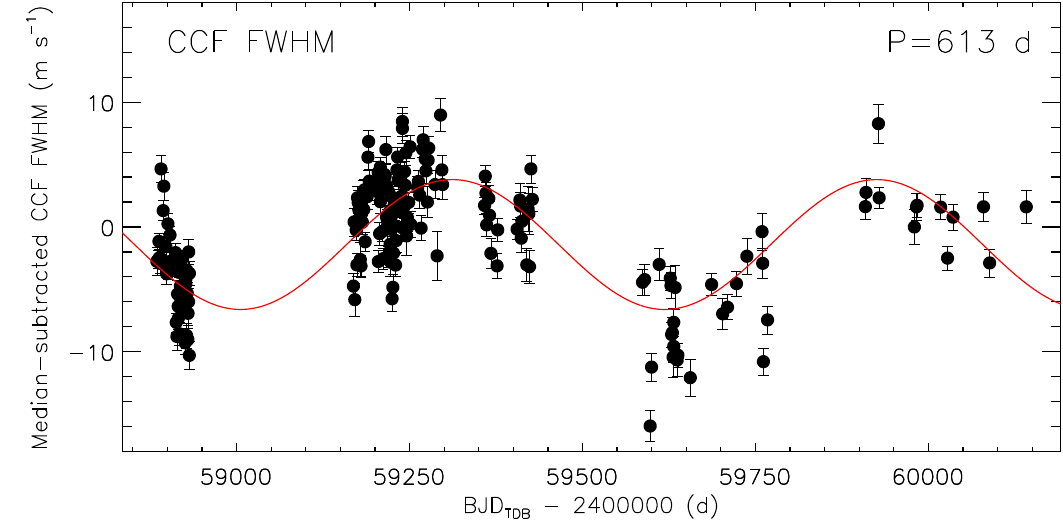}
    \caption{Frequency analysis of the activity indicator and line profile diagnostics of \sname extracted from the \harps spectra. \emph{Top}: Generalized Lomb-Scargle periodograms of: (1) the Ca\,{\sc ii} H\,\&\,K lines activity indicator \logrhk\ (upper panel); (2) the bisector inverse slope (BIS) of the \harps\ CCF (second panel); (3) the full width at half maximum (FWHM) of the \harps\ CCF (lower panel). The dashed horizontal red lines mark the 0.1\,\% false alarm probability (FAP) as computed using the bootstrap method. The orbital frequencies of the four planets are marked with dashed vertical blue lines. \emph{Bottom left}: Detail of the upper figure in the 0\,--\,0.020~d$^{-1}$ frequency range. From top to bottom: Periodograms of (1) \logrhk; (2) CCF BIS; (3) CCF FWHM; (4) CCF FWHM after subtracting the dominant frequency at $f_\mathrm{1}$; (5) window function. The orbital frequency of \sname~e ($f_\mathrm{e}$\,$\approx$\,0.00489~d$^{-1}$; $P_\mathrm{e}$\,$\approx$\,204.6~d), as well as the two significant peaks at $f_\mathrm{1}\,\approx\,0.00163$~d$^{-1}$ ($\sim$613~d) and $f_\mathrm{2}\,\approx\,0.00435$~d$^{-1}$ ($\sim$230~d) identified in the periodogram of the FWHM of the cross-correlation function, are marked with dashed vertical blue lines. The red arrow in the periodogram of the window function (lower panel) marks the peak at 0.00272~d$^{-1}$ ($\sim$368~d) presented in the main text, which is equal to the frequency spacing between the two peaks at $f_\mathrm{2}\,\approx\,0.00435$~d$^{-1}$ and $f_\mathrm{1}\,\approx\,0.00163$~d$^{-1}$. \emph{Bottom right}: Median-subtracted time series of the CCF FWHM and best-fitting sinusoidal model at the 613-day dominant signal.} 
    \label{fig:GLS_Activity}
\end{figure*}

\begin{figure}[!t]
    \centering
    \includegraphics[trim= 0cm 0cm 0cm 0cm,clip, width=\linewidth]{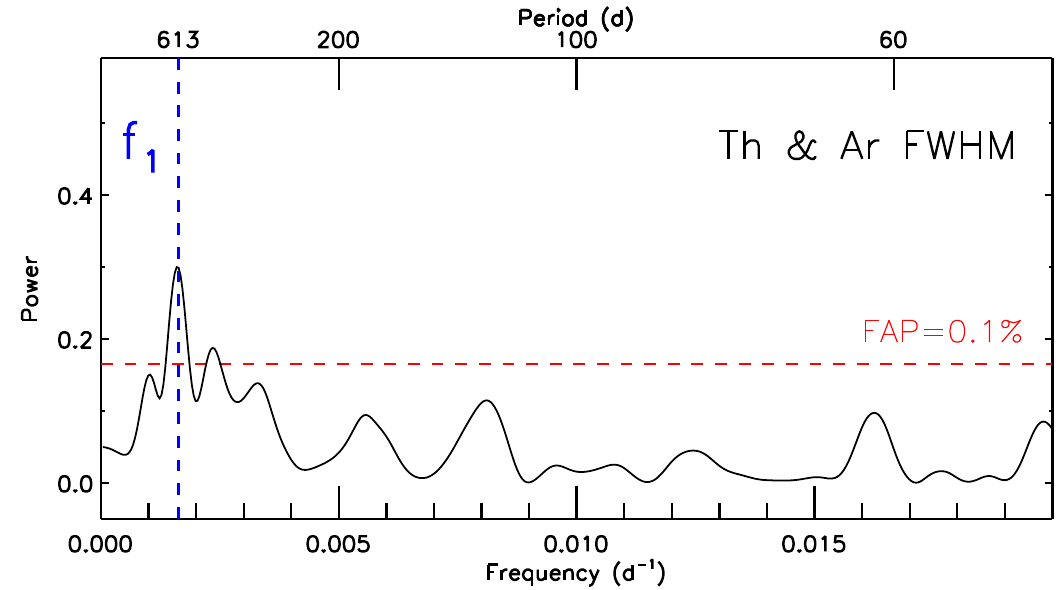}
    ~\\
    \includegraphics[trim= 0cm 0cm 0cm 0cm,clip, width=\linewidth]{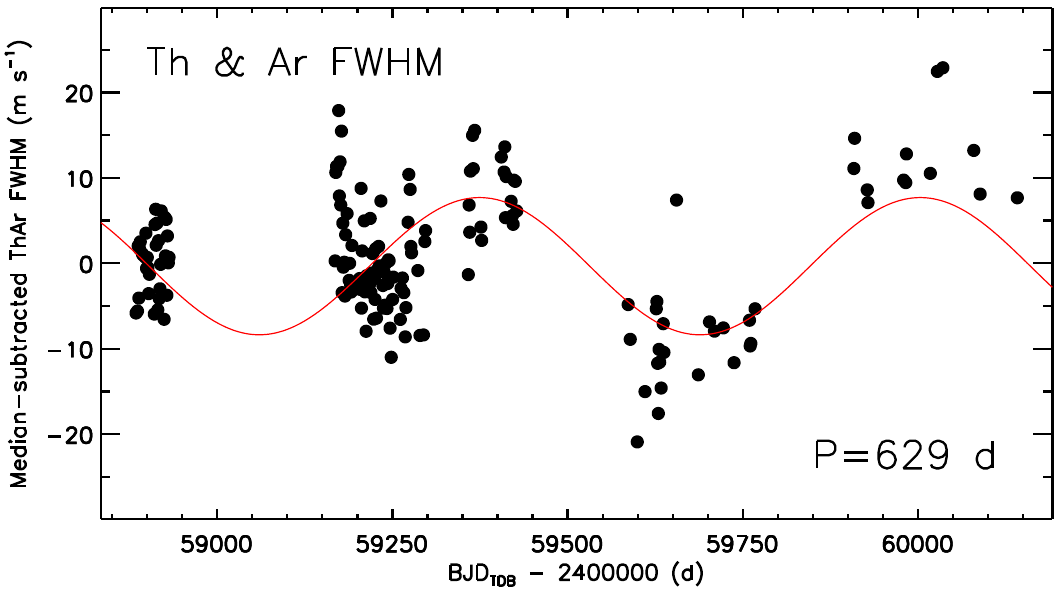}
    \caption{\emph{Upper panel:} Generalized Lomb-Scargle periodogram of the intensity-weighted average of the FWHM of the Th-Ar spectral lines. The \texttt{GLS} power of the FAP at 0.1\,\% is marked with a dashed horizontal red line. The dominant signal is found at 0.00159~d$^{-1}$ ($\sim$629~d). The dashed vertical blue line marks the position of the significant signal at $f_\mathrm{1}\,\approx\,0.00163$~d$^{-1}$ ($\sim$613~d) seen in the periodogram of the FWHM of the \harps\ cross-correlation functions (see Fig.~\ref{fig:GLS_Activity}). Given the frequency resolution of our time series ($\sim$0.0008~d$^{-1}$), the two frequencies above are indistinguishable. \emph{Lower panel}: Median-subtracted time series of the intensity-weighted average of the FWHM of the Th-Ar spectral lines and best-fitting sinusoidal model at the 629-day dominant signal.} 
    \label{fig:FWHM_ThAr}
\end{figure}

We carried out a frequency analysis of the \harps time series to search for the Doppler reflex motion induced by the 25.5-day warm sub-Neptune and detect possible signals indicative of other orbiting companions. The generalized Lomb-Scargle (\texttt{GLS}) periodogram \citep{Lomb1976,Scargle1982,Zechmeister2009} of the \harps \terra RV measurements (Fig.~\ref{fig:GLS_DopplerData}, first panel) shows its highest peak at $f_\mathrm{e}$\,$\approx$\,0.00489~d$^{-1}$, which corresponds to a period of~$\sim$204.6~d. To account for possible non-Gaussian noise in the data, we estimated the false alarm probability (FAP) of the signal at $f_\mathrm{e}$ using the bootstrap method described in \citet{Murdoch1993} and \citet{Kuerster1997}. Briefly, we kept the time stamps fixed and created 10$^6$ mock time series by randomly shuffling the measurements along with their uncertainties. We defined the FAP as the fraction of instances in which the periodogram of random data has a power higher than the periodogram of the real time series and regarded as significant those signals whose FAP\,<\,0.1\,\%. For the signal at $f_\mathrm{e}$\,$\approx$\,0.00489~d$^{-1}$, we found no false positive out of 10$^6$ bootstrap realizations, implying that its false-alarm probability is FAP\,$<$\,0.0001\,\%.

We searched the Doppler data for additional significant peaks by applying the pre-whitening technique \citep[see, e.g.,][]{Hatzes2010,Gandolfi2017}. We removed the dominant signal by performing a least-squares sine-fit to the amplitude, phase, and offset at $f_\mathrm{e}$\,$\approx$\,0.00489~d$^{-1}$, subtracted the fit from the \harps \terra RV time series, and computed the \texttt{GLS} periodogram of the residuals. The process was iterated until we reached our adopted significance threshold of FAP\,=\,0.1\,\%. 

The \texttt{GLS} periodogram of the RV residuals, after subtracting the signal at $f_\mathrm{e}$ (Fig.~\ref{fig:GLS_DopplerData}, second panel), displays its most significant peak at $f_\mathrm{d}$\,=\,0.03921\,d$^{-1}$ ($\sim$25.5~d), i.e., the transit frequency of the planet detected by \tess. By iterating the pre-whitening technique we identified a third significant peak at $f_\mathrm{c}$\,$\approx$\,0.07647~d$^{-1}$ ($\sim$13.1~d; Fig.~\ref{fig:GLS_DopplerData}, third panel) and a fourth one at $f_\mathrm{b}$\,$\approx$\,0.24054~d$^{-1}$ ($\sim$4.2~d; Fig.~\ref{fig:GLS_DopplerData}, fourth panel). Once the four signals are removed from the \harps \terra RV time series, the \texttt{GLS} periodogram of the residuals displays no peaks whose power exceeds the power at our FAP\,=\,0.1\,\% significance threshold (Fig.~\ref{fig:GLS_DopplerData}, fifth panel).

To investigate the sources of the four significant signals observed in the \harps \terra RV measurements, we undertook a frequency analysis of the \harps\ ancillary time series. Stellar rotation, magnetic cycles, stellar blends, and instrumental effects can indeed mimic Doppler planetary signals, resulting in the erroneous detection of nonexistent planets. Figure~\ref{fig:GLS_Activity} shows the \texttt{GLS} periodogram of the activity indicator \logrhk, along with those of the BIS and FWHM of the cross-correlation functions. We found that neither the Ca\,{\sc ii} H\,\&\,K lines activity indicator, nor the CCF profile diagnostics display significant peaks at $f_\mathrm{b}$, $f_\mathrm{c}$, and $f_\mathrm{d}$. The lack of a counterpart at $f_\mathrm{d}$ in the \texttt{GLS} periodograms of the \harps ancillary time series spectroscopically confirms the planetary nature of the transit signal detected by \tess at $\sim$25.5~d, which we hereby refer to as \sname~d. Similarly, the null detection of counterparts at $f_\mathrm{b}$ and $f_\mathrm{c}$ provides strong evidence that the signals at $\sim$4.2~d and $\sim$13.1~d are due to two additional orbiting planets, which we henceforth indicate with \sname~b and \sname~c, respectively.

The \texttt{GLS} periodogram of the FWHM of the cross-correlation functions shows a significant peak at $f_\mathrm{1}$\,$\approx$\,0.00163~d$^{-1}$ ($\sim$613~d; Fig.~\ref{fig:GLS_Activity}). As this peak has no counterpart either in the Doppler measurements or in the remaining HARPS ancillary time series, we investigated the possibility of an instrumental origin for the signal at $f_\mathrm{1}$. Our goal was to find possible changes in the FWHM of the instrumental profile that would result in variations of the FWHM of the CCF, while having no effect on the measurements of the BIS and RV. The FWHM of the Th-Ar calibration lines used to compute the wavelength solution of high-resolution spectrograph is a reliable proxy of the FWHM of the instrumental profile. For this purpose, we retrieved from the ESO archive\footnote{Available at \url{http://archive.eso.org/wdb/wdb/eso/repro/form}.} the Th-Ar line table files, which were produced by the \harps \texttt{DRS} in the afternoons preceding the nights when we observed \sname. The files include both the FWHM and the intensity of the Th and Ar emission lines, as measured from the Th-Ar spectrum taken in the afternoon before the start of observations, which is used by the \harps \texttt{DRS} to compute the nightly wavelength solution. For each observing night, we calculated the weighted average FWHM of the Th-Ar lines by taking into account the intensity of the lines when weighting the FWHM measurements.

The \texttt{GLS} periodogram of the FWHM of the instrument profile is shown in Fig.~\ref{fig:FWHM_ThAr} (upper panel), along with the median subtracted time series (lower panel). Our analysis revealed a significant peak (FAP\,$<$\,0.1\,\%) at 0.00159~d$^{-1}$ ($\sim$629~d), which, when taking into account the $\sim$0.0008~d$^{-1}$ frequency resolution of our \harps time series,\footnote{We recall that the frequency resolution is defined as the inverse of the time baseline. Our \harps time series cover a baseline of $\sim$1257~d, implying a frequency resolution of $1/1257$\,$\approx$\,0.0008~d$^{-1}$.} is in agreement with the periodicity identified at $f_\mathrm{1}$\,$\approx$\,0.00163~d$^{-1}$ ($\sim$613~d) seen in the periodogram of the FWHM of the \harps cross-correlation function (Fig.~\ref{fig:GLS_Activity}). We conclude that the CCF FWHM signal at $f_\mathrm{1}$ is not astrophysical in nature, but arises from a change in the \harps instrumental profile.

The \texttt{GLS} periodogram of the FWHM also shows a significant peak at $f_\mathrm{2}$\,$\approx$\,0.00435~d$^{-1}$ ($\sim$230~d; Fig.~\ref{fig:GLS_Activity}, bottom left, third panel), which is close to the frequency of the signal at $f_\mathrm{e}$\,$\approx$\,0.00489~d$^{-1}$ ($\sim$204.6~d) seen in the \harps\ RV measurements. Given the frequency resolution of our \harps\ time series ($\sim$0.0008~d$^{-1}$), the FWHM signal at $f_\mathrm{2}$\,$\approx$\,0.00435~d$^{-1}$ and the RV peak at $f_\mathrm{e}$\,$\approx$\,0.00489~d$^{-1}$ are indistinguishable and might potentially arise from the same source. Yet, an inspection of the periodogram of the window function reveals the presence of a peak at 0.00272~d$^{-1}$ (corresponding to a period of $\sim$1~year; Fig.~\ref{fig:GLS_Activity}, red arrow), which is equal to the frequency spacing between the two peaks at $f_\mathrm{2}$\,$\approx$\,0.00435~d$^{-1}$ and $f_\mathrm{1}$\,$\approx$\,0.00163~d$^{-1}$, implying that the former is a 1-year alias of the latter. We verified this hypothesis by removing the signal at $f_\mathrm{1}$ from the FWHM time series and recalculating the \texttt{GLS} periodogram of the residuals. We found that by subtracting the signal at $f_\mathrm{1}$, we also remove the signal at $f_\mathrm{2}$ (Fig.~\ref{fig:GLS_Activity}, bottom left, fourth panel), as expected from alias peaks, confirming our hypothesis. We note that, on the contrary, the power spectrum of the \harps \terra RV measurements (Fig.~\ref{fig:GLS_DopplerData}) does not display any significant peaks close to $f_\mathrm{1}$, implying that the Doppler signal at $f_\mathrm{e}$\,$\approx$\,0.00489~d$^{-1}$ ($\sim$204.6~d) is not an alias of an RV signal at lower frequencies. Moreover, it has no counterpart in the periodograms of the \logrhk activity indicator and of the CCF BIS (Fig.~\ref{fig:GLS_Activity}), providing supporting evidence that the signal at $f_\mathrm{e}$ is also due to an outer orbiting planet at $\sim$204.6~d, hereafter referred to as \sname~e.

\section{Transit search of the \tess\ light curve}
\label{Sec:TransitSearch}

\begin{figure*}[!t]
    \centering
    \includegraphics[trim= 0cm 0cm 0cm 0cm,clip, width=1.00\linewidth]{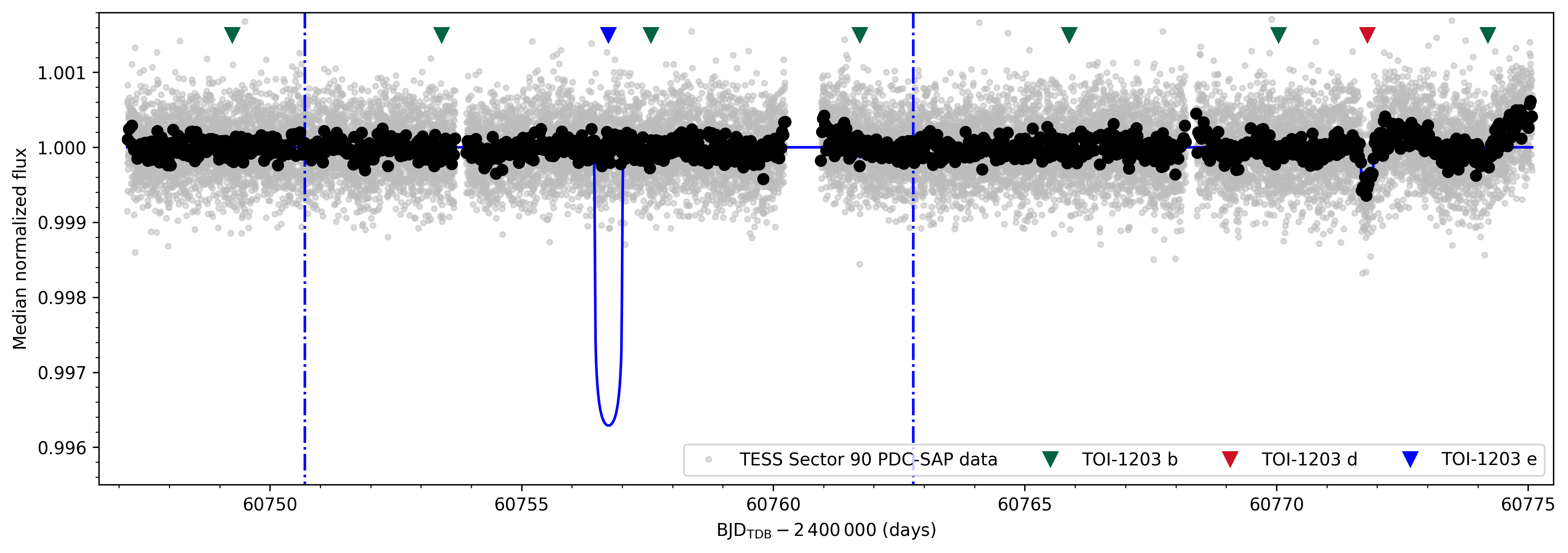}
    \caption{\tess Sector 90 PDC-SAP light curve of \sname. The 120~s data points are displayed with light gray circles, along with the 30-minute binned photometry (black circles). The mid-time of the expected transit of \sname~e, as predicted from the orbital period and reference time of inferior conjunction listed in Table~\ref{tab:modelparameters}, is marked with a light blue triangle, whereas the transit mid-times of \sname~b and d are marked with green and red triangles, respectively. The transit model of \sname~e, as derived from the orbital parameters in Tables~\ref{tab:modelparameters} and \ref{tab:derivedparameters}, assuming a planetary radius of 7.3~$R_\oplus$ and a central event ($b_\mathrm{e}$\,=\,0), is shown with a light blue thick line. The dash-dotted lines mark the 1$\sigma$ confidence interval of the expected transit mid-time of \sname~e.} 
    \label{fig:TESS_S90_LC_TOI-1203e}
\end{figure*}

We used the D\'etection Sp\'ecialis\'ee de Transits (\texttt{DST}) algorithm \citep{Cabrera2012} to independently search for transiting planets in the PDC-SAP flux time series from \tess Sectors 9, 10, 36, 63, and 90. The \texttt{DST} algorithm adopts a detection statistic threshold of 8. We distinctly found the signal at 25.5~d (\texttt{DST} statistic of 17.7) previously detected by the \tess SPOC team. We also identified an additional transit signal at 4.2~d, albeit with a DST statistic falling just short of the detection threshold (DST\,$\approx$\,7.1). No further periodic signal was found in the data. 

The SPOC pipeline measures the statistical significance of transit signals using a multiple event statistic \citep[MES;][]{Guerrero2021}. Events with MES\,$>$\,7.1 that pass further transit consistency tests would then be reported as candidates. In the case of \sname, the detection of a 25.5-day transit candidate with MES\,=\,18.5 was reported from a search for transit signals in the \tess Sector~9 and 10 light curves. The \tess pipeline did not yield any additional transit candidates, likely because their MES value fell below the detection threshold or they failed to meet specific consistency criteria. We note that, while the transit of the 4.2-day candidate is visible in \tess Sectors 9, 10, 63, and 90 light curves, we did not recover it in Sector 36. There were no instrumental systematics reported near the expected transit epochs of the candidate planet. Around a similar epoch, we recovered transit observations of the 4.2-day candidate using \cheops (Sect.~\ref{Sec:Cheops_Observations}), suggesting that its ``disappearance'' was not due to a change in the planet's orbit. The transits of the 4.2-day candidate have a low S/N. Any small changes to the background noise in the data set could hide the transit signal and evade detection, which we suspect may be the case for the \tess Sector~36 time series.

Triggered by the findings of the \harps spectroscopic follow-up (Sect.~\ref{Sec:HARPS_Observations}), we carried out an intensive search for transiting planets with periods around 13.1~d using injection of simulated transiting planets. We concluded that we would have confidently detected any transiting planets with transit depths of about 150~ppm (equivalent to $\sim$1.5~$R_\oplus$ for a central transit) at the period of \sname~c. Adopting the mass-radius relation for rocky planets ($\rho_\mathrm{p}$\,$>$\,3.3~\gccm) from \citet{Otegi2020} and assuming an orbital inclination close to 90$^\circ$, \sname~c's minimum mass of $M_\mathrm{c} \sin i_\mathrm{c}$\,=\,\mpc\ (Sect.~\ref{Sec:Joint_Analysis} and Table~\ref{tab:derivedparameters}) translates into a predicted planetary radius of $1.68\,\pm\,0.06$~$R_\oplus$. We are confident at about the 99.9\,\% level that the radius of \sname~c must be larger than our detection threshold of $\sim$1.5~$R_\oplus$ and that we would have detected the planet in the \tess light curves if it were transiting its host star.

We used the orbital period and the reference time of inferior conjunction for \sname~e, as determined from the \harps\ Doppler measurements (Table~\ref{tab:modelparameters}), to visually inspect the \tess light curves for potential transits of the planet. From the planet's minimum mass ($M_\mathrm{e} \sin i_\mathrm{e}$\,=\,\mpe), using \citet{Otegi2020}'s relation for volatile-rich planets ($\rho_\mathrm{p}$\,$<$\,3.3~\gccm) and assuming $i_\mathrm{e}$\,$\approx$\,90$^\circ$, we estimated a radius of $7.3^{+1.8}_{-1.5}$~$R_\oplus$. After accounting for limb-darkening effects and taking into account the stellar radius, the predicted planetary radius translates into a transit depth of $\sim$3600~ppm, with the orbital period and the scaled semimajor axis implying a duration of $\sim$13.7~hours for a central transit ($b_\mathrm{e}$\,=\,0). According to our ephemeris, a transit of \sname~e was expected to be observable in \tess Sector~90 (Fig.~\ref{fig:TESS_S90_LC_TOI-1203e}). Although there is a $\sim$4\,\% probability that the transit mid-point fell in the mid-sector gap (17.2~hours) and a 6\,\% chance that it occurred outside of the Sector 90 time window, we can exclude a transit detection with a confidence level of approximately 90\,\%, suggesting that \sname~e very likely does not transit its host star. We note that the forthcoming \tess observations in Sectors 99 and 100, currently scheduled from 5 January to 2 March 2026 (UTC), will not yield further evidence in this context, as they will fall between the next two potential transit opportunities, which are expected to occur on 12 October 2025 and 5 May 2026 (UTC).

\section{\cheops transit observations}
\label{Sec:Cheops_Observations}

\begin{table*}[!t]
         \centering
         \caption{Log of the \cheops observations of \sname.}
         \label{Table:Cheops_Visits}
              \scalebox{0.85}{\begin{tabular}{cccc}
                 \hline\hline
                 \noalign{\smallskip}
                 Planet & File key & Start date (UTC) & End date (UTC) \\
                 \noalign{\smallskip}
                 \hline
                 \noalign{\smallskip}
                 b & \texttt{CH\_PR110045\_TG001701\_V0200} & 2021-02-27 10:53:27 & 2021-02-28 08:13:07 \\
                 b & \texttt{CH\_PR110045\_TG002101\_V0200} & 2021-03-16 07:50:26 & 2021-03-16 17:27:26 \\
                 b & \texttt{CH\_PR100031\_TG002401\_V0200} & 2021-03-28 18:29:27 & 2021-03-29 04:16:48 \\
                 b & \texttt{CH\_PR110045\_TG002402\_V0200} & 2021-04-01 22:13:27 & 2021-04-02 07:53:36 \\
                 b & \texttt{CH\_PR110045\_TG002403\_V0200} & 2021-04-06 02:19:27 & 2021-04-06 12:06:49 \\
                 b & \texttt{CH\_PR110045\_TG002501\_V0200} & 2021-05-01 00:55:27 & 2021-05-01 10:42:49 \\
                 d & \texttt{CH\_PR110024\_TG014301\_V0200} & 2021-05-08 17:13:27 & 2021-05-09 13:23:18 \\
                 b & \texttt{CH\_PR110024\_TG014901\_V0200} & 2022-03-21 08:36:27 & 2022-03-21 18:07:09 \\
                 b & \texttt{CH\_PR110024\_TG014902\_V0200} & 2022-04-02 19:27:08 & 2022-04-03 05:14:30 \\
                 b & \texttt{CH\_PR110024\_TG014903\_V0200} & 2022-04-11 02:50:48 & 2022-04-11 11:59:26 \\
                 b & \texttt{CH\_PR110024\_TG014904\_V0200} & 2022-04-19 10:14:49 & 2022-04-19 19:27:29 \\
                 \noalign{\smallskip}
                 \hline
             \end{tabular}}
\end{table*}

In order to confirm the planetary signal at 4.2~d and refine the radius of the two transiting planets \sname~b and d, we performed, between February 2021 and April 2022, a high-precision photometric follow-up of \sname using the \cheops space telescope \citep{Benz2021,Fortier2024}. A log of the observations is given in Table~\ref{Table:Cheops_Visits}. Following the same methodology presented in Sect.~\ref{Sec:Joint_Analysis}, we derived a preliminary set of transit ephemeris for \sname~b by jointly modeling the \harps\ and \tess\ time series that were available at the time we started our \cheops\ follow-up in February 2021. To confirm the planet candidate at 4.2~d and improve its ephemeris, we first carried out a long (13 \cheops orbits, i.e., 21.4~h) single visit\footnote{A visit is a sequence of successive orbits of \cheops around the Earth that are dedicated to the observations of a given target. \cheops revolves about Earth every $\sim$99 minutes in a Sun-synchronous, low-Earth orbit ($\sim$700~km altitude).} centered around the expected transit of \sname~b and covering the 3$\sigma$ confidence interval. Once we confirmed the presence of the transit signal within the first \cheops visit, we updated the ephemeris and continued to observe \sname~b, collecting 9 additional shorter visits between March 2021 and April 2022. We also performed one visit of the transiting warm sub-Neptune \sname~d in May 2021. 

We extracted point spread function (PSF) photometry using the open source photometric extraction package \texttt{PIPE}\footnote{\url{https://github.com/alphapsa/PIPE}.} \citep{Brandeker2024}, resulting in a point-to-point photometric scatter between consecutive 27 s exposures of $\sim$230~ppm (median of absolute deviation). This is comparable and consistent with the aperture photometry provided by the \cheops Data Reduction Pipeline \citep[\texttt{DRP};][]{Hoyer2020}, but reduces the contamination modulation caused by nearby background stars. We detrended the eleven \cheops visits with the code \texttt{pycheops} \citep{Maxted2023}, which allows one to remove instrumental systematics \citep{Fortier2024} by decorrelating the data using the vectors that describe the short-term instrumental photometric trends: roll angle, centroid movements, smear curve, and contamination by background stars. We first applied a 3$\sigma$ clipping algorithm as implemented in \texttt{pycheops} to remove outliers from the \cheops data and analyzed each visit individually to check if it was affected by any of the known \cheops instrumental systematics. We excluded from the process the smear and contamination effects, because they are already accounted for by the \texttt{PIPE} extraction. We chose the detrending vectors via a Bayes factor pre-selection method \citep{Trotta2007}, which routinely fits each visit with a least mean squares method, accounting for the transit model and one of the possible decorrelation terms. We performed preliminary modeling of the \cheops transits of \sname~b and d using Markov chain Monte Carlo (MCMC) simulations implemented in \texttt{pycheops}, including the effects of both the transits and the decorrelation terms. We finally flattened the light curves, dividing the data by the inferred model of the instrumental systematics. Figure~\ref{fig:Transit_Light_curves} shows the folded \cheops transit light curves of \sname~b and d, as derived from the global analysis presented in Sect.~\ref{Sec:Joint_Analysis}.

\section{\soar speckle imaging observations}
\label{Sec:SoarObservations}

\begin{figure}[!t]
    \centering
    \includegraphics[trim= 0cm 0cm 0cm 0cm,clip, width=\linewidth]{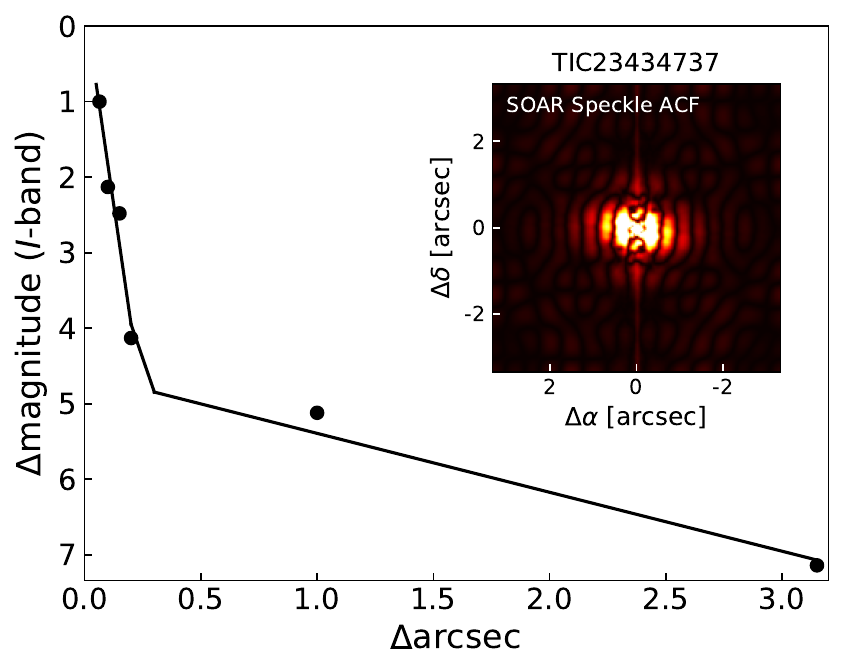}
    \caption{Contrast curve computed from the high-resolution speckle observations in Cousins I-band on the 4.1\,m Southern Astrophysical Research telescope. The inset shows the speckle autocorrelation function centered on the target star. No bright companions are detected within 3$\arcsec$ of \sname in this observation.}
    \label{fig:Speckle_Image}
\end{figure}

High angular resolution imaging is needed to search for nearby sources that can contaminate the \tess photometry, resulting in an underestimated planetary radius, or be the source of astrophysical false positives, such as background eclipsing binaries. We searched for stellar companions to \sname with speckle imaging on the 4.1\,m Southern Astrophysical Research (\soar) telescope \citep{Tokovinin2018} on 12 December 2019 UTC, observing in Cousins I-band, a visible bandpass similar to the \tess bandpass. This observation was sensitive to a 5.1-magnitude fainter star at an angular distance of 1$\arcsec$ from the target. More details of the observations within the \soar \tess survey are available in \citet{Ziegler2020}. The 5$\sigma$ detection sensitivity and speckle auto-correlation functions from the observations are shown in Fig.~\ref{fig:Speckle_Image}. No nearby stars were detected within 3$\arcsec$ of \sname in the \soar observations.

\section{Stellar parameters}
\label{Stellar_Parameters}

\subsection{Photospheric parameters and chemical abundances}

We co-added the individual \harps spectra prior to performing the spectroscopic analysis described in the present section. To this aim, we Doppler-shifted the individual spectra to a common reference wavelength by cross-correlating each spectrum against the one with the highest S/N. We finally carried out an S/N-weighted co-addition of the Doppler-shifted spectra, while applying a 3$\sigma$-clipping algorithm to remove possible cosmic ray hits and outliers. The co-added \harps spectrum has an S/N of $\sim$1730 per pixel at 5500~\AA. 

The stellar spectroscopic parameters of \sname, namely the effective temperature \teff, surface gravity \logg, microturbulent velocity \vmic, and iron abundance [Fe/H], were determined from the co-added \harps spectrum using the \texttt{ARES}+\texttt{MOOG} methodology described in \citet{Santos2013}, \citet{Sousa2014}, and \citet{Sousa2021}. We utilized the latest version of the Automatic Routine for line Equivalent widths in stellar Spectra\footnote{Publicly available at \url{https://github.com/sousasag/ARES}.} \citep[\texttt{ARES};][]{Sousa2007, Sousa2015} to measure the equivalent widths of selected iron (Fe) lines. The list of iron lines is the same as the one presented in \citet{Sousa2008}. We used a $\chi^2$ minimization procedure to find the ionization and excitation equilibrium and derive the best-fitting spectroscopic parameters. This process made use of a grid of Kurucz model atmospheres \citep{Kurucz-93} and the radiative transfer code \texttt{MOOG} \citep{Sneden-73}. We derived an effective temperature of \teff\,=\,5737\,$\pm$\,62~K, a surface gravity of log\,$g_\star$\,=\,4.34\,$\pm$\,0.10 (cgs), a microturbulent velocity of $v_\mathrm{mic}$\,=\,1.02\,$\pm$\,0.03~\kms, and an iron abundance of [Fe/H]\,=\,$-0.39$\,$\pm$\,0.04 (Table~\ref{tab:stellar_parameters}).  We also derived a more precise trigonometric \logg\,=\,4.26\,$\pm$\,0.03 using recent DR3 Gaia data \citep{GaiaCollaboration_DR3_2023}, following the procedure described in \citet{Sousa2021}, which provided a surface gravity in very good agreement with the spectroscopic one. The effective temperature and surface gravity imply that \sname is a G3\,V star, based on the compilation of \citet{Pecaut2012} and \citet{Pecaut2013}.

The photospheric abundances of magnesium (Mg), silicon (Si), and titanium (Ti) were also derived using the same tools and models as for the stellar parameter determination, under the assumption of local thermodynamic equilibrium. For the derivation of abundances, we closely followed the methods described in \citet{Adibekyan2012} and \citet{Adibekyan2015}. Taking into account the mean of the abundances of Mg, Si, and Ti as [$\alpha$/H], we calculated the $\alpha$-element enhancement of the star relative to iron and its metallicity $[\mathrm{M/H}]$ following \citet{Yi2001}. The derived abundances are listed in Table~\ref{tab:stellar_parameters}.

To ensure the validity of our findings, we also derived the spectroscopic parameters of \sname from the combined HARPS spectrum utilizing Spectroscopy Made Easy\footnote{Publicly available at \url{https://github.com/AWehrhahn/SME}.} \citep[\texttt{SME};][]{Valenti1996,Piskunov2017}. This spectral analysis tool computes synthetic spectra and fits them to high-resolution observed spectra via a $\chi^2$ minimization procedure. The analysis was conducted using the non-local thermodynamic equilibrium (non-LTE) version 5.2.2 of \texttt{SME}, in conjunction with \texttt{MARCS} model atmospheres \citep{Gustafsson2008}. 

We assumed a microturbulence and a macroturbulence velocity of \vmic\,=\,1.0\,$\pm$\,0.1,~\kms and \vmac\,=\,3.2\,$\pm$\,0.7~\kms, respectively, based on the empirical calibration equations for Sun-like stars from \citep{Bruntt2010} and \citep{Doyle2014}. The effective temperature \teff\ was determined by fitting the wings of the H$\alpha$ and H$\beta$ lines, in addition to those of the Na I doublet at 5890 and 5896~$\AA$ \citep[see, e.g.,][]{Fuhrmann1993,Axer1994,Fuhrmann1994}. The surface gravity \logg\ was measured from the wings of the Ca I $\lambda$ 6102, $\lambda$ 6122, $\lambda$ 6162~$\AA$ triplet, and the Ca I $\lambda$ 6439~$\AA$ line, as well as from the Mg~I $\lambda$\,5167, $\lambda$\,5173, $\lambda$\,5184~$\AA$ triplet. The determination of iron abundance [Fe/H] and projected rotational velocity \vsini\ was achieved through the simultaneous fitting of unblended iron lines located in the spectral region from 5880 to 6600~$\AA$. Our \texttt{SME} analysis provides an effective temperature of \teff\,=\,5740\,$\pm$\,45~K, a surface gravity of \logg\,=\,4.39\,$\pm$\,0.06 (cgs), an iron abundance of [Fe/H]\,=\,$-$0.32\,$\pm$\,0.05, and a projected rotational velocity of \vsini\,=\,1.4\,$\pm$\,0.5~\kms, which are in good agreement with those derived using \texttt{ARES}+\texttt{MOOG}, corroborating the robustness of our findings.

\subsection{Stellar radius, mass, and age}

Using the spectroscopic parameters derived with \texttt{ARES}+\texttt{MOOG} and the observed broadband fluxes and uncertainties retrieved from the most recent data releases in the following bandpasses: {\it Gaia} G, G$_{\rm BP}$, and G$_{\rm RP}$, 2MASS J, H, and Ks, and {\it WISE}\ W1 and W2 \citep{Skrutskie2006,Wright2010,GaiaCollaboration_DR3_2023}, we determined the stellar radius via an MCMC infrared flux method \citep{Blackwell1977,Schanche2020}. By constraining stellar atmospheric models with our $T_\mathrm{eff}$, log\,$g_\star$, and $[\mathrm{M/H}]$ we produced synthetic photometry in the {\it Gaia}, 2MASS, and {\it WISE} bandpasses that are compared to the broadband photometry to obtain the stellar angular diameter that is converted to the stellar radius using the {\it Gaia} DR3 offset-corrected parallax \citep{Lindegren2021}. To account for uncertainties in stellar atmospheric modeling, we conducted a Bayesian modeling averaging of the posterior distributions from the individual \texttt{ATLAS} \citep{Kurucz1993,Castelli2003} and \texttt{PHOENIX} \citep{Allard2014} catalogs to produce a value of $R_\star$\,=\,1.179\,$\pm$\,0.011\,$R_{\odot}$ (Table~\ref{tab:stellar_parameters}).

Adopting \teff, $[\mathrm{Fe/H}]$, and $R_\star$ as input parameters, we finally derived the stellar mass $M_\star$ and age $t_\star$ using two different sets of stellar evolutionary models. The first pair of mass and age estimates was computed via the isochrone placement algorithm \citep{bonfanti15,bonfanti16}, which interpolates the input values within pre-computed grids of \texttt{PARSEC}\footnote{\textsl{PA}dova and T\textsl{R}ieste \textsl{S}tellar \textsl{E}volutionary \textsl{C}ode: \url{http://stev.oapd.inaf.it/cgi-bin/cmd}.} v1.2S isochrones and evolutionary tracks \citep{marigo17}. We also derived a second pair of mass and age values employing the \texttt{CLES} \citep[Code Liègeois d'Évolution Stellaire;][]{scuflaire08} code, which generates the best-fitting evolutionary track following the Levenberg-Marquadt minimization scheme, as described in \citet{salmon21}. After checking the consistency of the two respective pairs of outcomes via the $\chi^2$-based criterion described in \citet{bonfanti21}, we finally merged our results and obtained a mass of $M_\star$\,=\,0.886\,$\pm$\,0.036~$M_{\odot}$ and an age of $t_\star$\,=\,12.5$_{-2.4}^{+1.3}$~Gyr (Table~\ref{tab:stellar_parameters}).

\begin{figure*}[!t]
    \centering
    \includegraphics[trim=0cm 0cm 0cm 0cm,clip, width=0.475\linewidth]{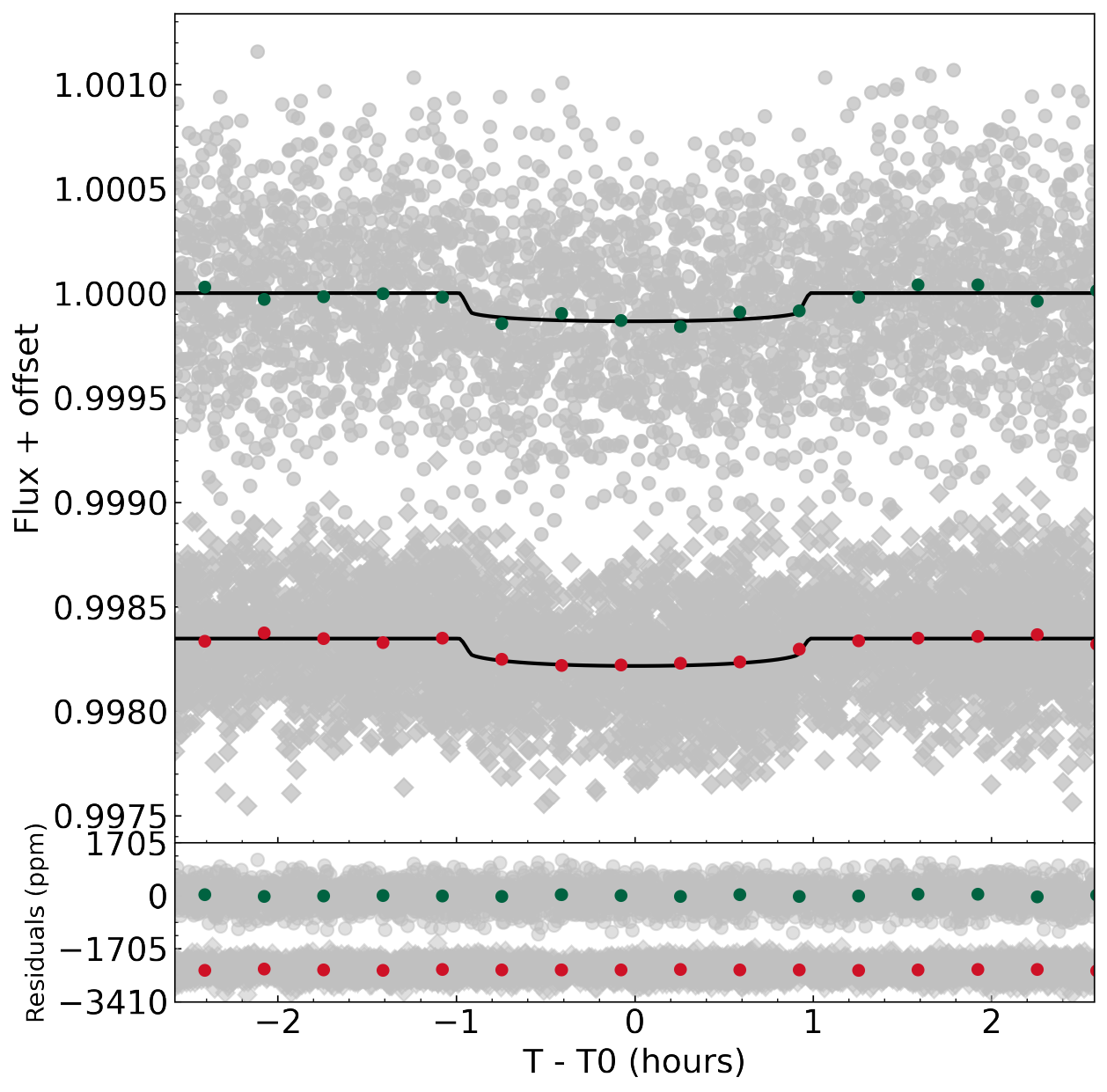}
    \hfill
    \includegraphics[trim= 1cm 0cm 0cm 0cm,clip, width=0.45\linewidth]{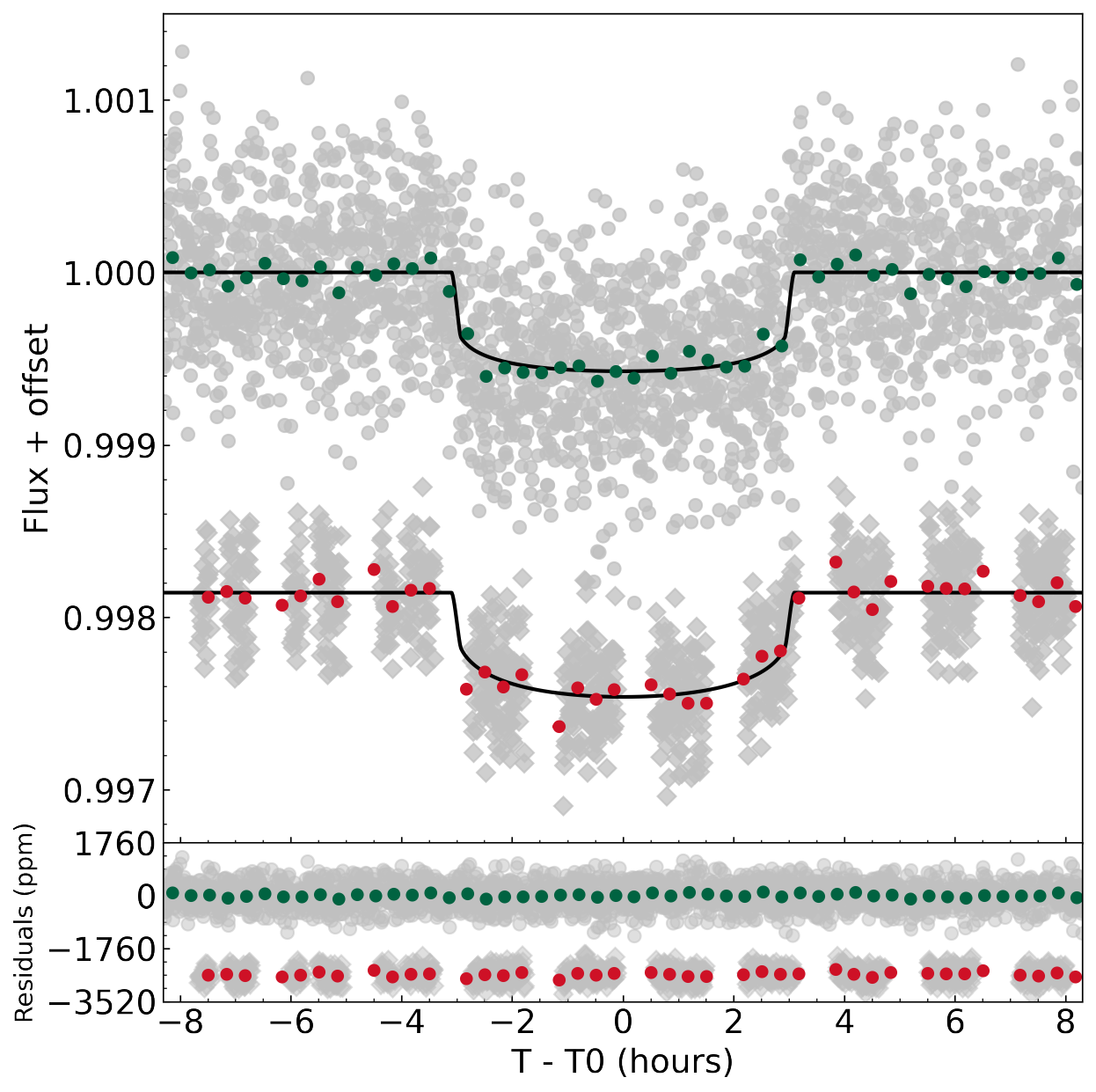}
    \caption{Folded transit light curves of \sname~b (left panel) and \sname~d (right panel). The \tess\ (top) and \cheops\ (bottom) data points are plotted with gray circles. The 20-minute binned photometry is displayed as green and red circles for \tess\ and \cheops, respectively. The thick black lines mark the transit models derived from the medians of the marginalized posterior distributions of the transit parameters (Table~\ref{tab:modelparameters}).}
    \label{fig:Transit_Light_curves}
\end{figure*}

\begin{figure*}[!h]
    \centering
    \includegraphics[width=\linewidth]{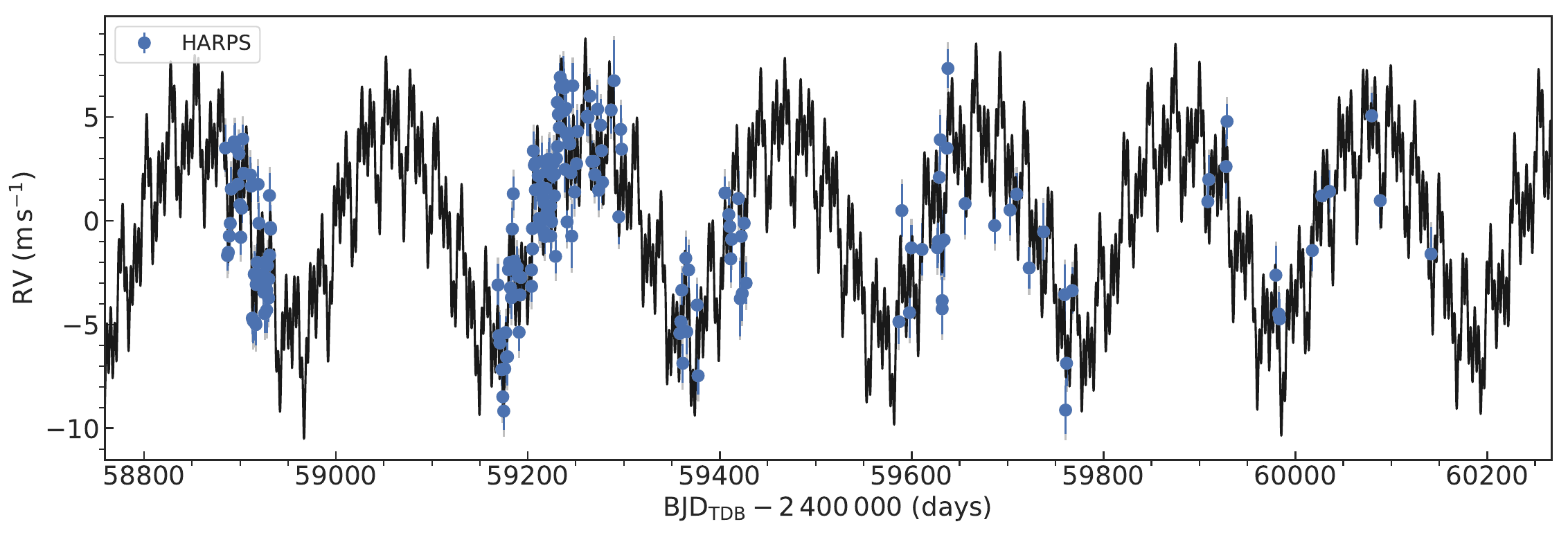}
    \includegraphics[width=0.495\linewidth]{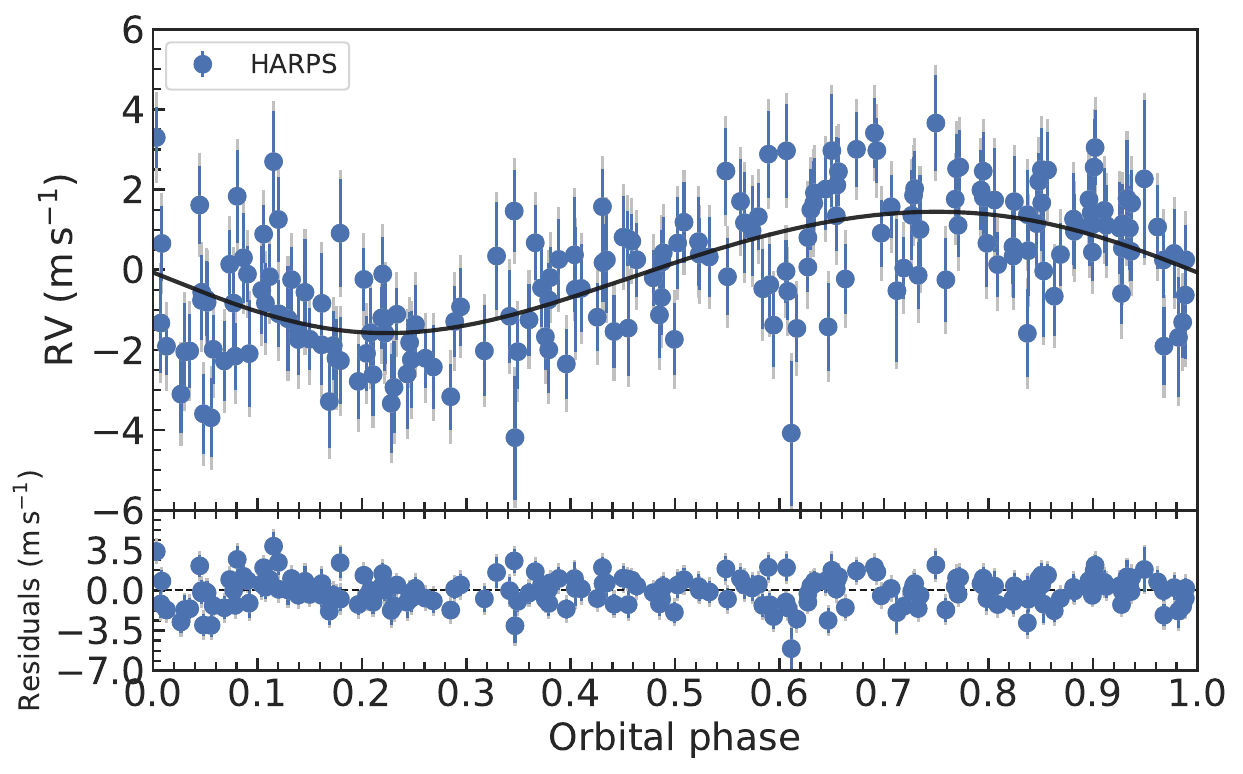}
    \includegraphics[width=0.495\linewidth]{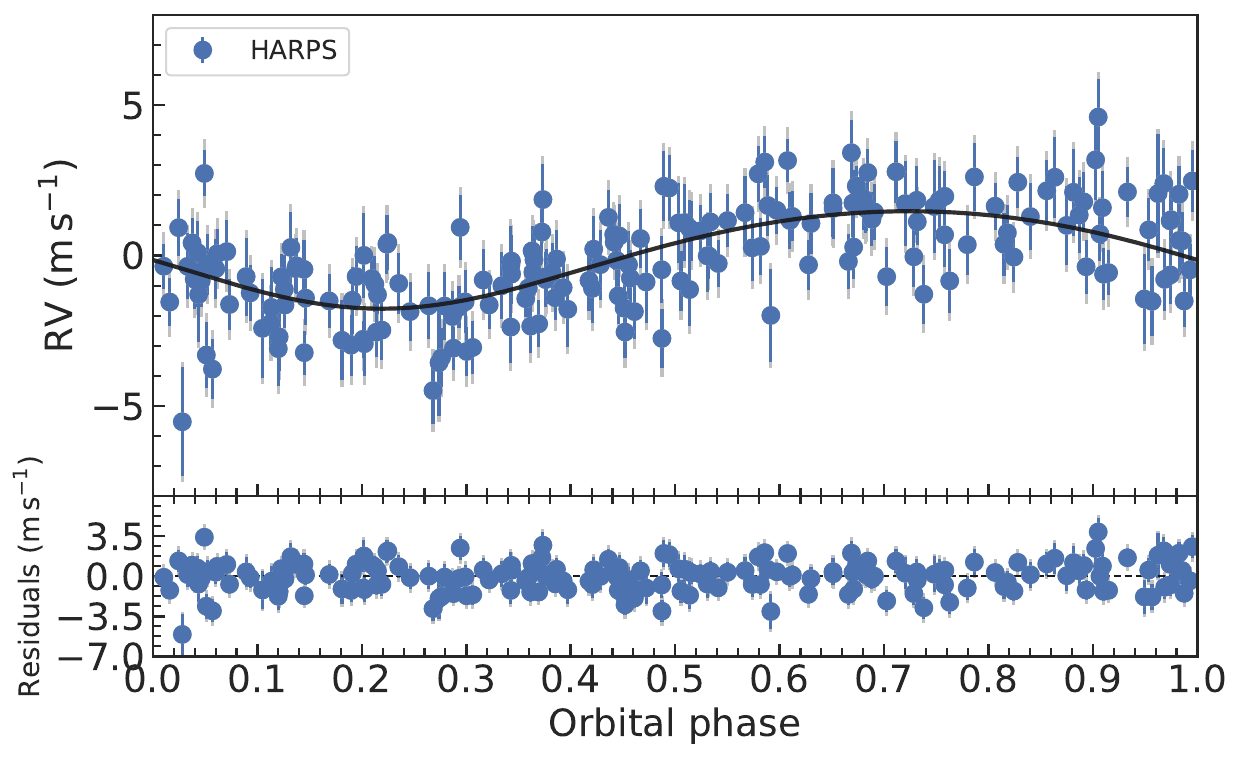}
    \includegraphics[width=0.495\linewidth]{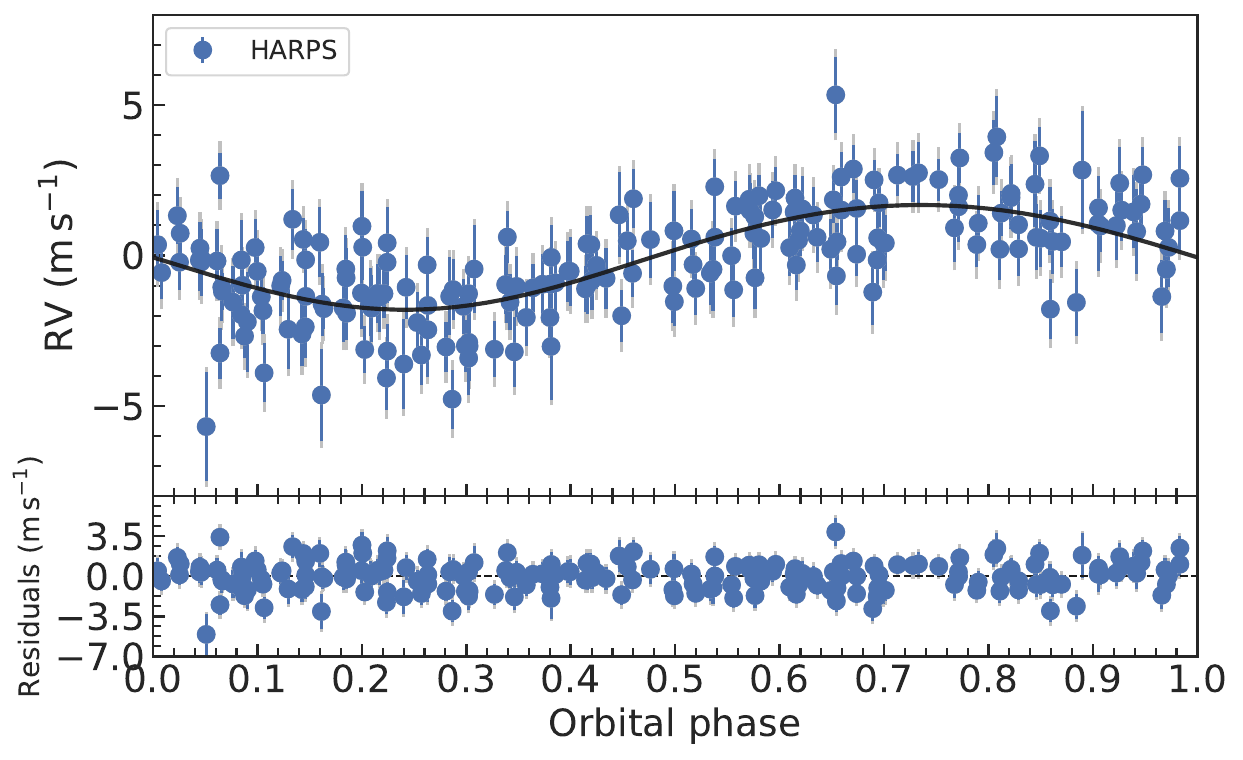}
    \includegraphics[width=0.495\linewidth]{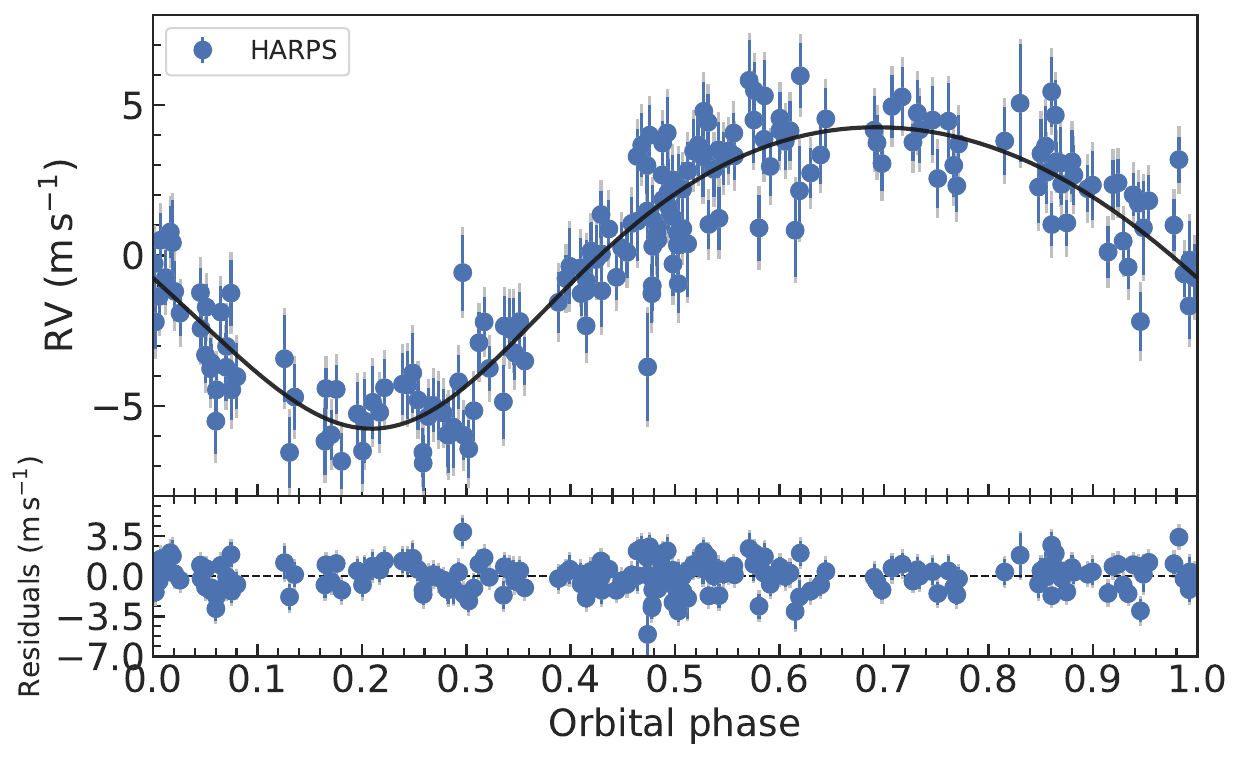}
    \caption{\emph{Upper panel}: \harps \terra RV time series. The thick black line marks the Keplerian model derived from the medians of the marginalized posterior distributions of the RV parameters (Table~\ref{tab:modelparameters}). \emph{Lower panels}: Phase-folded RV curve of \sname~b (middle left), c (middle right), d (bottom left), and e (bottom right) and Keplerian model (thick black line). The error bars include the RV jitter contribution in gray.}
    \label{fig:RV_curves}
\end{figure*}

\subsection{Stellar rotation and activity}

In an attempt to measure the rotation period of \sname, we masked out the transits of \sname~b and d from the \tess SAP light curves and searched the residuals for periodic and quasi-periodic photometric variations induced by the presence of photospheric active regions (spots and plages) coupled with stellar rotation. The \tess SAP light curves of \sname are affected by momentum dumps and thermal ramping \citep[see, e.g.,][]{Feinstein2019,Lund2021,Hartman2025}. After masking out the latter and modeling the former with quadratic regressions, we found that the periodogram of the residuals shows no significant peaks with bootstrap FAPs (see Sect.~\ref{Sec:HARPS_FrequencyAnalysis}) lower than $\sim$10\,\%.

We also performed a periodogram analysis of ground-based archival photometry from the Wide Angle Search for Planets (\wasp) project \citep{2006PASP..118.1407P}. The field of \sname was monitored by \wasp in 2007 and then from 2011 to 2014. In the years 2007 and 2011 \wasp-South was equipped with 200\,mm, f/1.8 lenses and a filter spanning 400--700~nm, while from 2012 to 2014 it used 85\,mm, f/1.2 lenses with an SDSS-$r$ filter. Observing with a typical 10-minute cadence over spans of 160 nights in each observing season, \wasp-South accumulated a total of 50\,000 photometric data points for \sname. We searched the accumulated data for any rotational modulation using the methods from \citet{2011PASP..123..547M}, but found no significant periodicity. For periods from 2~d out to $\sim$100~d we estimate a 95\,\%-confidence upper limit of 2~mmag.

Based on our measurement of the stellar radius ($R_{\star}$\,=\,1.179\,$\pm$\,0.011\,$R_{\odot}$) and projected rotational velocity (\vsini\,=\,1.4\,$\pm$\,0.5\,\kms), we derived an upper limit for the rotation period of $P_\mathrm{rot,max}$\,=\,$43^{+23}_{-11}$~d. Using the gyrochronology method proposed by \citet{Barnes2007}, along with the rotation--activity--age relation proposed by \citet{2008ApJ...687.1264M} and \citet{2015MNRAS.450.1787A}, and based on our age determination ($t_{\star}$\,=\,$12.5^{+1.3}_{-2.4}$\,Gyr) and the $B-V$\,=\,0.620\,$\pm$\,0.042 color index (Table~\ref{tab:stellar_parameters}), we estimated a stellar rotation period of $P_\mathrm{rot}$\,=\,39\,$\pm$\,5~d. Using the \logrhk--rotation period empirical relation found by \citet{SuarezMascareno2015} for metal poor stars ($-0.5$\,$\leq$\,[Fe/H]\,$\leq$\,$-0.1$), along with the median value of the Ca\,{\sc ii} H\,\&\,K lines activity indicator \logrhk\,=\,$-4.998\,\pm\,0.001$ (Table~\ref{tab:stellar_parameters}), we estimated a rotation period of $P_\mathrm{rot}$\,=\,31$^{+12}_{-9}$~d, in agreement with the previous estimate.

We conclude that \sname is a slowly rotating, inactive, old star with a low level of magnetic activity. This is further supported by the lack of periodic and quasi-periodic activity-induced signals in the Ca\,{\sc ii} H\,\&\,K lines activity indicator \logrhk\ and line profile variation diagnostics (Sect.~\ref{Sec:HARPS_FrequencyAnalysis}). This finding is not unexpected given the low metallicity of \sname ([Fe/H]\,=\,$-0.39\,\pm\,0.04$) and reinforces previous studies showing that metal-poor stars are generally less magnetically active \citep{See2021} and flare less frequently \citep{See2023}.

\subsection{Binarity}\label{sec_binarity}

Analysis of stellar proper motions by \gaia DR2 shows that \sname is likely a visual binary star system \citep{Mugrauer20}. In fact, the primary star has a very faint ($\Delta$G\,=\,$-$11.45) co-moving M-dwarf companion. The companion is located at an angular separation of 11.74\arcsec\ from the primary, which translates into a projected binary separation of $\sim$765~au \citep{Mugrauer20}. The effective temperature and mass of the secondary star are estimated to be $\sim$2500~K and $0.08\pm0.01$~M$_{\odot}$, respectively \citep{Mugrauer20}. 

\subsection{Membership probability to galactic dynamical populations}

We investigated the membership probability of \sname to the various galactic dynamical populations. Using \gaia DR3 coordinates, proper motions, and parallax, along with our measured stellar absolute RV (Table~\ref{tab:stellar_parameters}), we computed the galactic U, V, and W velocities in the local standard of rest (LSR). Adopting the LRS determination of \citet{Schonrich2010}, we found: U$_\mathrm{LSR}$\,=\,106.54\,$\pm$\,0.10~km\,s$^{-1}$, V$_\mathrm{LSR}$\,=\,$-$33.49\,$\pm$\,0.10~km\,s$^{-1}$ and W$_\mathrm{LSR}$\,=\,59.96\,$\pm$\,0.07~km\,s$^{-1}$. Using the method given by \citet{Reddy2006}, we computed the membership probabilities of \sname in the galactic thin disk, thick disk, and halo as: P$_\mathrm{thin}$\,=\,5.6\,$\pm$\,0.1\,\%, P$_\mathrm{thick}$\,=\,93.2\,$\pm$\,0.5\,\%, and P$_\mathrm{halo}$\,=\,1.182\,$\pm$\,0.003\,\%. When we combine this kinematic classification of the star with our abundance determinations of [Fe/H]\,=\,$-$0.39\,$\pm$\,0.04 and [$\alpha$/Fe]\,=\,0.21\,$\pm$\,0.04, and with our derived stellar age of $\sim$12.5~Gyr, we can convincingly classify \sname as a member of the old $\alpha$-element enhanced galactic thick disk stellar population.

\section{Joint analysis}
\label{Sec:Joint_Analysis}

We performed a joint analysis of the \tess\ and \cheops\ transit light curves (Sect.~\ref{Sec:TESS_Observations} and Sect.~\ref{Sec:Cheops_Observations}) and \harps \terra RV measurements (Sect.~\ref{Sec:HARPS_Observations}) using the software suite \texttt{pyaneti} \citep{Barragan2019,Barragan2021}, which allows one to determine the model parameters from posterior distributions derived using MCMC methods. To speed up the computation process, for the \tess transit photometry, we extracted $\sim$6 and 18 hours of segments from the flattened, 120-second cadence, \tess PDC-SAP light curves (Sect.~\ref{Sec:TESS_Observations}), with each segment centered around the transits of \sname~b~and~d, respectively. Due to the lack of a detectable transit signal for planet~b in \tess Sector~36 (Sect.~\ref{Sec:TransitSearch}), we opted to exclude the transit photometry of \sname~b collected during this sector. The total number of modeled transits of \sname~b is 32, with 22 transits observed by \tess (Sector 9, 10, 63, and 90) and 10 by \cheops. For \sname~d, we modeled 6 transits, of which 5 were observed by \tess (Sectors 9, 10, 36, 63, and 90) and 1 by \cheops. The transit photometry covers a baseline of about 6 years.

The RV model includes four Keplerians, to account for the Doppler reflex motion induced by \sname~b, c, d, and e. Any instrumental noise not included in the nominal RV uncertainties was accounted for by adding an RV jitter term. We modeled the \tess and \cheops transit light curves using the limb-darkened quadratic law of \citet{Mandel2002} and adopted the parametrization proposed by \citet{Kipping2013} for the linear and quadratic limb-darkening coefficients $u_1$ and $u_2$. We set Gaussian priors on the \citet{Kipping2013}'s parameters $q_1$ and $q_2$ using the $u_1$ and $u_2$ theoretical values derived by \citet{Claret2017} and \citet{Claret2021} for the \tess and \cheops bandpasses, and imposed conservative error bars of 0.1 for both $q_1$ and $q_2$. For the eccentricity ($e$) and the argument of periastron of the stellar orbit ($\omega_\star$) we adopted the parametrization proposed by \citet{Anderson2011}, i.e., $\sqrt{e}\sin{\omega_\star}$ and $\sqrt{e}\cos{\omega_\star}$. A preliminary analysis showed that the transit light curve poorly constrains the scaled semimajor axis ($a/R_\star$). We therefore sampled the stellar density $\rho_\star$ setting a Gaussian prior on the mass and radius of the star (Table~\ref{tab:stellar_parameters}), and recovered the scaled semimajor axis $a/R_\star$ via Kepler's third law \citep{Winn2010}. We imposed uniform priors for the remaining model parameters. Details of the modeled parameters and adopted priors are given in Table~\ref{tab:modelparameters}. 

We used 500 independent Markov chains initialized randomly inside the prior ranges. Once all chains converged, we used the last 500 iterations and saved the chain states every 10 iterations. This approach generates a posterior distribution of 250\,000 points for each model parameter. We initially performed a preliminary analysis with 100 Markov chains to estimate the orbital periods and times of inferior conjunctions\footnote{Mid-times of reference transit for \sname~b and d.} (along with their uncertainties) for the four planets, while adopting wide uniform priors for the two model parameters. We then centered the priors around the estimated values, narrowed the ranges to $\sim$30--60$\sigma$ to speed up convergence, and conducted the final analysis using 500 Markow chains. Upon completion of the MCMC modeling, we verified that none of the final posterior distributions was truncated. Tables~\ref{tab:modelparameters} and \ref{tab:derivedparameters} list the values and their uncertainties of the model and the derived planetary parameters, respectively. They are defined as the median and 68.3\,\% region of the credible interval of the marginalized posterior distributions for each inferred parameter. The transit and RV curves are shown in Fig.~\ref{fig:Transit_Light_curves} and Fig.~\ref{fig:RV_curves}, along with the models derived from the medians of the marginalized posterior distributions of the inferred parameters (Table~\ref{tab:modelparameters}).

\section{Dynamical analysis}
\label{sec:Dynamical_Analysis}

The planetary system orbiting \sname is notably compact, particularly with respect to the three innermost planets (namely, \sname~b, c, and d), which display period ratios of $P_\mathrm{c}/P_\mathrm{b}$\,$\approx$\,3.14 and $P_\mathrm{d}/P_\mathrm{c}$\,$\approx$\,1.95 suggesting the existence of potential resonant interactions between the planets. Furthermore, the period ratio $P_\mathrm{e}/P_\mathrm{d}$\,$\approx$\,8.02 further suggests an 8:1 commensurability between the outermost planets d and e. This complex configuration raises several interesting questions regarding the dynamics of the system, such as whether the observed system is stable over long timescales, if the planets are locked in two-body or three-body mean motion resonances (MMRs), and how the uncertainties of the orbital elements, such as orbital periods, eccentricities, and inclinations, impact the overall stability. To address these questions, we performed a detailed analysis of the dynamics and stability of the system based on the orbital parameters listed in Tables~\ref{tab:modelparameters} and \ref{tab:derivedparameters}. 

\subsection{Stability analysis}

\begin{figure*}[thp!]
    \centering
    \includegraphics[width=\textwidth]{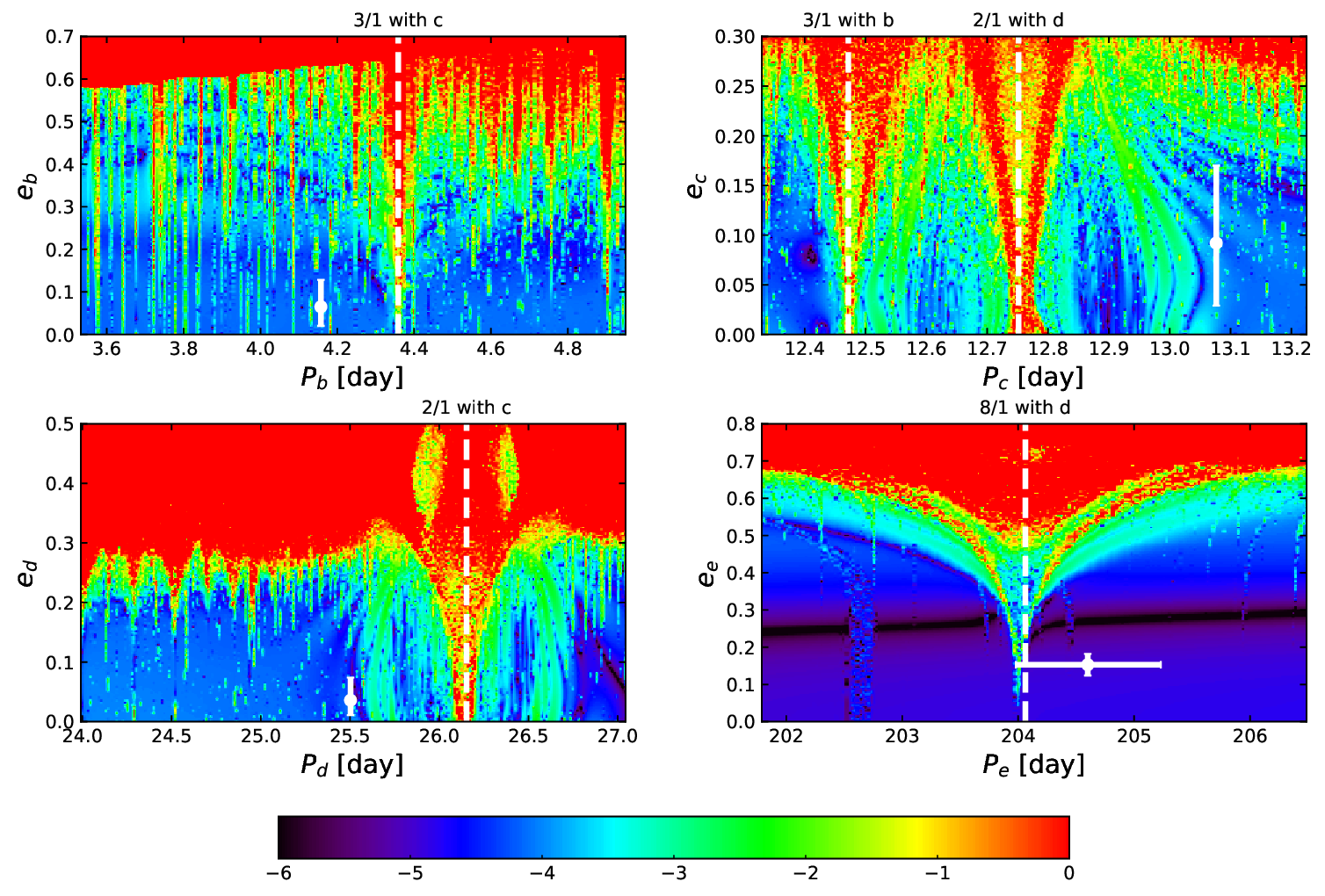}
    \caption{Stability analysis of the \sname system in the orbital period-eccentricity domain: planet~b (top left), c (top right), d (bottom left), and e (bottom right). For fixed initial conditions, the phase space of the system is explored by varying the orbital period  $P_i$  and eccentricity $e_i$ of each planet independently. For each initial condition, the system is integrated over 40~kyr, and a stability criterion is derived with the frequency analysis of the mean longitude. The chaotic diffusion is measured by the variation in the frequencies and the color scale corresponds to the decimal logarithm of the stability index $D$. The red zone corresponds to highly unstable orbits, while the dark blue region can be assumed to be stable on a billion-year timescale. The current positions of the planets are marked with circles in each plot, and error bars represent the observational uncertainties (Tables~\ref{tab:modelparameters} and \ref{tab:derivedparameters}). The white dashed vertical lines correspond to the main MMRs in the system.}
    \label{fig:stability_analysis}
\end{figure*}

To analyze the stability of the system, we performed a global frequency analysis \citep{Laskar1990,LASKAR1993} in the vicinity of the inferred solution (Tables~\ref{tab:modelparameters} and \ref{tab:derivedparameters}), using the methodology described in \citet{Correia2005} and \citet{Correia2010}. The system was integrated on a regular 2D mesh of initial conditions, varying the orbital period and eccentricity of each planet individually, while keeping all other orbital elements fixed at their nominal values. For the non-transiting planets~c and e, we fixed their orbital inclinations at $90^\circ$ relative to the plane of the sky. Although \tess light curves rule out with high confidence levels an exact $90^\circ$ inclination for \sname~c and e -- since this would produce observable transits (Sect.~\ref{Sec:TransitSearch}) -- this assumption is a reasonable choice for a primary dynamical study, as we expect the true inclinations to be close to $90^\circ$. Moreover, this value provides minimum values for the true masses for the planets and corresponds to weaker planet-planet interactions, likely resulting in more stable (or ``optimistic'') stability maps for the system. The integrations were carried out using the symplectic integrator \texttt{SABA1064} of \cite{Farres2013}, with a time step of $5 \times 10^{-3}$~years, and included general relativity corrections. Each initial condition was integrated over 40~kyr, and a stability indicator was derived using the frequency analysis of the mean longitude. The stability indicator is defined as $D = \left|\,n_i - n_i^{\prime} \,\right|$, where $n_i$ and $n_i^{\prime}$ represent the mean motions of the planets (in degrees per year) calculated over the first and second halves of the integration interval, respectively \citep[see][for more details]{Couetdic2010}. For regular motion, there is no significant variation in the mean motion along the trajectory, while it can vary significantly for chaotic trajectories. 

Figure~\ref{fig:stability_analysis} shows our results in the orbital period-eccentricity domain for the four planets, where the red area marks strongly chaotic orbits, while the dark blue one represents configurations typically considered extremely stable over a billion-year timescale. The maps show many ``islands'' of MMRs, appearing as long V-shaped structures, several of which are unstable. None of the planets orbiting \sname lies within these resonances. Instead, they are located in more stable regions (dark blue areas) either between, or near resonant zones. We hence conclude that the planetary system TOI-1203 is dynamically stable. In particular, the pair b-c is close to a 3:1 MMR, c-d near a 2:1 MMR, and d-e near an 8:1 MMR. We also note that the inner pair b-c is on the correct side of the resonance predicted by tidal evolution models \citep{Delisle2012,Delisle2014a}. The close proximity of \sname~b and c to their host star results in significant tidal interactions between the planets and the star, causing the period ratio to exceed the exact resonance value \citep{Lissauer2011,Delisle2014b}.

\begin{figure}[t!]
\centering
\includegraphics[width=\linewidth]{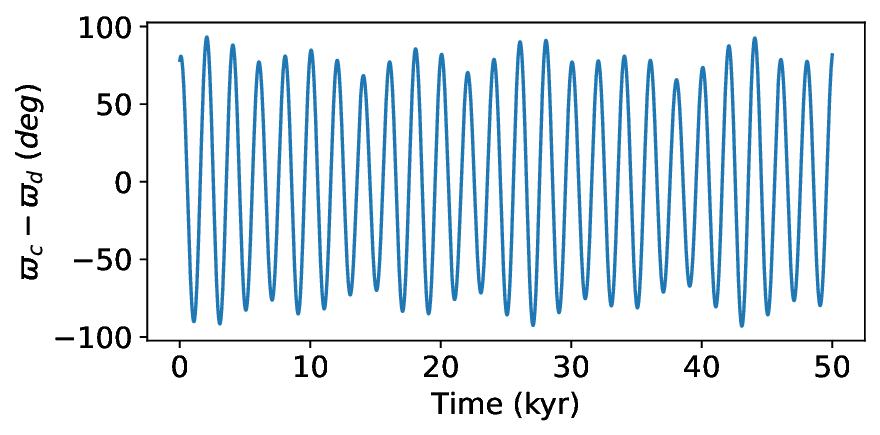}
\caption{Evolution of the angle $\varpi_c - \varpi_d$ in the \sname system, starting with the orbital solution listed in Tables~\ref{tab:modelparameters} and \ref{tab:derivedparameters}.}
\label{fig:evolution_varpi}
\end{figure}

We further tested directly the stability of the system by performing a numerical integration over 100~Myr, confirming that the system remains stable with regular orbital behavior throughout the simulation. In addition, we performed frequency analysis of the orbital solution over 1~Myr to determine the system's fundamental frequencies. These include the mean motions $n_\mathrm{b}$, $n_\mathrm{c}$, $n_\mathrm{d}$, and $n_\mathrm{e}$, the secular precession frequencies of the pericenters $g_1$, $g_2$, $g_3$, and $g_4$, and the secular frequencies of the nodes $s_1$, $s_2$, and $s_3$ as listed in Table~\ref{tab:frequencies}. The compact configuration of the three innermost planets results in a strong coupling within the secular system, particularly between planets c and d. The analysis shows that both planets precess with the same frequency $g_2$. The two pericenters are thus locked and $\Delta \varpi = \varpi_\mathrm{c} - \varpi_\mathrm{d}$ oscillates around $180^\circ$, with a maximum semi-amplitude of about $70^\circ$ (Fig.~\ref{fig:evolution_varpi}). This indicates that these planets maintained an anti-aligned configuration throughout the system's evolution. It should be noted that this behavior is not a dynamical resonance but rather results from a linear secular coupling.

To present the solution more clearly, it is useful to make a linear change of variables into eccentricity proper modes (see \citealp{Laskar1990}). Due to the proximity of the planets to MMRs and the eccentricity of the outer planet \sname~e ($e_\mathrm{e}$\,=\,\ee; Table~\ref{tab:derivedparameters}), this transformation is obtained numerically through frequency analysis of the solutions. Using the classical complex notation $z_p = e_p e^{i\varpi_p}$, where  $p$ = b, c, d, e, we expressed the linear Laplace-Lagrange solution as
\begin{equation}
\begin{pmatrix}
z_\mathrm{b} \\
z_\mathrm{c} \\
z_\mathrm{d} \\
z_\mathrm{e}
\end{pmatrix}
=
10^{-6} \times
\begin{pmatrix}
39316 & -15612 & 9648 & 3259 \\
-1038 & 63564 & 46694 & 6511 \\
-6362 & -54107 & 29238 & 8909 \\
17 & 21 & -419 & 15161
\end{pmatrix}
\begin{pmatrix}
u_1 \\
u_2 \\
u_3 \\
u_4
\end{pmatrix}.
\label{eq:matrix}
\end{equation}

The proper modes $u_k$ (with k=1, 2, 3, 4) are obtained from $z_p$ by inverting the above linear relation. To good approximation, we have $u_k \approx e^{i(g_kt+\phi_k)}$, where $g_k$ and the phases $\phi_k$ are given in Table~\ref{tab:frequencies}. From Eq.~\ref{eq:matrix}, one can better understand the observed libration between the pericenters $\varpi_c$ and $\varpi_d$. Indeed, for both planets c and d, the dominant term in the decomposition is $u_2$, with frequency $g_2$, and thus both orbits precess with the same rate of $g_2$ but with opposite phase.

\begin{table}[t!]
\centering
\caption{Fundamental frequencies, periods, and phases for the orbital solution in Tables~\ref{tab:modelparameters} and \ref{tab:derivedparameters}.}
\renewcommand{\arraystretch}{1.2}  
\begin{tabular}{l r r r}
\hline
\hline
\noalign{\smallskip}
& {Frequency} & {Period} & {Phase} \\
& {($^\circ$/yr)} & {(yr)} & {($^\circ$)}  \\
\noalign{\smallskip}
\hline
\noalign{\smallskip}
$n_\mathrm{b}$ & 31628.9 & 0.011382 &  -88.426\\
$n_\mathrm{c}$ & 10054.2 & 0.035806 &  -65.018\\
$n_\mathrm{d}$ & 5157.03 & 0.069807 & 125.732\\
$n_\mathrm{e}$ & 642.807 & 0.560044 & -19.980\\
$g_1$ & 0.109694 & 3281.84 & -175.394\\
$g_2$ & 0.255985 & 1406.33 & -95.101\\
$g_3$ & 0.045349 & 7938.39 & -154.429\\
$g_4$ & 0.001195 & 301170 & 14.578\\
$s_1$ &-0.101517 & 3546.21 & -179.369\\
$s_2$ &-0.008353 & 43099.9 & -179.868\\
$s_3$ &-0.270741 & 1329.68 & -174.254  \\
\noalign{\smallskip}
\bottomrule
\end{tabular}
\label{tab:frequencies}
\tablefoot{
    $n_\mathrm{b}$, $n_\mathrm{c}$, $n_\mathrm{d}$, and $n_\mathrm{e}$ are the mean motions, $g_1$, $g_2$, $g_3$, and $g_4$ the secular frequencies of the pericenters, $s_1$, $s_2$, and $s_3$ the secular frequencies of the nodes. The conservation of angular momentum implies that only three independent nodal frequencies of the pericenters are present (namely, $s_1$, $s_2$, and $s_3$), with $s_4$\,=\,0.
    }
\end{table}

\begin{figure}[t!]
\centering
\includegraphics[width=\linewidth]{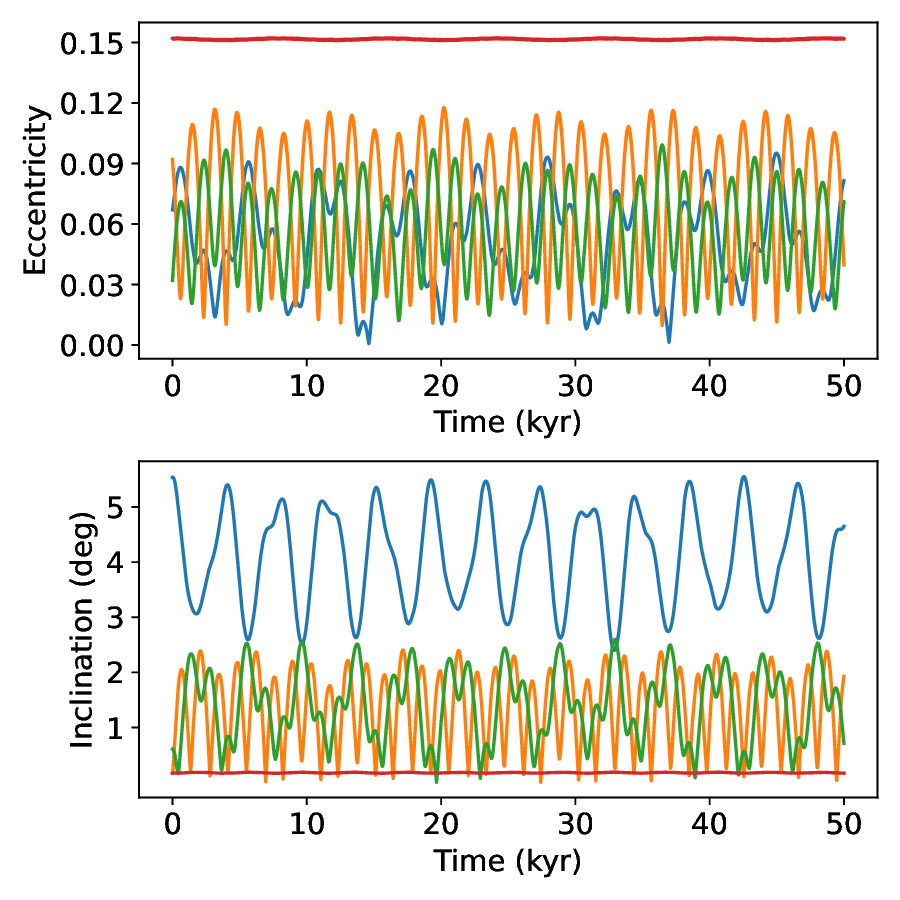}
\caption{Secular evolution of the eccentricities and inclinations with respect to the invariant plane in the \sname system, starting with the orbital solution from Tables~\ref{tab:modelparameters} and \ref{tab:derivedparameters}. The curves represent the different planets: blue (\sname~b), orange (\sname~c), green (\sname~d), and red (\sname~e). The orientation of the invariant plane is set by the total angular momentum vector of the system and is therefore strongly influenced by the orbital planes of the three outer and more massive planets (c, d, and e). Planet~d has an orbit that is nearly edge-on, while planets~c and e were initialized at an inclination of $90^\circ$ relative to the plane of the sky. As a result, the invariant plane lies close to the orbital planes of these three planets, i.e., close to $90^\circ$ relative to the plane of the sky.}
\label{fig:evolution}
\end{figure}

\begin{figure}[t!]
\centering
\includegraphics[width=\linewidth]{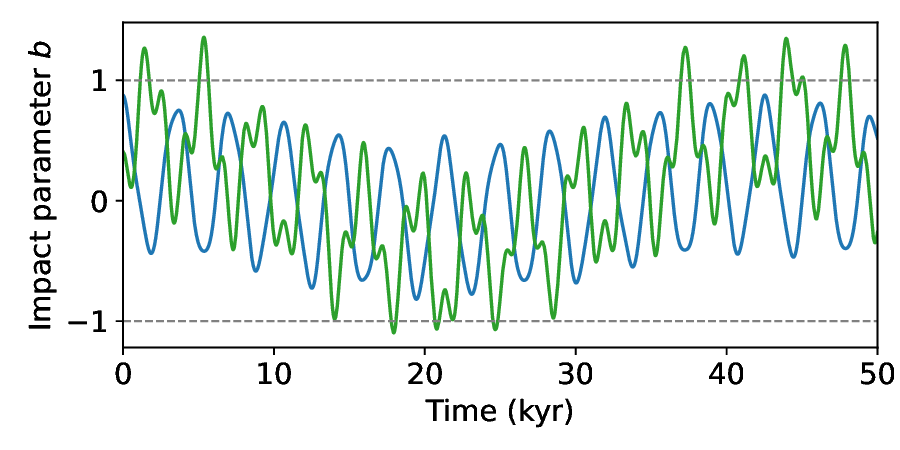}
\caption{Secular evolution of the impact parameter $b$ for \sname~b (blue) and \sname~d (green), starting with the orbital solution listed in Tables~\ref{tab:modelparameters} and \ref{tab:derivedparameters}.}
\label{fig:impactparam}
\end{figure}

Figure~\ref{fig:evolution} displays the evolution of the eccentricities and inclinations with respect to the invariant plane of the planets over $50$~kyr. The inner three planets exhibit considerable variations in both eccentricity and inclination. In contrast, \sname~e displays minimal variation, due to its higher mass and greater distance from the inner planets, which make it less susceptible to their perturbations, while still influencing their evolution. Figure~\ref{fig:impactparam} shows the evolution of the impact parameter $b$ for the transiting planets \sname~b and~d over the same time interval. Planet~b remains within the transiting regime throughout the integration, while planet~d shifts in and out of a transiting configuration over secular timescales.

\subsection{Resonant chain}

\begin{figure}[t!]
    \centering
    \includegraphics[width=\linewidth]{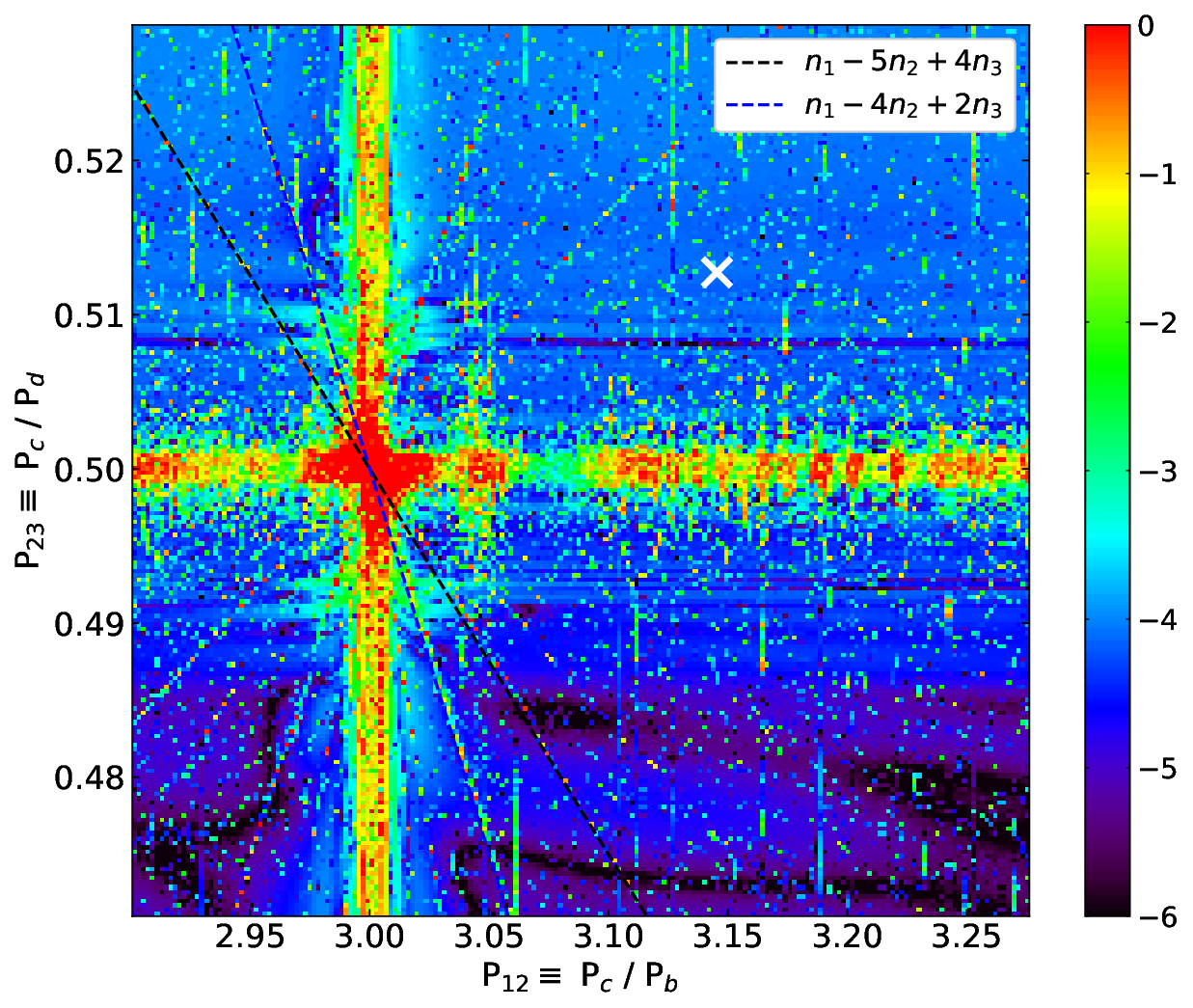}
    \caption{Stability analysis of the \sname system in the $(P_{12}, P_{23})$ plane, where $P_{12} = P_\mathrm{c} / P_\mathrm{b}$ and $P_{23} = P_\mathrm{c} / P_\mathrm{d}$. For fixed initial conditions, the phase space is explored by varying the period ratios of the inner three planets, while all other parameters are kept at their nominal values. As in Fig. \ref{fig:stability_analysis}, the color scale corresponds to the decimal logarithm of the stability index $D$. The red zones correspond to highly unstable orbits, while the dark blue regions can be assumed to be stable on a billion-year timescale. The dashed lines highlight the location of two low-order three-body MMRs, corresponding to a Laplace resonance (order 0) and a first-order resonance. The cross marks the location of the system, which lies outside any identified three-body resonances.}
    \label{fig:stability_analysis_periods}
\end{figure}

A three-planet MMR occurs when there exist integers  $k_1$, $k_2$, and $k_3$ such that $k_1 n_1 + k_2 n_2 + k_3 n_3 = 0$, where $n_1$, $n_2$, and $n_3$ are the mean motions of the planets. The resonance is classified according to the sum $|\,k_1 + k_2 + k_3\,|$, which defines the order of the resonance. To investigate whether the three innermost planets could be part of a resonant chain, we studied a different section of the phase space in the plane defined by the period ratios $P_{12} = P_\mathrm{c}/P_\mathrm{b}$ and $P_{23} = P_\mathrm{c}/P_\mathrm{d}$. Figure~\ref{fig:stability_analysis_periods} explores the stability in the $(P_{12}$,$P_{23})$ domain, while keeping other orbital parameters fixed at their nominal values. In this plane, two-body MMRs appear as vertical or horizontal bands, with the prominent ones corresponding to the 3:1 resonance between planets b and c, and the 2:1 resonance between planets c and d. Three-body MMRs, on the other hand, are represented by thinner, oblique lines. Here, we highlight two low-order resonances, corresponding to a Laplace resonance (order 0) and an order 1 resonance. Higher-order resonances appear increasingly thin, making them harder to detect and thus have smaller influence on the system’s dynamics. The cross in Fig.~\ref{fig:stability_analysis_periods} marks the location of our system, which lies outside any significant three-body MMRs, confirming that the planets are not currently locked in a resonant chain.

\subsection{Additional constraints}

For the non-transiting planets~c and e, we cannot directly determine their orbital inclinations $i$; so far, we have been assuming that their inclinations relative to the plane of the sky is $90^\circ$, which gives minimum values for the true planetary masses. Additionally, the longitude of the ascending node ($\Omega$) for all planets remains undetermined, as astrometry would be required to establish the orientation of the projected orbit on the plane of the sky. For simplicity, we have set all longitudes of ascending nodes to $0^\circ$, assuming a common reference direction in the plane of the sky. Thus, for planets~c and e, we can build a 2D stability map for the two unknown parameters to determine how dynamics can constrain their possible values.

\begin{figure}[t!]
\centering
\includegraphics[width=\linewidth]{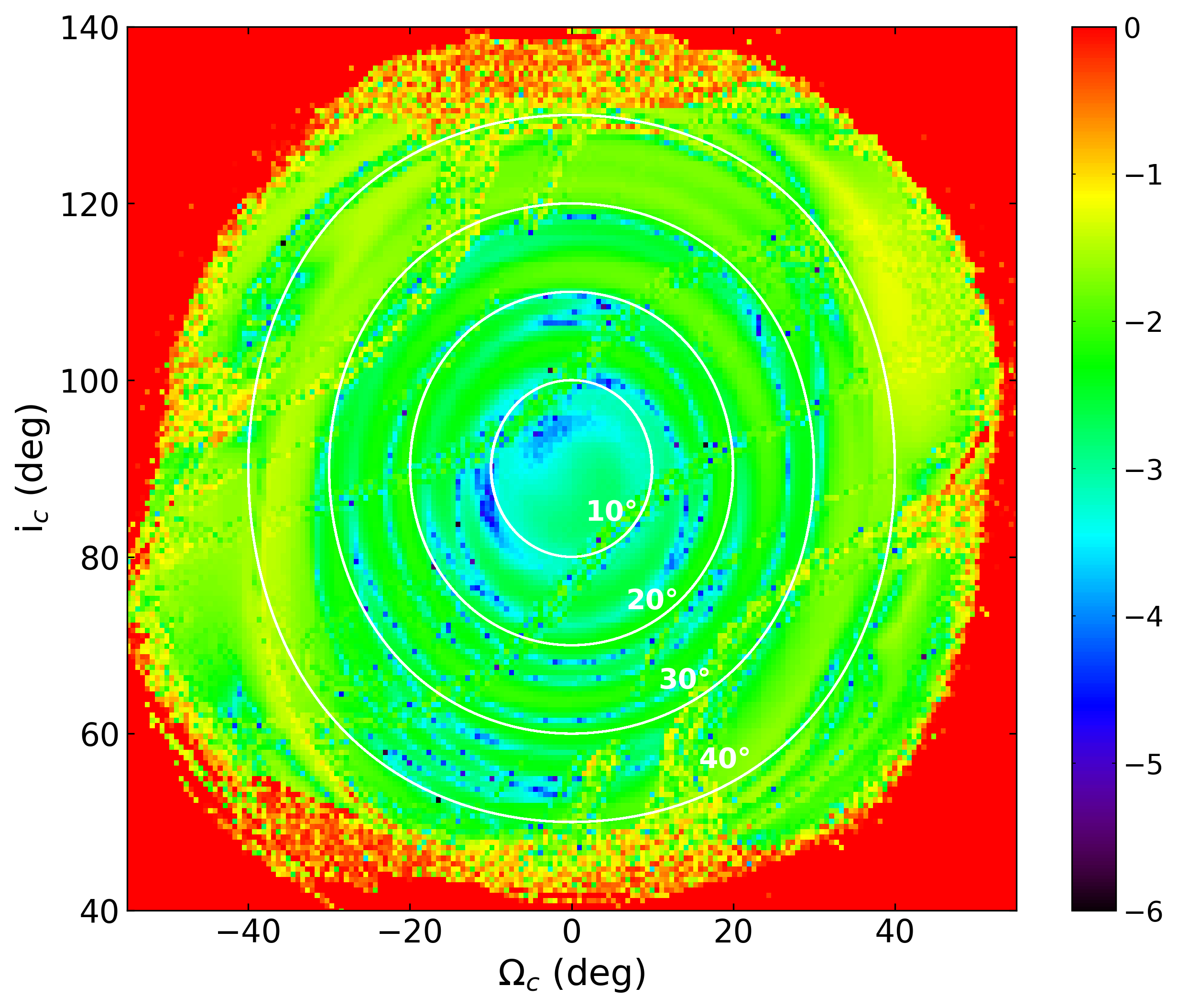}
\includegraphics[width=\linewidth]{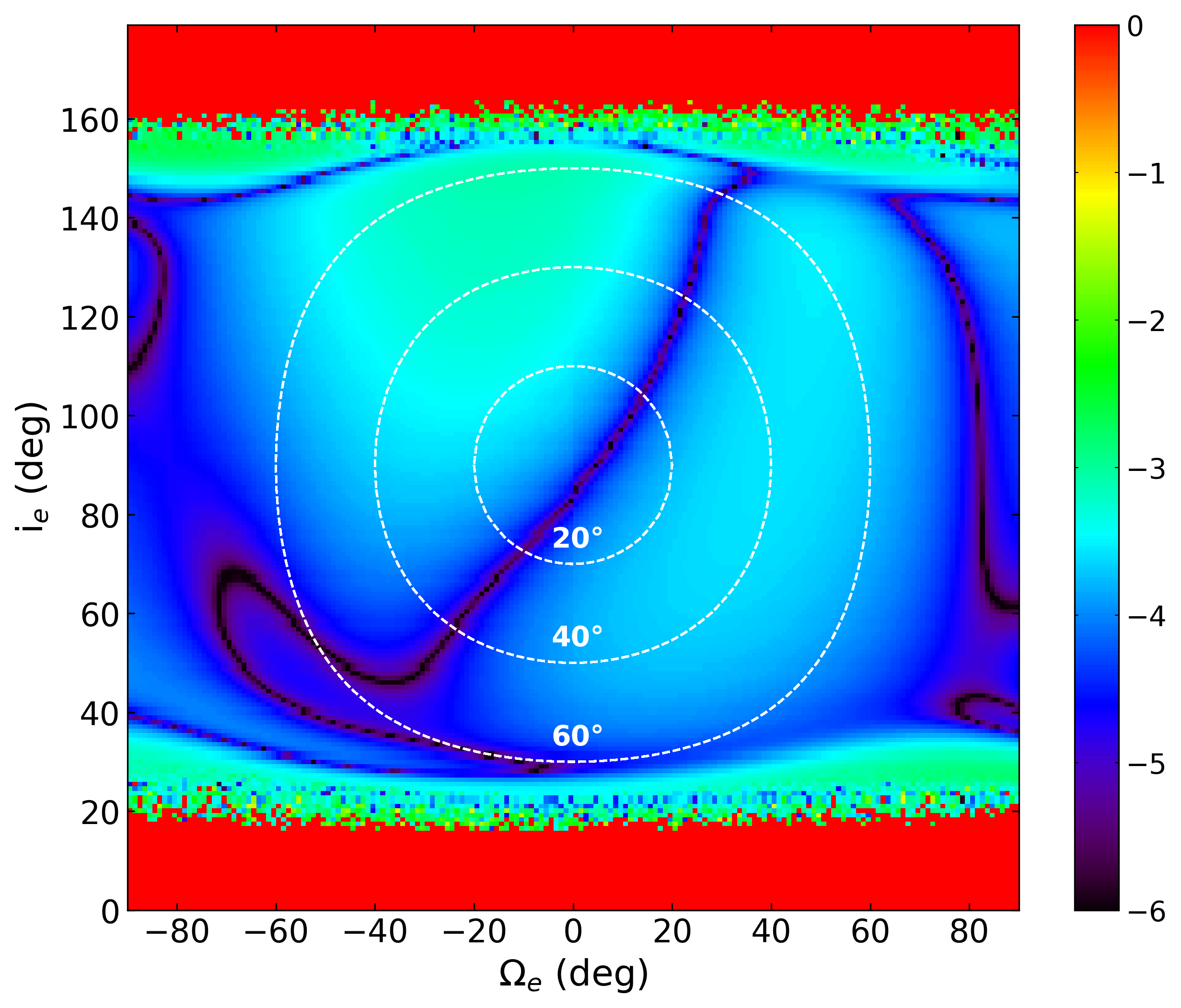}
\caption{Stability analysis of the \sname system in the $(\Omega, i)$ plane for planets~c and e, where $\Omega$ and $i$ represent the longitude of ascending node and inclination of each planet, respectively. For each planet, the phase space is explored by varying these two parameters, while keeping all other orbital elements fixed at their nominal values. As in Figs.~\ref{fig:stability_analysis}~and~\ref{fig:stability_analysis_periods}, the color scale corresponds to the decimal logarithm of the stability index $D$. The red zones correspond to highly unstable orbits, while the dark blue regions can be assumed to be stable on a billion-year timescale. The white curves denote levels of constant mutual inclination between each planet and the plane of the sky, with contours labeled at $10^\circ$, $20^\circ$, $30^\circ$, and $40^\circ$ for planet c, and at  $20^\circ$, $40^\circ$, and $60^\circ$ for planet e. }
\label{fig:stability_analysis_inclination}
\end{figure}

Figure~\ref{fig:stability_analysis_inclination} explores the stability in the $(\Omega, i)$ domain for planets~c and e, while keeping the remaining parameters fixed at the values listed in Tables~\ref{tab:modelparameters} and \ref{tab:derivedparameters}. For planet~c, the results indicate that the system remains stable only within certain ranges of $(\Omega_\mathrm{c},i_\mathrm{c})$ values, specifically for $50^\circ \lesssim i_\mathrm{c} \lesssim 130^\circ$ and $\vert\,\Omega_\mathrm{c}\,\vert \lesssim 40^\circ$. These dynamical constraints allow us to further limit the true mass of planet c, despite the absence of transit detection. Given the minimum mass of $M_\mathrm{c} \sin i_\mathrm{c} \approx 5.5~M_\oplus$ and our inclination constraint of $50^\circ \lesssim i_\mathrm{c} \lesssim 130^\circ$, we derived an upper limit for the mass of \sname~c of $M_\mathrm{c} \lesssim 7.2~M_\oplus$. For planet~e, the stability region extends over a wider inclination range, specifically between $30^\circ$ and $150^\circ$. Using the minimum mass of $M_\mathrm{e} \sin i_\mathrm{e} \approx 42.1~M_\oplus$ and our inclination constraint of $30^\circ \lesssim i_\mathrm{e} \lesssim 150^\circ$, we limited the mass of \sname~e to be $M_\mathrm{e} \lesssim 84.2~M_\oplus$. In Fig.~\ref{fig:stability_analysis_inclination}, we also plot the contours of mutual inclination between each planet and the line-of-sight, defined by $\cos I = \sin i \cos \Omega$. We observe that for \sname~c all stable areas correspond to $ I \lesssim 40^\circ$, while for \sname~e, stability is maintained for mutual inclinations up to $60^\circ$. This effectively sets an upper bound for the inclination of each planet's orbital plane relative to the line-of-sight. 

\begin{figure*}[t!]
    \centering
    \includegraphics[width=18cm]{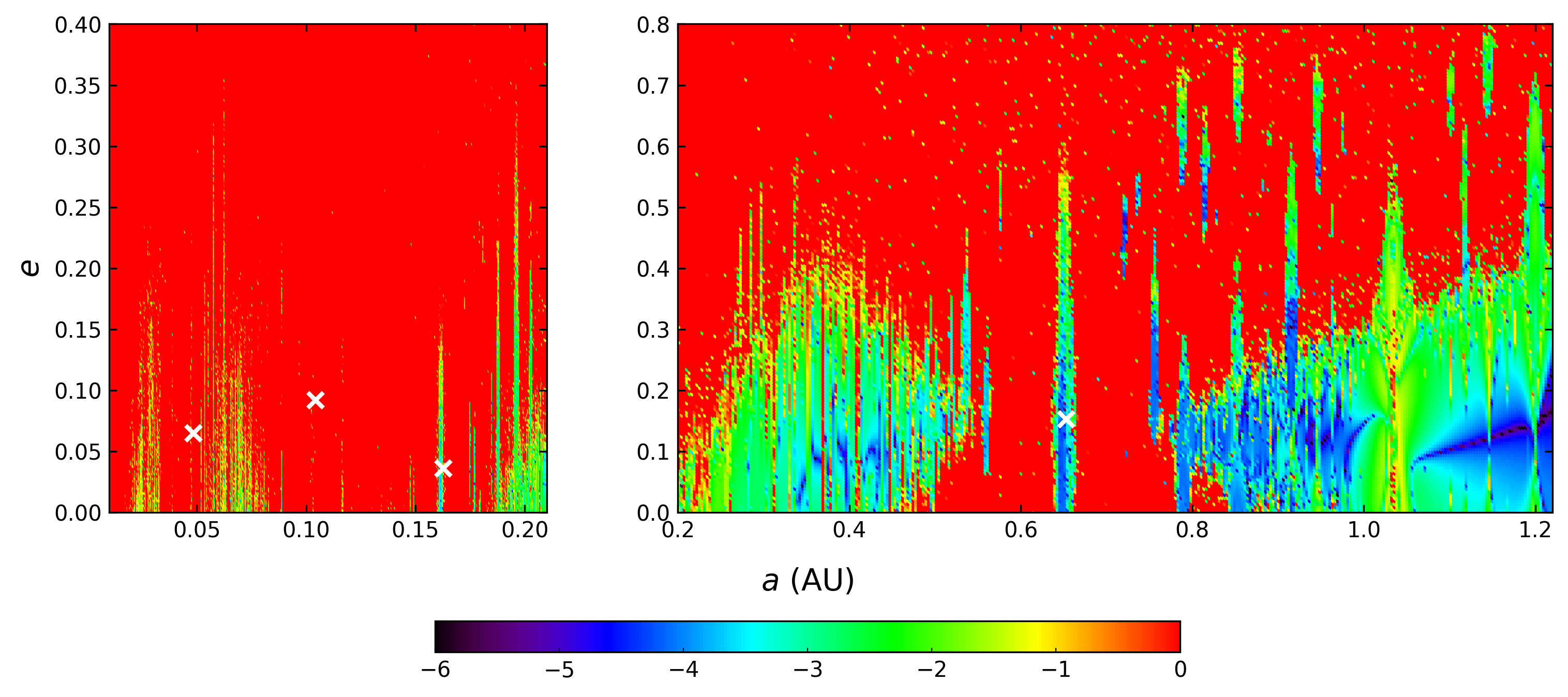}
    \caption{Possible location of an additional fifth planet in the \sname system. The stability of an Earth-size planet ($K=0.1$~\ms) is analyzed for various semimajor axes vs. eccentricity. All the angles of the additional fifth planet are set to $0^\circ$, except for the inclination ($90^\circ$). As in Figs.~\ref{fig:stability_analysis}, \ref{fig:stability_analysis_periods}, and \ref{fig:stability_analysis_inclination} the color scale corresponds to the decimal logarithm of the stability index $D$. The red zones correspond to highly unstable orbits, while the dark blue regions can be assumed to be stable on a billion-year timescale. The stable zones where additional planets could be found are the dark blue regions. The white crosses mark the position of the four planets.}
    \label{fig:stability_analysis_mass}
\end{figure*}

Finally, we tested the possibility of an additional fifth planet in the \sname system, assuming that the dynamics of the four known planets would not be perturbed much by the presence of this extra planet. To investigate this, we varied its semimajor axis and the eccentricity over a wide range, and performed a stability analysis (Fig.~\ref{fig:stability_analysis_mass}). The test was performed with a fixed value of $K=0.1$~\ms, meaning that as we varied the semimajor axis, the mass of the fifth planet adjusted accordingly. For example, with this $K$ value, an object at a distance of $1$~au, would have a mass similar to that of Earth. A similar analysis was also conducted for a lower-mass object ($K=10^{-3}$~\ms) and a higher-mass object ($K=1$~\ms), finding no significant changes in the dynamics of the system. The stability map (Fig.~\ref{fig:stability_analysis_mass}), reveals that stable orbits are possible beyond 1~au (outside the outermost planet’s orbit), which corresponds to orbital periods longer than $400$~d. Additional stable zones appear around $0.3$~au, corresponding to periods between approximately $100$ and $200$~d, located between planets d and e. A smaller region of stability is also evident near $0.08$~au, or around 6~d, located between planets b and c. Interestingly, we also observe two narrow regions of stability around $0.16$~au and $0.65$~au, which coincide with the orbital distances of planets d and e respectively: they correspond to possible co-orbital resonances with these planets.

\subsection{Impact of the stellar companion on the dynamics}

The star \sname hosts a co-moving stellar companion at a projected separation of $\sim$765~au with an estimated mass of $\sim$0.08~$M_\odot$ \citep{Mugrauer20}. The two stars share the same motion in the sky and are relatively close, indicating that they could be gravitationally bound or could have been so in the past. To investigate the potential influence of the stellar companion on the system’s stability and the observed non-zero eccentricity of planet e, we performed numerical simulations of the entire \sname system with and without the companion star.

In these simulations, the eccentricities of all planets were initially set to zero while the remaining parameters were kept fixed at the values listed in Tables~\ref{tab:modelparameters} and \ref{tab:derivedparameters}. The stellar companion was placed at a projected separation of 765~au with a high orbital eccentricity (e\,=\,0.5) and an orbital inclination of $40^\circ$ relative to the plane of the sky (corresponding to a high mutual inclination of $\sim$50$^\circ$ with respect to the planetary orbits). This configuration was chosen to maximize potential perturbations from the companion star, which could amplify the eccentricity oscillations of the planets. The system was integrated over 10~Myr, and the evolution of the eccentricities of all planets was monitored. Table~\ref{tab:eccentricity_amplitudes} presents the maximum eccentricity amplitudes recorded for each planet in both scenarios.

\begin{table}[!t]
\centering
\caption{Maximum eccentricity amplitudes for the planets in the \sname system over a 10~Myr integration, for simulations with and without the stellar companion.}
\label{tab:eccentricity_amplitudes}
\begin{tabular}{c c c}
\hline
\hline
\noalign{\smallskip}
Planet & No companion star & With companion star\\
\noalign{\smallskip}
\hline
\noalign{\smallskip}
\sname~b & 0.00043 & 0.00048 \\
\sname~c & 0.00167 & 0.00182 \\
\sname~d & 0.00077 & 0.00085 \\
\sname~e & 0.00044 & 0.00047 \\
\noalign{\smallskip}
\hline
\end{tabular}
\end{table}

The results in Table~\ref{tab:eccentricity_amplitudes} show that for all planets, the maximum amplitude of eccentricity oscillations increased only marginally with the addition of the stellar companion. This suggests that the stellar companion has minimal impact on the system’s dynamics and thus on its stability. It also cannot account for the observed non-zero eccentricity of \sname~e, which likely originates from other mechanisms, such as disk interactions during the system’s formation, planet-planet interactions, or migration effects. 

Finally, it is reasonable to expect that in the past the stellar companion could have been closer to the primary star. To test the outcome of this possibility, we ran an additional simulation with the companion star placed at a closer separation of 200~au instead of 765~au. Even at this close distance, the companion star exhibited no significant influence on the dynamics of the planets. This suggests that the stellar companion likely played a negligible role in shaping the system’s current architecture through gravitational interactions with the planets, due to its relatively low mass ($\sim$0.08\,$M_\odot$) and wide separation ($\sim$765~au). The planets in the system likely formed farther out and then migrated inwards due to their interactions with the primordial disk, and it is unlikely that the stellar companion played a role in driving this migration. During the early stages, the companion may have slightly excited the eccentricities of the planets, but any such excitation would have been efficiently damped by the disk. Its potential impact on the protoplanetary disk and planetary formation processes, however, requires further dedicated studies.

\section{Internal structure analysis of \sname~b and d}
\label{sec:InternalStructureAnalysis}

The high precision we achieved on the masses and radii of \sname~b and d allows us to infer possible internal compositions and structures for the two transiting planets. To this end, we applied the internal structure modeling framework \texttt{plaNETic} \citep{Egger2024}, which uses the planetary structure model of the \texttt{BICEPS} code \citep{Haldemann2024} as a forward model. However, instead of applying the forward model directly, \texttt{plaNETic} uses neural networks as a replacement in the adopted full-grid accept-reject inverse scheme, thereby significantly speeding up the computation time and allowing us to probe different prior options. In this framework, each planet is modeled as a combination of an inner core (iron with up to 19\,\% sulfur), mantle (oxidized silicon, magnesium, and iron), and a volatile layer (water and hydrogen-helium uniformly mixed).

We ran six different models, assuming different priors. First, we assumed a prior compatible with a high-metallicity atmosphere (option~A), which would, for instance, be plausible if the planets formed outside the ice line of the system. Alternatively, we also considered a prior compatible with a scenario where the planets formed inside the ice line and water could only be accreted through the accreted gas (option~B). For both of these cases, we used three different priors for the planetary Si/Mg/Fe ratio, assuming it to be stellar \citep[options~A1 and B1; see, e.g.,][]{Thiabaud2015}, iron-enriched compared to the host star \citep[options~A2 and B2; see, e.g.,][]{Adibekyan2021}, and allowing uniform sampling from a simplex with an upper limit of 0.75 for iron (options~A3 and B3). A detailed description of \texttt{plaNETic} and the adopted priors can be found in \cite{Egger2024}.

\begin{figure*}[th!]
\centering
\includegraphics[width=\textwidth]{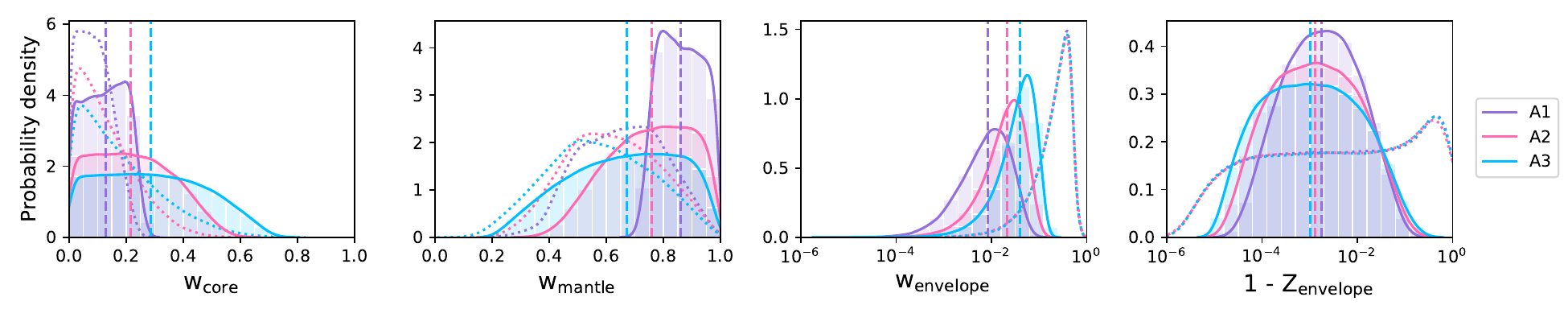}
\caption{Posterior distributions for the inferred internal structure of \sname~b. Shown are the mass fractions of the inner core (w$_\mathrm{core}$), mantle (w$_\mathrm{mantle}$), and envelope (w$_\mathrm{envelope}$), as well as the mass fraction of hydrogen and helium in the volatile layer ($1\,-$~Z$_\mathrm{envelope}$), where Z$_\mathrm{envelope}$ refers to the mass fraction of water in the envelope. The distributions were generated with a prior assuming a water-rich composition (option~A). The density of \sname~b is not compatible with a water-poor prior and envelope mass fractions larger than 10$^{-6}$ (option~B, not depicted). The different colors showcase models that were run assuming different priors on the planetary Si/Mg/Fe ratio. The dashed vertical lines mark the medians of the posteriors, while the dotted lines show the adopted priors.
\label{fig:TOI-1203b_intstruct}
}
\end{figure*}

\begin{figure*}[th!]
\centering
\includegraphics[width=\textwidth]{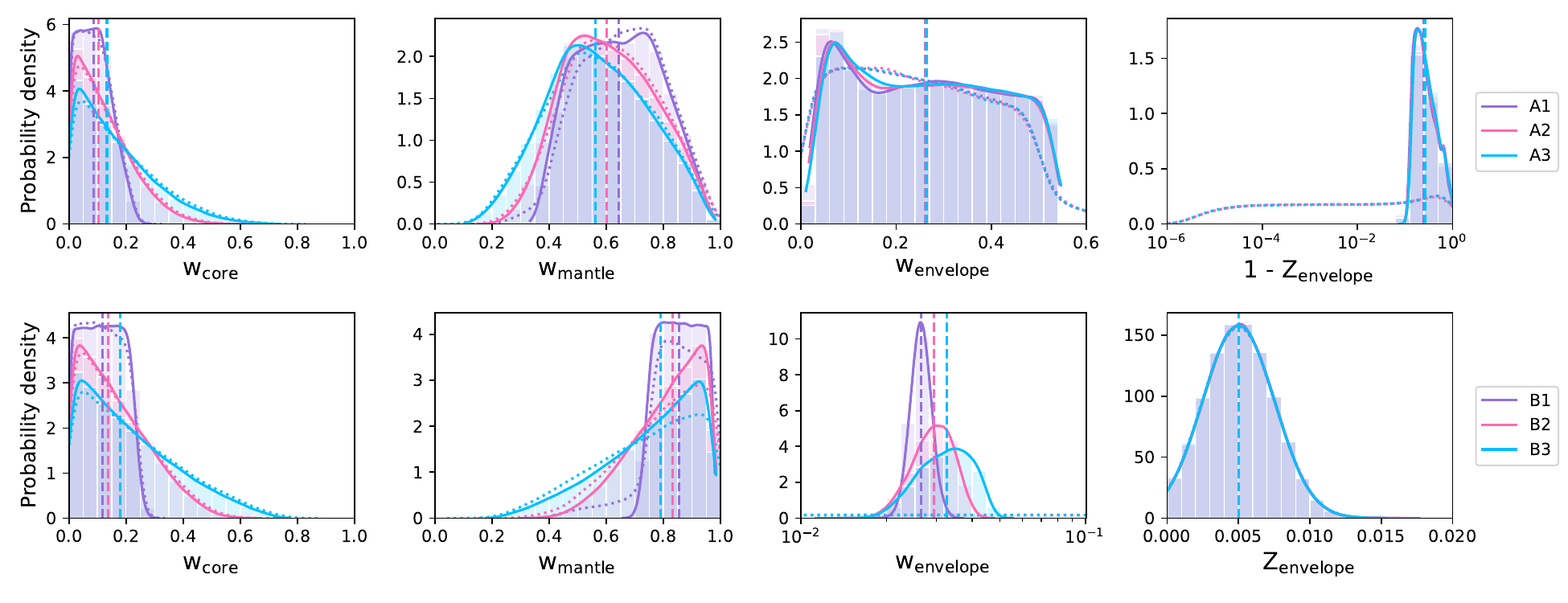}
\caption{Same as Fig.~\ref{fig:TOI-1203b_intstruct}, but for \sname~d. The bottom panels shows models that were generated assuming a water-poor prior, consistent with a formation inside the iceline (option~B).
\label{fig:TOI-1203d_intstruct}
}
\end{figure*}

The results of our internal structure analysis are presented in Figs.~\ref{fig:TOI-1203b_intstruct} and \ref{fig:TOI-1203d_intstruct}. For the water-rich case (option~A), our models predict that \sname~b might possess a water-rich volatile layer of up to a few percent of the planetary mass (Fig.~\ref{fig:TOI-1203b_intstruct}). Assuming a water-poor composition (option~B; not depicted), we found that the high mean density of the planet implies a hydrogen-helium mass fraction that does not exceed 10$^{-6}$. As a result of photoevaporation processes, such a low-mass H and He envelope would not remain stable throughout the system's lifespan, and \sname~b would ultimately be a bare core. On the other hand, \sname~d is compatible with a wide range of possible envelope mass fractions and envelope water mass fractions up to $\sim$90\,\% (option~A). For a water-poor composition (option~B), we found that the planet's density is compatible with envelope mass fractions of a few percent (Fig.~\ref{fig:TOI-1203d_intstruct}).

\section{Conclusions}
\label{Sec:Conclusions}

\begin{figure}[ht!]
\centering
\includegraphics[width=\linewidth]{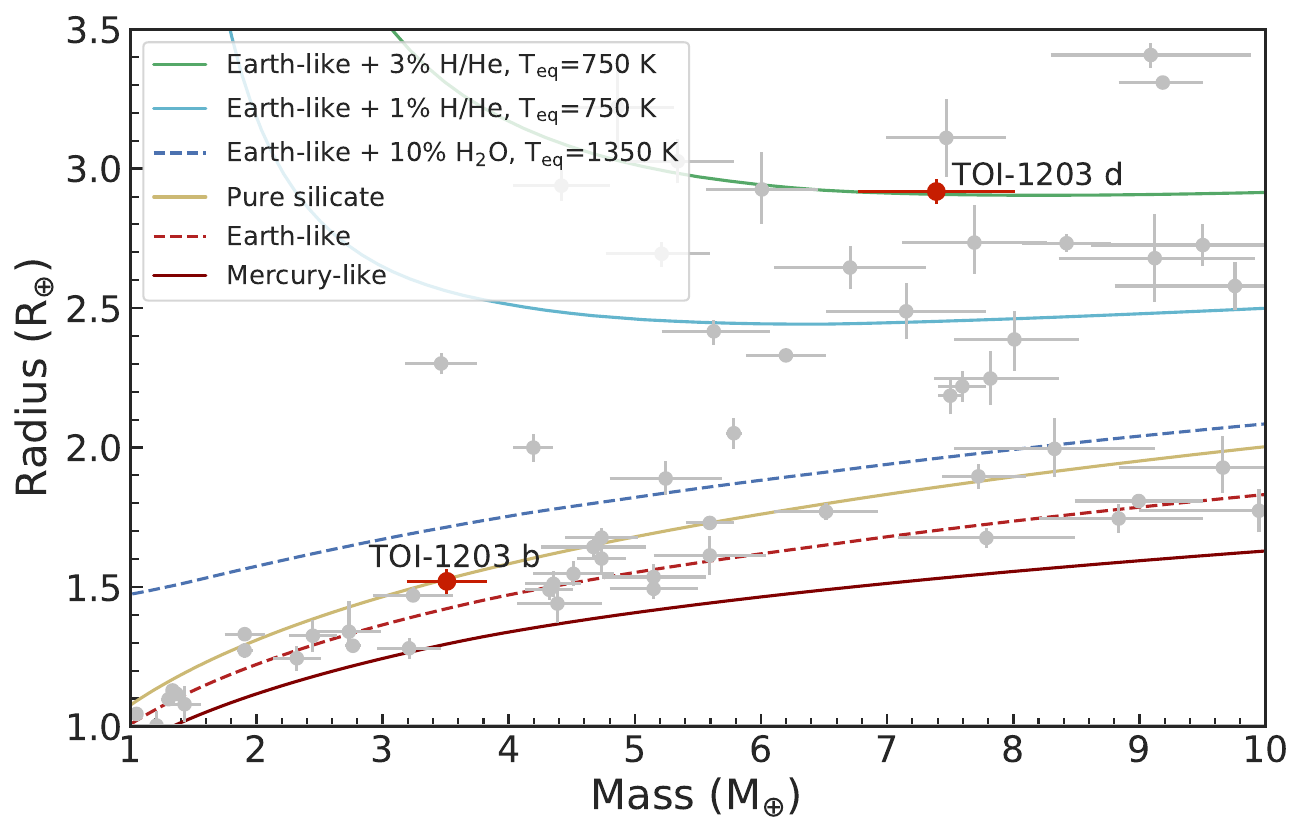}
    \caption{Mass-radius diagram for small planets (1\,$\le$\,$R_\mathrm{p}$\,$\le$\,3.5\,$R_{\oplus}$, 1\,$\le$\,$M_\mathrm{p}$\,$<$\,10\,$M_{\oplus}$) with mass and radius determinations better than 10\,\%, as retrieved from the Transiting Extrasolar Planet Catalogue \citep[TEPCat,][as of July 2025]{2011Southworth}. \sname~b and d are highlighted with red circles. The solid and dashed curves are \texttt{BICEPS} theoretical models \citep{Haldemann2024} for planets with a Mercury-like composition (brown thick line); Earth-like composition (red dashed line); 100\,\% silicate composition (beige thick line); Earth-like composition + 10\,\% H$_2$O with an equilibrium temperature of $T_\mathrm{eq}$\,=\,1350~K (blue dashed line); Earth-like composition + 1\,\% H/He with an equilibrium temperature of $T_\mathrm{eq}$\,=\,750~K (light blue thick line); Earth-like composition + 3\,\% H/He with an equilibrium temperature of $T_\mathrm{eq}$\,=\,750~K (green thick line).}
    \label{Fig:Mass_Radius_Diagram}
\end{figure}

In the present paper, we report the discovery and characterization of a four-planet system orbiting the old, thick disk, G-type star \sname. Our intensive Doppler follow-up conducted with the \harps spectrograph led to:
\begin{itemize}
    \item The spectroscopic confirmation of the transiting warm sub-Neptune \sname~d ($P_\mathrm{orb,d}$\,$\approx$\,25.5~d), which was previously identified in the \tess\ light curve by \citet{Guerrero2021} and subsequently validated by \citet{Giacalone2021}.
    
    \item The discovery of \sname~b, an inner low-mass small planet on a $\sim$4.2-day orbit. An independent transit search of the \tess\ light curve allowed us to detect a potential transit signal at the period found by \harps. Our high-precision photometric follow-up with the \cheops\ space telescope led to the confirmation of the transit signal at a very high significance level, while refining the radius of the planet.
    
    \item The discovery of a third low-mass planet (\sname~c) with an orbital period of $P_\mathrm{orb,c}$\,$\approx$\,13.1~d and a minimum mass of $M_\mathrm{c}\,\mathrm{sin\,i}_\mathrm{c}$\,=\,\mpc.

    \item The discovery of a fourth outer planet (\sname~e) on a long-period ($P_\mathrm{orb,e}$\,$\approx$\,204.6~d), eccentric ($e$\,=\,\ee) orbit with a minimum mass of $M_\mathrm{e}\,\mathrm{sin\,i}_\mathrm{e}$\,=\,\mpe.
\end{itemize}

The close-in transiting planet \sname~b ($P_\mathrm{orb,b}$\,$\approx$\,4.2~d) has a mass of $M_\mathrm{b}$\,=\,\mpb\ ($\sim$9\,\% relative precision) and a radius of $R_\mathrm{b}$\,=\,\rpb\ ($\sim$\,3\% relative precision), implying a bulk density of $\rho_\mathrm{b}$\,=\,\denpb ($\sim$13\,\% relative precision). Figure~\ref{Fig:Mass_Radius_Diagram} shows the position of \sname~b in the mass-radius diagram compared to the subsample of small (1\,$\le$\,$R_\mathrm{p}$\,$\le$\,3.5\,$R_\oplus$), low-mass (1\,$\le$\,$M_\mathrm{p}$\,$\le$\,10\,$M_\oplus$) planets whose radii and masses are known with a precision better than 10\,\% (as of July 2025). \texttt{BICEPS} theoretical models \citep{Haldemann2024} are overplotted with different lines and colors. \sname~b joins the relatively small number of small low-mass transiting planets with a likely rocky composition, whose masses and radii have been precisely measured. The properties of these small rocky planets were recently shown to depend on stellar age, with planets orbiting older stars having a larger fraction of Mg and Si relative to younger stars, which host more Fe-rich planets \citep{Weeks2025}. With an age of $12.5^{+1.3}_{-2.4}$~Gyr, \sname~b is a particularly old planet. According to \citet{Weeks2025}, the core of such planets should consist mainly of Mg and Si, with only about 20\,\% Fe. As illustrated in Fig.~\ref{Fig:Mass_Radius_Diagram}, the position of \sname~b in the mass-radius diagram suggests that its composition is close to that of a pure silicate planet, a scenario that agrees with the posterior distributions shown in Fig.~\ref{fig:TOI-1203b_intstruct}. This is consistent with the trend identified in \citet{Weeks2025} that older stars host planets with a relatively large Mg/Si fraction. With a mass of $M_\mathrm{d}$\,=\,\mpd\ ($\sim$8\,\% relative precision) and a radius of $R_\mathrm{d}$\,=\,\rpd\ ($\sim$1.5\,\% relative precision), the transiting warm mini-Neptune \sname~d ($P_\mathrm{orb,d}$\,$\approx$\,25.5~d) has a mean density of $\rho_\mathrm{d}$\,=\,\denpd\ ($\sim$10\,\% relative precision), which suggests a structure comprising a solid core surrounded by a volatile layer (Fig.~\ref{Fig:Mass_Radius_Diagram}).

Intriguingly, \sname is likely a binary star system (Sect.~\ref{sec_binarity}) that contains two transiting S-type planets that orbit the primary star and span the radius valley. By analyzing \kepler \citep{Borucki2010} S-type planet candidates, \citet{Sullivan24} have recently reported that the location of the radius valley decreases compared to single stars for systems with binary separations larger than 300~au. Hence, characterizing systems like \sname, which has a sky-projected binary separation of $\sim$765~au, is key to confirm or disprove the trend found in \kepler data.

\begin{figure}[t!]
\centering
\includegraphics[width=\linewidth]{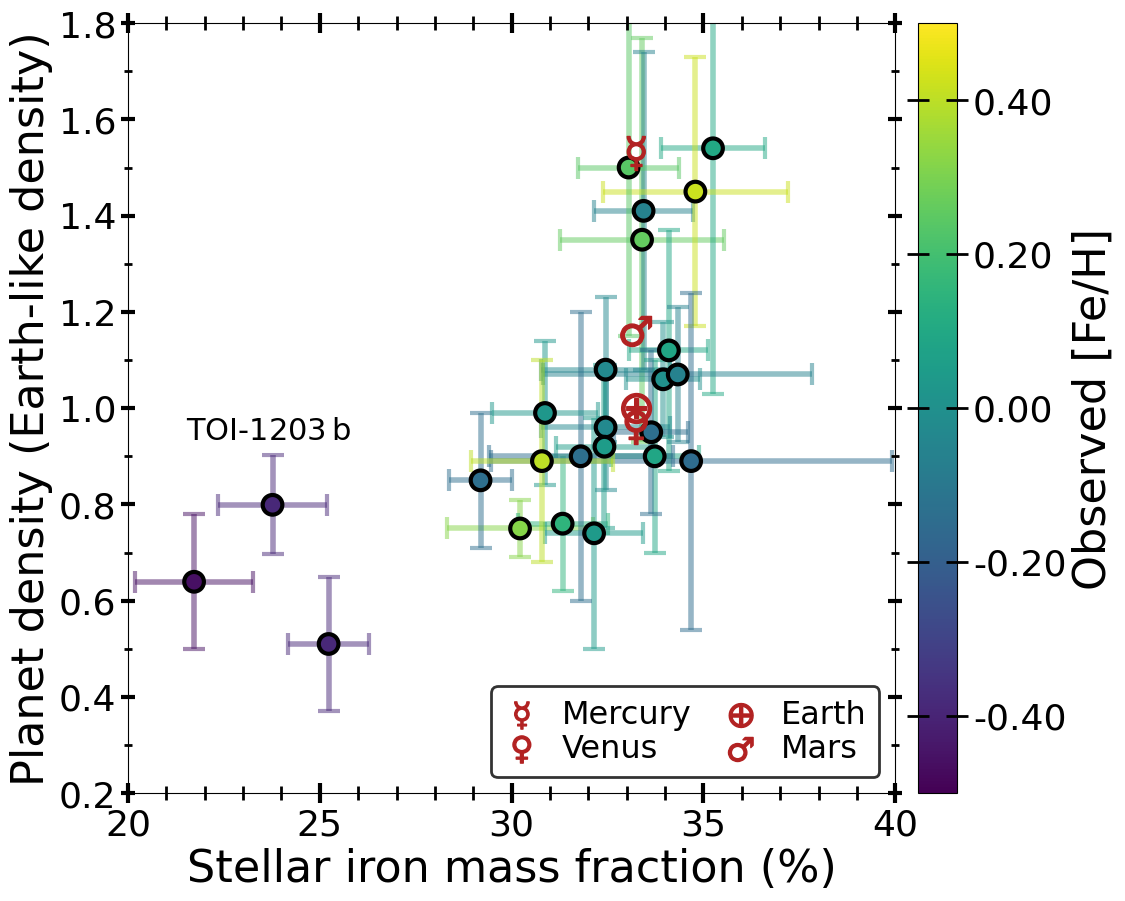}
    \caption{Planetary density vs. stellar iron mass fraction for the planetary systems presented in \citet{Adibekyan2021} with the inclusion of K2-111~b \citep{Mortier2020}. The positions of the Solar System's rocky planets are highlighted with different symbols.} \label{Fig:PlanetDensity_vs_StellarIronMassFraction}
\end{figure}

\sname is an extremely interesting old benchmark planetary system. Having an age of $\sim$12.5~Gyr, this system was formed very early in the history of our galaxy. With [Fe/H]\,=\,$-0.39\,\pm\,0.04$ and $[\alpha/\mathrm{Fe}]$\,=\,0.21\,$\pm$\,0.04, this star was probably among the most metal rich and $\alpha$-element rich stars at that cosmic epoch. Thus, \sname may well be one of the earliest planetary systems to have formed in the Milky Way. The large abundance of $\alpha$ elements would have formed the minerals for the planetary mantles. This is quite consistent with the small core and large mantle mass fraction we found in our interior modeling of \sname~b (Fig.~\ref{fig:TOI-1203b_intstruct}). This type of interior structure is probably typical for rocky planets found around $\alpha$-element rich thick disk and halo stars \citep{Santos2017,Adibekyan2021}. To highlight this, in Fig.~\ref{Fig:PlanetDensity_vs_StellarIronMassFraction} we plot the normalized Earth-like planet density of \sname~b vs. the stellar iron fraction of \sname, alongside the sample of planetary systems presented in \citet{Adibekyan2021} with the inclusion of \object{K2-111~b} \citep{Mortier2020}. This host star metallicity-planet composition link may also extend to watery worlds \citep{Adibekyan2021} and gaseous mini-Neptunes \citep{Wilson2022}, and could influence the atmospheric evolution of giant planets \citep{Mantovan2024b}. Interestingly, combined chemical Galactic and proto-planetary simulations hint that rocky Earth-mass planets should be more common around $\alpha$-element enhanced thick disk stars, as the low [Fe/H] content of these disks would inhibit giant Neptune and Jupiter-sized planet formation \citep{Nielsen2023}.\\

An important result of our study is the apparent instability of the \harps spectrograph. Both the FWHM of the CCF and the line widths of Th-Ar emission lines show a periodic signal at 615-630~d (Sect.~\ref{Sec:HARPS_FrequencyAnalysis}). The most likely source is a slight change in the instrumental profile (IP). \harps was specifically designed to search for extrasolar planets orbiting FGKM stars \citep{Mayor2003} and therefore it delivers exquisite long-term RV stability and precision. Over its more than 20 years of use, \harps has proven its stability \citep[see, e.g.,][]{Pepe2004,Bouchy2009,Dumusque2011,Lovis2011,LoCurto2013,Anglada-Escude2016,Udry2019}, but even the most stable spectrographs have instabilities at some level. Fortunately, in this case the IP variations do not seem to have a detectable effect on the measured RVs of \sname. This indicates that the changes in the IP are largely symmetric, since a change in the asymmetry of the IP is needed to produce an instrumental RV shift.

For \harps the changing IP may not influence the RV shifts at the precision of $\sim$1~\ms, but it might at the $\sim$10~\cms. The latter is the measurement precision needed to detect an Earth-mass planet in the habitable zone of G-type main sequence stars. Significant efforts have focused on improving the wavelength calibration needed to achieve this precision in new generation spectrographs \citep[see, e.g.,][]{Pepe2021}. It should be emphasized, however, that the RV measurement uncertainty that stems from photon noise and wavelength calibration is just one component; a changing IP can represent a significant contribution to the error budget. Radial velocity instrumental shifts caused by changes in the IP not only decrease the measurement precision, but they could also mimic a planet signal. Using our case as an example, if the changes in the \harps IP produced an RV instrumental shift of 10~\cms\ with a period of $\sim$620~d, this would be interpreted as arising from a $\sim$1.3~$M_\oplus$-mass object at $\sim$1.42~au from a 1~$M_\odot$-mass star. If these changes were also seen in the IP then the planet hypothesis could be immediately ruled out.

Currently, the iodine method is the most extensively used technique for measuring the IP \citep{Valenti1995}. Although the \harps spectrograph was initially equipped with an iodine cell to measure the IP, it unfortunately broke shortly after \harps was commissioned and was not replaced. This has made it difficult to assess any changes of the \harps IP. Alternatively, one can use the laser peaks of the Laser Frequency Comb, as these are unresolved by most high-resolution spectrographs and can be treated as delta functions. They thus carry information on the shape of the IP. In any case, it is essential for any spectrograph designed for ultra-precise RV measurements to have the ability to monitor the IP to ensure that its temporal changes are not introducing instrumental shifts. For the detection of an Earth analog, this requires superb instrument stability over a decade, which may be hard to achieve in practice. Thus, monitoring the IP shape will be essential.

\bibliographystyle{aa} 
\bibliography{biblio}

\begin{appendix}

\section{Acknowledgments}
\small{
We thank the anonymous referee for their thorough review, favorable feedback, and constructive comments and remarks, which have improved the quality of our manuscript.
We are extremely grateful to the ESO staff members for their unique and superb support during the observations.
D.G. would like to acknowledge the inspiring discussions with Gaspare Lo Curto, Xavier Dumusque, and Guillem Anglada.
\cheops\ is an ESA mission in partnership with Switzerland with important contributions to the payload and the ground segment from Austria, Belgium, France, Germany, Hungary, Italy, Portugal, Spain, Sweden, and the United Kingdom. The \cheops Consortium would like to gratefully acknowledge the support received by all the agencies, offices, universities, and industries involved. Their flexibility and willingness to explore new approaches were essential to the success of this mission. The \cheops raw data analyzed in this article are made available to the public and the astronomical community through the \cheops mission archive (\url{https://cheops.unige.ch/archive_browser/}). 
This paper made use of data collected by the TESS mission and are publicly available from the Mikulski Archive for Space Telescopes (MAST) operated by the Space Telescope Science Institute (STScI). Funding for the \tess\ mission is provided by the NASA's Science Mission Directorate. We acknowledge the use of public \tess data from pipelines at the \tess Science Office and at the \tess Science Processing Operations Center (SPOC). Resources supporting this work were provided by the NASA High-End Computing (HEC) Program through the NASA Advanced Supercomputing (NAS) Division at Ames Research Center for the production of the SPOC data products.
This work was granted access to the HPC resources of MesoPSL financed by the Region Île-de-France and the project Equip@Meso (reference ANR-10-EQPX-29-01) of the  programme Investissements d’Avenir supervised by the Agence Nationale pour la Recherche. 
This work has been carried out within the framework of the National Centre of Competence in Research (NCCR) PlanetS supported by the Swiss National Science Foundation under grants 51NF40\_182901 and 51NF40\_205606.
D.Ga. and L.M.Se. gratefully acknowledges financial support from the CRT foundation under Grant No. 2018.2323 ``Gaseous or rocky? Unveiling the nature of small worlds''. 
K.W.F.La. was supported by Deutsche Forschungsgemeinschaft grants RA714/14-1 within the DFG Schwerpunkt SPP 1992, Exploring the Diversity of Extrasolar Planets. 
M.Fr. and C.M.Pa. gratefully acknowledge the support of the Swedish National Space Agency (DNR 65/19, 174/18). 
T.Wi. acknowledges support from the UKSA and the University of Warwick. 
A.Br. was supported by the SNSA. 
S.G.So. acknowledge support from FCT through FCT contract nr. CEECIND/00826/2018 and POPH/FSE (EC). 
The Portuguese team thanks the Portuguese Space Agency for the provision of financial support in the framework of the PRODEX Programme of the European Space Agency (ESA) under contract number 4000142255. 
J.Ve. acknowledges support from the Swiss National Science Foundation (SNSF) under grant PZ00P2\_208945. 
Y.Al. acknowledges support from the Swiss National Science Foundation (SNSF) under grant 200020\_192038. 
A.C.M.Co. acknowledges support from the FCT, Portugal, through the CFisUC projects UIDB/04564/2020 and UIDP/04564/2020, with DOI identifiers 10.54499/UIDB/04564/2020 and 10.54499/UIDP/04564/2020, respectively. 
The Belgian participation to \cheops\ has been supported by the Belgian Federal Science Policy Office (BELSPO) in the framework of the PRODEX Program, and by the University of Liège through an ARC grant for Concerted Research Actions financed by the Wallonia-Brussels Federation. 
L.De. thanks the Belgian Federal Science Policy Office (BELSPO) for the provision of financial support in the framework of the PRODEX Programme of the European Space Agency (ESA) under contract number 4000142531. 
D.Ba., H.J.De, E.Pa., and I.Ri. acknowledge financial support from the Agencia Estatal de Investigaci\'on of the Ministerio de Ciencia e Innovaci\'on MCIN/AEI/10.13039/501100011033 and the ERDF ``A way of making Europe'' through projects PID2019-107061GB-C61, PID2019-107061GB-C66, PID2021-125627OB-C31, and PID2021-125627OB-C32, from the Centre of Excellence ``Severo Ochoa'' award to the Instituto de Astrofísica de Canarias (CEX2019-000920-S), from the Centre of Excellence ``Mar\'ia de Maeztu'' award to the Institut de Ci\`encies de l’Espai (CEX2020-001058-M), and from the Generalitat de Catalunya/CERCA programme. 
S.C.C.Ba. acknowledges the support from Fundação para a Ciência e Tecnologia (FCT) in the form of work contract through the Scientific Employment Incentive program with reference 2023.06687.CEECIND. 
L.Bo., V.Na., I.Pa., G.Pi., R.Ra., G.Sc., and T.Zi. acknowledge support from \cheops\ ASI-INAF agreement n. 2019-29-HH.0.
C.Br. and A.Si. acknowledge support from the Swiss Space Office through the ESA PRODEX program.
T.Zi. acknowledges NVIDIA Academic Hardware Grant Program for the use of the Titan V GPU card and the Italian MUR Departments of Excellence grant 2023-2027 ``Quantum Frontiers''.
F.Mu. acknowledges the financial support from the Agencia Estatal de Investigaci\'{o}n del Ministerio de Ciencia, Innovaci\'{o}n y Universidades (MCIU/AEI) through grant PID2023-152906NA-I00.
A.Co.Ca. acknowledges support from STFC consolidated grant number ST/V000861/1, and UKSA grant number ST/X002217/1. 
P.E.Cu. is funded by the Austrian Science Fund (FWF) Erwin Schroedinger Fellowship, program J4595-N. 
This project was supported by the CNES. 
This work was supported by FCT - Funda\c{c}\~{a}o para a Ci\^{e}ncia e a Tecnologia through national funds and by FEDER through COMPETE2020 through the research grants UIDB/04434/2020, UIDP/04434/2020, 2022.06962.PTDC. 
O.D.S.De. is supported in the form of work contract (DL 57/2016/CP1364/CT0004) funded by national funds through FCT. 
B.-O.De. acknowledges support from the Swiss State Secretariat for Education, Research and Innovation (SERI) under contract number MB22.00046. 
A.De., B.Ed., K.Ga., and J.Ko. acknowledge their role as ESA-appointed \cheops\ Science Team Members. 
This project has received funding from the Swiss National Science Foundation for project 200021\_200726. The authors acknowledge the financial support of the SNSF. 
M.Gi. is an F.R.S.-FNRS Senior Research Associate. 
M.N.G\"u. is the ESA \cheops Project Scientist and Mission Representative. B.M.M. is the ESA \cheops Project Scientist. K.G.Is. was the ESA \cheops Project Scientist until the end of December 2022 and Mission Representative until the end of January 2023. All of them are/were responsible for the Guest Observers (GO) Programme. None of them relay/relayed proprietary information between the GO and Guaranteed Time Observation (GTO) Programmes, nor do/did they decide on the definition and target selection of the GTO Programme.
C.He. acknowledges the European Union H2020-MSCA-ITN-2019 under GrantAgreement no. 860470 (CHAMELEON), and the HPC facilities at the Vienna Science Cluster (VSC). 
J.Ko. and A.Le. acknowledge support of the Swiss National Science Foundation under grant number TMSGI2\_211697.
M.Le. acknowledges support of the Swiss National Science Foundation under grant number PCEFP2\_194576. 
P.F.L.Ma. acknowledges support from STFC research grant number ST/R000638/1.
This work was also partially supported by a grant from the Simons Foundation (PI: Queloz; grant number 327127). 
N.C.Sa. acknowledges funding by the European Union (ERC, FIERCE, 101052347). Views and opinions expressed are however those of the author(s) only and do not necessarily reflect those of the European Union or the European Research Council. Neither the European Union nor the granting authority can be held responsible for them. 
Gy.M.Sz. acknowledges the support of the Hungarian National Research, Development and Innovation Office (NKFIH) grant K-125015, a PRODEX Experiment Agreement No. 4000137122, the Lend\"ulet LP2018-7/2021 grant of the Hungarian Academy of Science and the support of the city of Szombathely. 
V.V.Gr. is an F.R.S-FNRS Research Associate. 
E.Vi. acknowledges support from the ``DISCOBOLO'' project funded by the Spanish Ministerio de Ciencia, Innovación y Universidades undergrant PID2021-127289NB-I00. 
NAW acknowledges UKSA grant ST/R004838/1.} 

\onecolumn
\section{Additional tables}

\begin{table}[!h]
\centering
\caption{\harps \texttt{DRS} and \terra RV measurements, full width at half maximum (FWHM) and bisector inverse slope (BIS) of the cross-correlation function (CCF), Ca\,{\sc ii} H\,\&\,K lines activity indicator (\logrhk), exposure time, and S/N ratio per pixel at 550\,nm.}
\label{tab:RV-Table}
\begin{tabular}{cccrccrcccc}
\hline
\hline
\noalign{\smallskip}
BJD$_\mathrm{TDB}$ & RV$_\mathrm{DRS}$  & eRV$_\mathrm{DRS}$ &  RV$_\mathrm{TERRA}$  & eRV$_\mathrm{TERRA}$ & CCF FWHM  & CCF BIS & \logrhk & $\sigma$\,\logrhk & T$_\mathrm{exp}$ & S/N \\
$-$2\,400\,000  &       (\kms)       &       (\kms)       &          (\kms)       &           (\kms)     &    (\kms) &  (\kms) &            &        &        (s)       &          \\
\noalign{\smallskip}
\hline
\noalign{\smallskip}
58884.720544 & 72.6570 & 0.0010 &  0.0036 & 0.0011 &  6.7434 & -0.0034 & -4.9818 & 0.0053 & 1500 & 102.2 \\
58886.843220 & 72.6527 & 0.0006 & -0.0016 & 0.0007 &  6.7436 & -0.0005 & -4.9998 & 0.0038 & 1500 & 162.4 \\
58887.760047 & 72.6513 & 0.0006 & -0.0014 & 0.0006 &  6.7451 & -0.0046 & -4.9895 & 0.0032 & 1500 & 156.1 \\
58888.780719 & 72.6540 & 0.0006 & -0.0006 & 0.0007 &  6.7437 & -0.0034 & -5.0009 & 0.0027 & 1500 & 173.5 \\
58889.777891 & 72.6544 & 0.0007 & -0.0000 & 0.0008 &  6.7433 & -0.0022 & -5.0051 & 0.0035 & 1500 & 136.6 \\
58890.788187 & 72.6553 & 0.0008 &  0.0016 & 0.0011 &  6.7509 & -0.0027 & -5.0001 & 0.0044 & 1500 & 113.2 \\
58893.885001 & 72.6579 & 0.0006 &  0.0039 & 0.0008 &  6.7475 & -0.0027 & -5.0135 & 0.0044 & 1500 & 166.8 \\
58894.794273 & 72.6579 & 0.0008 &  0.0037 & 0.0011 &  6.7495 & -0.0003 & -5.0053 & 0.0050 & 1500 & 114.2 \\
58897.859783 & 72.6564 & 0.0008 &  0.0019 & 0.0009 &  6.7446 & -0.0019 & -5.0172 & 0.0055 & 1500 & 129.5 \\
58898.829822 & 72.6576 & 0.0006 &  0.0033 & 0.0008 &  6.7425 & -0.0038 & -5.0096 & 0.0036 & 1500 & 175.2 \\
\ldots & \ldots & \ldots & \ldots & \ldots & \ldots & \ldots & \ldots & \ldots & \ldots & \ldots \\ 
\ldots & \ldots & \ldots & \ldots & \ldots & \ldots & \ldots & \ldots & \ldots & \ldots & \ldots \\ 
\noalign{\smallskip}
\hline
\end{tabular}
\tablefoot{Barycentric Julian dates are given in barycentric dynamical time \citep[BJD$_\mathrm{TDB}$;][]{Eastman2010}. The entire RV data set is available in a machine-readable table at the Strasbourg astronomical Data Center (CDS).}
\end{table}

\begin{table*}[!t]
\begin{center}
\begin{threeparttable}
  \caption{\sname\ model parameters. \label{tab:modelparameters}}  
  \begin{tabular}{lcc}
  \hline\hline
  \noalign{\smallskip}
  Parameter & Prior$^{(\mathrm{a})}$ & Derived value \\
  \noalign{\smallskip}
  \hline
    \noalign{\smallskip}
    \multicolumn{3}{l}{\emph{\bf{Model parameters of \sname~b}}} \\
    \noalign{\smallskip}
    Orbital period $P_{\mathrm{orb,\,b}}$ (d) &  $\mathcal{U}[4.1571 , 4.1575]$ & \Pb[] \\
    \noalign{\smallskip}
    Transit epoch $T_\mathrm{0,\,b}$ (BJD$_\mathrm{TDB}\,-\,$2\,400\,000) & $\mathcal{U}[58545.8385, 58545.9416]$ & \Tzerob[]  \\ 
    \noalign{\smallskip}
    Planet-to-star radius ratio $R_\mathrm{b}/R_{\star}$ & $\mathcal{U}[0.00,0.03]$ & \rrb[]  \\
    \noalign{\smallskip}
    Impact parameter $b_\mathrm{b}$  & $\mathcal{U}[0,1.15]$  & \bb[] \\
    \noalign{\smallskip}
    Radial velocity semi-amplitude variation $K_\mathrm{b}$ (\ms) & $\mathcal{U}[0,10]$ & \kb[] \\
    $\sqrt{e_\mathrm{b}} \sin \omega_\mathrm{\star,\,b}$ &  $\mathcal{U}[-1,1]$ & ~~~\esinb[] \\
    \noalign{\smallskip}
    $\sqrt{e_\mathrm{b}} \cos \omega_\mathrm{\star,\,b}$ &  $\mathcal{U}[-1,1]$ & \ecosb[] \\
    \noalign{\smallskip}
    \hline
    \noalign{\smallskip}
    \multicolumn{3}{l}{\emph{\bf{Model parameters of \sname~c}}} \\
    \noalign{\smallskip}
    Orbital period $P_{\mathrm{orb,\,c}}$ (d) &  $\mathcal{U}[12.90,13.25]$ & \Pc[] \\
    \noalign{\smallskip}
    Time of inferior conjunction $T_\mathrm{inf,\,c}$ (BJD$_\mathrm{TDB}\,-\,$2\,400\,000) & $\mathcal{U}[59193.4, 59204.2]$ & \Tzeroc[]  \\
    \noalign{\smallskip}
    Radial velocity semi-amplitude variation $K_\mathrm{c}$ (\ms) & $\mathcal{U}[0,10]$ & \kc[] \\
    \noalign{\smallskip}
    $\sqrt{e_\mathrm{c}} \sin \omega_\mathrm{\star,\,c}$ &  $\mathcal{U}[-1,1]$ & ~~~\esinc[] \\
    \noalign{\smallskip}
    $\sqrt{e_\mathrm{c}} \cos \omega_\mathrm{\star,\,c}$ &  $\mathcal{U}[-1,1]$ & \ecosc[] \\
    \noalign{\smallskip}
    \hline
    \noalign{\smallskip}
    \multicolumn{3}{l}{\emph{\bf{Model parameters of \sname~d}}} \\
    \noalign{\smallskip}
    Orbital period $P_{\mathrm{orb,\,d}}$ (d) &  $\mathcal{U}[ 25.5015, 25.5040]$ & \Pd[] \\
    \noalign{\smallskip}
    Transit epoch $T_\mathrm{0,\,d}$ (BJD$_\mathrm{TDB}\,-\,$2\,400\,000) & $\mathcal{U}[58553.02, 58553.12]$ & \Tzerod[]  \\ 
    \noalign{\smallskip}
    Planet-to-star radius ratio $R_\mathrm{d}/R_{\star}$ & $\mathcal{U}[0.00,0.05]$ & \rrd[]  \\
    \noalign{\smallskip}
    Impact parameter $b_\mathrm{d}$  & $\mathcal{U}[0,1.15]$  & \bd[] \\
    \noalign{\smallskip}
    Radial velocity semi-amplitude variation $K_\mathrm{d}$ (\ms) & $\mathcal{U}[0,10]$ & \kd[] \\
    \noalign{\smallskip}
    $\sqrt{e_\mathrm{d}} \sin \omega_\mathrm{\star,\,d}$ &  $\mathcal{U}[-1,1]$  & \esind[] \\
    \noalign{\smallskip}
    $\sqrt{e_\mathrm{d}} \cos \omega_\mathrm{\star,\,d}$ &  $\mathcal{U}[-1,1]$  & \ecosd[] \\
    \noalign{\smallskip}
    \hline
    \noalign{\smallskip}
    \multicolumn{3}{l}{\emph{\bf{Model parameters of \sname~e}}} \\
    \noalign{\smallskip}
    Orbital period $P_{\mathrm{orb,\,e}}$ (d) &  $\mathcal{U}[190.6, 218.6]$ & \Pe[] \\
    \noalign{\smallskip}
    Time of inferior conjunction $T_\mathrm{inf,\,e}$ (BJD$_\mathrm{TDB}\,-\,$2\,400\,000) & $\mathcal{U}[58871.0, 58960.0]$ & \Tzeroe[]  \\
    \noalign{\smallskip}
    Radial velocity semi-amplitude variation $K_\mathrm{e}$ (\ms) & $\mathcal{U}[0,10]$ & \ke[] \\
    \noalign{\smallskip}
    $\sqrt{e_\mathrm{e}} \sin \omega_\mathrm{\star,\,e}$ &  $\mathcal{U}[-1,1]$ & \esine[] \\
    \noalign{\smallskip}
    $\sqrt{e_\mathrm{e}} \cos \omega_\mathrm{\star,\,e}$ &  $\mathcal{U}[-1,1]$ & \ecose[] \\
    \noalign{\smallskip}
    \hline    
    \noalign{\smallskip}
    \multicolumn{3}{l}{\emph{\bf{Additional model parameters}}} \\
    \noalign{\smallskip}
    Stellar mean density $\rho_\star$ (\gccm) & $\mathcal{N}[0.763,0.038]$ & \denstrb[] \\ 
    \noalign{\smallskip}
    Parameterized limb-darkening coefficient $q_1$ (\tess)  & $\mathcal{N}[0.32,0.10]$ & \qoneTESS[] \\
    \noalign{\smallskip}    
    Parameterized limb-darkening coefficient $q_2$ (\tess) & $\mathcal{N}[0.25,0.10]$ & \qtwoTESS[] \\
    \noalign{\smallskip}
    Parameterized limb-darkening coefficient $q_1$ (\cheops)  & $\mathcal{N}[0.43,0.10]$ & \qoneCHEOPS[] \\
    \noalign{\smallskip}    
    Parameterized limb-darkening coefficient $q_2$ (\cheops) & $\mathcal{N}[0.30,0.10]$ & \qtwoCHEOPS[] \\
    \noalign{\smallskip}
    Systemic velocity $\gamma$  (m s$^{-1}$) & $\mathcal{U}[-100.0, 100.0]$ & \gammaHARPS[] \\
    \noalign{\smallskip}
    RV jitter term $\sigma_{\mathrm{HARPS}}$  (\ms) & $\mathcal{U}[0,10]$ & \jHARPS[]  \\
    \noalign{\smallskip}
    \hline 
    \end{tabular}
    \begin{tablenotes}\footnotesize
    \item \emph{Note} -- 
                       $^{(\mathrm{a})}$ $\mathcal{U}[a,b]$ refers to uniform priors between $a$ and $b$, whereas $\mathcal{N}[a,b]$ refers to Gaussian priors with mean $a$ and standard deviation $b$.
    \end{tablenotes}
\end{threeparttable}
\end{center}
\end{table*}

\begin{table*}[!t]
\begin{center}
\begin{threeparttable}
  \caption{\sname\ derived parameters. \label{tab:derivedparameters}}  
  \begin{tabular}{lc}
  \hline\hline
  \noalign{\smallskip}
  Parameter & Derived value \\
  \noalign{\smallskip}
  \hline
    \noalign{\smallskip}
    \multicolumn{2}{l}{\emph{\bf{Derived parameters of \sname~b}}} \\
    \noalign{\smallskip}
    Planet mass $M_\mathrm{b}$ ($M_\oplus$) &  \mpb[]  \\
    \noalign{\smallskip}
    Planet radius $R_\mathrm{b}$ ($R_\oplus$) & \rpb[] \\
    \noalign{\smallskip}
    Planet mean density $\rho_\mathrm{b}$ ($\mathrm{g\,cm^{-3}}$)  & \denpb[] \\
    \noalign{\smallskip}
    Scaled semimajor axis $a_\mathrm{b}/R_{\star}$  & \arb[] \\ 
    \noalign{\smallskip}
    Semi-major axis of the planetary orbit $a_\mathrm{b}$ (au)  & \ab[]  \\
    \noalign{\smallskip}
    Orbit inclination $i_\mathrm{b}$ (deg)  & \ib[] \\
    \noalign{\smallskip}
    Orbit eccentricity $e_\mathrm{b}$  & \eb[]  \\
    \noalign{\smallskip}
    Argument of periastron of stellar orbit $\omega_\mathrm{\star,\,b}$ (deg)  & \wb[] \\
    \noalign{\smallskip}
    Time of periastron passage $T_\mathrm{per,\,b}$ (BJD$_\mathrm{TDB}\,-\,$2\,400\,000)  & \Tperib[]  \\
    \noalign{\smallskip}
    Equilibrium temperature$^{(\mathrm{a})}$  $T_\mathrm{eq,\,b}$ (K)   &  \Teqb[] \\
    \noalign{\smallskip}
    Stellar insolation $F_{\star,\mathrm{b}}$ ($F_\oplus$) & \insolationb[] \\
    \noalign{\smallskip}
    Total transit duration (1$^\mathrm{st}$ to 4$^\mathrm{th}$ contact) $\tau_\mathrm{14,\,b}$ (h)  & \ttotb[] \\
    \noalign{\smallskip}
    Full transit duration (2$^\mathrm{nd}$ to 3$^\mathrm{rd}$ contact) $\tau_\mathrm{23,\,b}$ (h)  & \tfulb[] \\
    \noalign{\smallskip}
    \hline 
    \noalign{\smallskip}
    \multicolumn{2}{l}{\emph{\bf{Derived parameters of \sname~c}}} \\
    \noalign{\smallskip}
    Planet minimum mass $M_\mathrm{c} \sin i_\mathrm{c}$ ($M_\oplus$) & \mpc[]  \\
    \noalign{\smallskip}
    Orbit eccentricity $e_\mathrm{c}$ & \ec[]  \\
   \noalign{\smallskip}
    Argument of periastron of stellar orbit $\omega_\mathrm{\star,\,c}$ (deg) & \wc[] \\
    \noalign{\smallskip}
    Time of periastron passage $T_\mathrm{per,\,c}$ (BJD$_\mathrm{TDB}\,-\,$2\,400\,000) & \Tperic[]  \\
    \noalign{\smallskip}
    \hline
    \noalign{\smallskip}
    \multicolumn{2}{l}{\emph{\bf{Derived parameters of \sname~d}}} \\
    \noalign{\smallskip}
    Planet mass $M_\mathrm{d}$ ($M_\oplus$) &  \mpd[]  \\
    \noalign{\smallskip}
    Planet radius $R_\mathrm{d}$ ($R_\oplus$) & \rpd[] \\
    \noalign{\smallskip}
    Planet mean density $\rho_\mathrm{d}$ ($\mathrm{g\,cm^{-3}}$)  & \denpd[] \\
    \noalign{\smallskip}
    Scaled semimajor axis $a_\mathrm{d}/R_{\star}$  & \ard[] \\ 
    \noalign{\smallskip}
    Semi-major axis of the planetary orbit $a_\mathrm{d}$ (au)  & \ad[]  \\
    \noalign{\smallskip}
    Orbit inclination $i_\mathrm{d}$ (deg)  & \id[] \\
    \noalign{\smallskip}
    Orbit eccentricity $e_\mathrm{d}$  & \ed[]  \\
    \noalign{\smallskip}
    Argument of periastron of stellar orbit $\omega_\mathrm{\star,\,d}$ (deg)  & \wdd[] \\
    \noalign{\smallskip}
    Time of periastron passage $T_\mathrm{per,\,d}$ (BJD$_\mathrm{TDB}\,-\,$2\,400\,000)  & \Tperid[]  \\
    \noalign{\smallskip}
    Equilibrium temperature$^{(\mathrm{a})}$  $T_\mathrm{eq,\,d}$ (K)   &  \Teqd[] \\
    \noalign{\smallskip}
    Stellar insolation $F_{\star,\mathrm{d}}$ ($F_\oplus$) & \insolationd[] \\
    \noalign{\smallskip}
    Total transit duration (1$^\mathrm{st}$ to 4$^\mathrm{th}$ contact) $\tau_\mathrm{14,\,d}$ (h)  & \ttotd[] \\
    \noalign{\smallskip}
    Full transit duration (2$^\mathrm{nd}$ to 3$^\mathrm{rd}$ contact) $\tau_\mathrm{23,\,d}$ (h)  & \tfuld[] \\
    \noalign{\smallskip}
    \hline 
    \noalign{\smallskip}
    \multicolumn{2}{l}{\emph{\bf{Derived parameters of \sname~e}}} \\
    \noalign{\smallskip}
    Planet minimum mass $M_\mathrm{e} \sin i_\mathrm{e}$ ($M_\oplus$) & \mpe[]  \\
    \noalign{\smallskip}
    Orbit eccentricity $e_\mathrm{e}$ & \ee[]  \\
    \noalign{\smallskip}
    Argument of periastron of stellar orbit $\omega_\mathrm{\star,\,e}$ (deg) & \we[] \\
    \noalign{\smallskip}
    Time of periastron passage $T_\mathrm{per,\,e}$ (BJD$_\mathrm{TDB}\,-\,$2\,400\,000) & \Tperie[]  \\
    \noalign{\smallskip}
    \hline
    \end{tabular}
    \begin{tablenotes}\footnotesize
    \item \emph{Note} -- 
                       $^{(\mathrm{a})}$ Assuming zero albedo and uniform redistribution of heat.
    \end{tablenotes}
\end{threeparttable}
\end{center}
\end{table*}

\end{appendix}

\end{document}